\pdfoutput=1
\documentclass[bibyear]{aa}
\usepackage[varg]{txfonts}
\usepackage{multicol}
\usepackage{natbib}
\usepackage{rotating}
\usepackage{graphicx}
\usepackage{rotating}
\usepackage{multirow}
\usepackage{url}
\usepackage{amssymb}
\usepackage{comment}
\usepackage{adjustbox}
\usepackage[graphicx]{realboxes}
\usepackage{rotating}
\usepackage{pdflscape}
\bibpunct{(}{)}{;}{a}{}{,} 
\usepackage[flushleft]{threeparttable}
\usepackage{caption}

\newcommand{\zabs}{\ensuremath{z_{\rm abs}}}

\newcommand{\HH}{\mbox{H$\rm _2$}}

\newcommand{\lya}{\mbox{${\rm Ly}\alpha$}}

\newcommand{\ArI}{\ion{Ar}{i}}
\newcommand{\CI}{\ion{C}{i}}
\newcommand{\CII}{\ion{C}{ii}}
\newcommand{\CIV}{\ion{C}{iv}}
\newcommand{\CaII}{\ion{Ca}{ii}}
\newcommand{\NaI}{\ion{Na}{i}}
\newcommand{\ClI}{\ion{Cl}{i}}

\newcommand{\CrII}{\ion{Cr}{ii}}

\newcommand{\FeII}{\ion{Fe}{ii}}
\newcommand{\HI}{\ion{H}{i}}
\newcommand{\MgI}{\ion{Mg}{i}}
\newcommand{\MgII}{\ion{Mg}{ii}}
\newcommand{\MnII}{\ion{Mn}{ii}}
\newcommand{\NI}{\ion{N}{i}}
\newcommand{\NV}{\ion{N}{v}}
\newcommand{\NiII}{\ion{Ni}{ii}}
\newcommand{\OI}{\ion{O}{i}}
\newcommand{\OVI}{\ion{O}{vi}}
\newcommand{\PII}{\ion{P}{ii}}

\newcommand{\SII}{\ion{S}{ii}}
\newcommand{\SiII}{\ion{Si}{ii}}
\newcommand{\SiIV}{\ion{Si}{iv}}

\newcommand{\TiII}{\ion{Ti}{ii}}
\newcommand{\ZnII}{\ion{Zn}{ii}}

\newcommand{\OIII}{\ion{O}{iii}}
\newcommand{\kms}{\ensuremath{{\rm km\,s^{-1}}}}

\newcommand{\deltav}{\ensuremath{{\Delta\,v_{90}}}}
\newcommand{\delv}{\ensuremath{{\Delta\,v}}}

\begin{document}

\title{Multi-phase gas properties of extremely strong intervening DLAs towards quasars \thanks{Based on observations performed with the Very Large Telescope of the European Southern Observatory under Prog.~ID 095.A-0224(A) and 0101.A-0891(A) using the X-shooter spectrograph.}}
\titlerunning{Multi-phase gas properties of intervening QSO-ESDLAs}

\author{A.~Ranjan\inst{1,2,3}
   \and R.~Srianand\inst{3}
   \and P.~Petitjean\inst{1}
   \and G.~Shaw\inst{6}
   \and Y.-K.~Sheen\inst{2}
   \and S.~A.~Balashev\inst{4}
   \and N.~Gupta\inst{3}
   \and C.~Ledoux\inst{5}
   \and K.~N.~Telikova\inst{4}
}

\institute{
Institut d'Astrophysique de Paris, Sorbonne Universit\'e \& CNRS, 98bis boulevard Arago, 75014 Paris, France
  \and
Korea Astronomy and Space Science Institute, 776, Daedeokdae-ro, Yuseong-gu, Daejeon, 34055, Korea
  \and
  Inter-University Centre for Astronomy and Astrophysics, Post Bag 4, Ganeshkhind, 411 007, Pune, India
\and Ioffe Institute, Politekhnicheskaya 26, 194021 Saint Petersburg, Russia
\and European Southern Observatory, Alonso de C\'ordova 3107, Vitacura, Casilla 19001, Santiago, Chile
\and Tata Inst. of Fundamental Research, Homi Bhabha Road, Colaba Mumbai 400005, India
}

\date{\today}

\abstract{We present the results of a spectroscopic analysis of extremely strong damped \lya\, absorbers (ESDLAs; log $N$(\HI)$\gtrsim$21.7) observed with the medium resolution spectrograph, X-shooter at the Very Large Telescope (VLT). Recent studies in the literature indicate that ESDLAs probe gas from within the star-forming disk of the associated galaxies and thus ESDLAs provide a unique opportunity to study the interstellar medium of galaxies at high redshift. We report column densities ($N$), equivalent widths ($w$, for \MgII\ and \CaII\ transitions), and the kinematic spread (\deltav) of species from neutral (namely \OI, \ArI, \ClI, \NI,\,  and \NaI), singly ionised (\MgII, \CaII, \SII, \NiII, \MnII, \TiII,\, and \PII), and higher ionisation (\CIV, \SiIV, \NV,\, and \OVI) species.  We estimate the dust-corrected metallicity measured using different singly ionised gas species such as \PII, \SII, \SiII, \MnII\, and \CrII,\, and \ZnII. We find that, using the dust correction prescription, the measured metallicities are consistent for all mentioned species in all ESDLAs within 3-$\rm \sigma$ uncertainty. We further perform a quantitative comparison of column densities, equivalent widths, and kinematic spreads of ESDLAs with other samples that are associated with galaxies and detected in absorption along the line of sight towards high-redshift quasars (QSOs). We find that the distributions of the \ArI\, to \HI\, column density ratio (N(\ArI)/N(\HI)) in DLAs and ESDLAs are similar. We further report that ESDLAs do not show a strong deficiency of \ArI\, relative to other $\rm \alpha$-capture elements as is seen in DLAs. This supports the idea that the mentioned under-abundance of \ArI\, in DLAs is possibly caused by the presence of background UV photons that penetrate the low $N$(\HI) clouds to ionise \ArI, but they cannot penetrate deep enough in the high $N$(\HI) ESDLA environment. The $w$(\MgII$\lambda$2796) distribution in ESDLAs is found to be similar to that of metal-rich \CI-selected absorbers, but the velocity spread of their \MgII\, profile is different. The dust content (measured by modelling the quasar extinction) and $w(\CaII\lambda3934)$ distributions are similar in ESDLAs and \CaII-selected absorbers, yet we do not see any correlation between $w(\CaII\lambda3934)$ and dust content. The \deltav\, velocity spread of singly ionised species in ESDLAs is statistically smaller than that of DLAs. For higher ionisation species (such as \CIV\, and \SiIV) that trace the warm ionised medium, \deltav\, is similar in the two populations. This suggests that the ESDLAs sample a different \HI\, region of their associated galaxy compared to the general DLA population. We further study the $N$(\ClI) distribution in high-redshift DLA and ESDLA sightlines, as \ClI\, is a good tracer of \HH\, gas. The $N$(\ClI)$-N$(\HH) correlation is followed by all the clouds (ESDLAs and otherwise) having log $N$(\HH)$<$22.
}

\maketitle

\section{Introduction}

Damped \lya\, absorption systems (DLAs) correspond to a particular class of \HI\, absorption systems, with a \HI\, column density, log $N$(\HI) (cm$^{-2}$) $\geq 20.3$ \footnote{In the following text, column densities, $N$, are given in $\mbox{cm}^{-2}$.} \citep[see e.g.][]{Wolfe1986DLA}. These absorbers can be easily recognised through their damped \lya\, absorption feature in the spectra of bright background sources such as quasars (QSOs) and $\rm \gamma$-ray bursts (GRBs). Statistically, DLAs have been shown to dominate the neutral gas mass density in the high-redshift universe, providing the primary fuel for star formation \citep[][]{wolfe2005, Prochaska2005, Noterdaeme2009, Noterdaeme2012b, Zafar2013, Crighton2015, Sanchez-Ramirez2016}. In some cases, DLAs are found to be associated with the halo and/or the circumgalactic medium (CGM) of galaxies at high redshift ($z\rm \sim$2) \citep[see e.g.][]{Noterdaeme2012a, Peroux2012, Fynbo2013, Peroux2018} as well as low redshifts (z$<$1) \citep[see e.g.][and references therein]{Rahmani2016}. Furthermore, an anticorrelation between the \HI\, column density and the impact parameter of the host galaxy, $\rm \rho$, is observed \citep[see e.g.][]{Pontzen2008, Noterdaeme2012a, RahmatiandSchaye2014, Krogager+17}.

Hence, it is expected that the highest end of the $N$(\HI) distribution can preferentially probe the star-forming disk of the associated galaxies. Following this argument, \citet[][]{Noterdaeme2014} studied $\sim$100 extremely strong DLAs (or ESDLAs) defined as systems with \HI\, column density, log $N$(\HI) $\geq$ 21.7 at high redshift ($z\rm \sim2-4$). Stacking the low resolution SDSS spectra led to the detection of \lya\, emission within the SDSS fibres (i.e. within an impact parameter of $\rm \sim$8 kpc from the quasar sight line at the redshift of the DLAs, i.e. z$\rm \sim$2.5). Additionally, numerical simulations \citep[e.g.][]{Altay+13} show that the highest end of the \HI\, column density distribution function is sensitive to the effects of stellar feedback and \HH\, formation. Indeed, it is seen in high-$z$ observations that ESDLAs have a significantly enhanced fraction of diffuse \HH\, gas observed in comparison with regular DLAs \citep[e.g.][]{Noterdaeme2015b, Balashev2018}.

This motivated a medium spectral resolution, wide wavelength range follow-up study of 11 ESDLAs  using the X-shooter spectrograph at the Very Large Telescope (VLT, Paranal, Chile). These ESDLAs were identified using SDSS spectra \citep{Noterdaeme2014}. The analysis of one of these ESDLAs towards SDSS\,J1513$+$0352 \citep[see][]{Ranjan+2018} led to the \HH\, detection with the highest \HH\, column density ever observed in QSO absorption line studies along with \lya\, emission detected at a small impact parameter ($\rm \rho$ = 1.4 kpc) relative to the quasar line of sight.  \\

The detailed analysis of the initial sample of 11 ESDLAs (along with ESDLAs taken from literature) is presented in \citealt[][]{Ranjan2020}. The study notes that ESDLAs probe transitioning \HI-\HH\, gas clouds with much higher frequency ($\sim$50\% of the cases have \HH) as compared to the general DLA population, where only 5-10\% of the systems are detected with \HH\, \citep{Petitjean2000, Ledoux2003,noterdaeme2008, Balashev2018}. They also show that faint emission lines are detected in proximity to the absorbers which indicates that ESDLAs arise from the star-forming disk of their associated galaxy. Although, the direct study of the galaxy morphology is not possible due to the non-detection of the faint stellar continuum of these galaxies. Hence, we need to probe alternative methods to understand the nature of ESDLAs and their associated galaxies. The galaxies associated with ESDLAs are detected independently of their intrinsic luminosity. In \citet{Ranjan2020}, we focus primarily on the \HH\, detection and the proximity of ESDLAs with their associated galaxy. In this paper, we intend to gain additional information about the physical and chemical properties of ESDLAs and the associated galaxies by studying the absorption lines of neutral (such as \ClI\, and \ArI), singly ionised (such as \MgII\, and \CaII), as well as highly ionised (such as \CIV, \SiIV, \NV,\, and \OVI) gas species. The comprehensive study probes features of gas in different ionisation states that are associated with the ESDLA host galaxy. We calculated the column density (and/or the equivalent width in specific cases) and quantified the gas kinematics (using \deltav, defined as the velocity interval that contains 90\% of the area under the apparent optical depth spectrum) of differently ionised species. We study the distribution of column density and velocity spread of gas species in different ionisation states and further compare them to other DLA sub-sets associated with galaxies such as the metal-rich \CI-selected absorbers \citep[see e.g.][]{Ledoux2015, Zou2018, Noterdaeme2018}, dust rich DLAs selected based on their \CaII\, equivalent width \citep[see e.g.][]{Wild_and_Hewett2005}, \MgII-selected strong absorbers \citep[see e.g.][]{Rao2005}, and absorbers associated with outflowing gas clouds \citep[see e.g.][]{Fox2007, Fox2008}. The comparison will help us to further understand the ESDLA gas clouds and their associated galaxies. \\

Details of the observations and data reduction are presented in Section~\ref{observations}. The analysis of the absorption lines is described in Section~\ref{absorption_analysis}. The results of our study are presented in Section~\ref{sec:Results} and discussed in Section~\ref{discussion}. In Section~\ref{Conclusion} we summarise our findings.


\section{Observations and data reduction \label{observations} }

Here, we use a sample of 11 ESDLAs observed with the VLT/X-shooter spectrograph. Spectra of 11 quasars from the first programme (ESO programme ID 095.A-0224(A)) were analysed in \citet{Ranjan2020} and \citet{Ranjan+2018}. The observations were carried out in service mode under good seeing conditions (typically 0.7-0.8$\arcsec$) between April 2015 and July 2016 with the multiwavelength medium-resolution spectrograph X-shooter \citep{Vernet2011} mounted at the Cassegrain focus of the Very Large Telescope (VLT-UT2) at Paranal, Chile. A two-step nodding mode with an offset of 4 arcsec between the two integrations was used. We reduced the data using the standard X-shooter pipeline \citep{Modigliani2010} and combined individual exposures by weighting the measured flux in each pixel by the inverse of its variance to obtain the combined 2D and 1D spectra. The target spectra were obtained at medium spectral resolution, $R$ $\sim$ 5000-10000, depending on the arm of the spectrograph corresponding to the typical width of the instrument function, $\rm FWHM\sim 30-60$\, \kms. We found that the resolution in most of the individual spectra are seeing-dominated. The detailed list of resolution for individual spectra can be found in \citet{Ranjan2020}. For the absorption line analysis presented here, we have only used the combined 1D spectra of each arm of the X-shooter spectrograph. Further details about the observations and data reduction can be found in \citet{Ranjan2020}. The long form QSO names (with precise RA, DEC information) with the QSO and the absorbers' redshifts are given in Table.~\ref{Column_density_table_3}. Throughout this paper, we use a short notation for the quasar names, for example SDSS~J\,223250.98+124225.29 is referred to as J2232+1242.   \\

We additionally use information about seven new ESDLAs observed with VLT X-shooter using similar observing conditions from \citet{Telikova2022article} \citep[initially discussed in][]{Telikova2020proceedings}. We further add information about five ESDLAs from literature, found towards the QSOs HE0027$-$1836 \citep[from][]{Noterdaeme2007}, J0843$+$0221 \citep[from][]{Balashev2017}, J113$-$0010 \citep[from][]{Noterdaeme+12}, and J0230$-$0334 and Q0743$+$1421 \citep[from][]{Kulkarni2015}. These ESDLAs form a part of the high-$z$ ESDLA sample that meets the N(\HI) criterion and were observed using medium and/or high resolution spectrographs.

\section{Absorption-line analysis \label{absorption_analysis}}

We used multi-component Voigt-profile fitting to derive column densities from absorption features. The fitting was performed using VPFIT\footnote{\url{https://www.ast.cam.ac.uk/~rfc/vpfit}} \citep{Carswell2014}. The VPFIT website\footnote{\url{https://www.ast.cam.ac.uk/~rfc/vpfit}} also provides a data file for the atomic parameters that was compiled primarily from \citealt{Morton2003}, but taking updated references from many other works in the literature \citep[such as][]{Petitjean2004, Abgrall2006, Salumbides2006, Ivanov2008, baillya2010a, Berengut2011}. We primarily used the atomic parameters from this data file for our fitting along with the others mentioned directly in the text. \\

The initial fit for \CI\, and singly ionised species for the ESDLA sample was reported in \citet{Ranjan+2018} and \citet{Ranjan2020}. We took advantage of the wide wavelength range of X-shooter spectra to include many transition lines of \FeII\, and \SiII, such as '\SiII$\lambda$1193 ([$\rm \AA$]\footnote{All wavelengths in this article are given in terms of $\rm \AA$ unless specified otherwise})', '\SiII$\lambda$1304', '\SiII$\lambda$1526', '\SiII$\lambda$1808', '\FeII$\lambda$1608', '\FeII$\lambda$1611', '\FeII$\lambda$2249', '\FeII$\lambda$2260', '\FeII$\lambda$2344', '\FeII$\lambda$2374', '\FeII$\lambda$2382', '\FeII$\lambda$2586', and '\FeII$\lambda$2600'. Using so many transitions with varying oscillator strengths\footnote{Transitions with high oscillator strength produce strong absorption signature that might sometimes be intrinsically saturated depending of the sub-component column density, yet they might be unsaturated in transitions with lower oscillator strength.} is advantageous as it helps in resolving the highest number of low column density sub-components (in transitions with strong absorption) that can be identified separately given the spectral resolution. In addition, we can also constrain the column density for sub-components that appear saturated in transitions with a strong absorption signature by looking at the transitions with a weaker absorption signature. Hence, fitting a multi-component absorption model for all transitions mentioned above (in addition to other transitions from \CrII, \ZnII, \MgI, \NiII,\, and \TiII) with their $b$-values and redshifts tied together helps us obtain a robust estimate on the total column density of the mentioned species for these ESDLAs. For \CaII,\, we report the total column density as well rest-frame equivalent width ($w$) for its two prominent transitions, '\CaII$\lambda$3934' and '\CaII$\lambda$3969'. The \MgII\, profiles ('\MgII$\lambda$2796' and '\MgII$\lambda$2803') are strongly saturated. Hence, we report only the rest-frame equivalent width ($w$) for both the transitions. We applied a similar fitting approach for higher ionisation species, such as \SiIV, \CIV,\, and \NV\, (i.e. tying the $b$-values and redshifts for multiple transitions of the mentioned ionisation), but the fitting was performed separately from that of the low-ionisation species. We found that the redshifts for individual components derived from fitting the higher ionisation lines (\CIV, \SiIV,\, and \NV) are only slightly separated in velocity space ($\rm \lesssim\,50\,km s^{-1}$) from the components of the lower ionisation lines.   \\

\section{Results}
\label{sec:Results}

In \citet{Ranjan2020}, we studied only \CI\, and singly ionised species (e.g. \FeII, \ZnII,\, and \SiII) and the \HI-\HH\, transition for 11 ESDLAs mentioned in Table~\ref{Column_density_table_3}. The table shows basic information about the 11 ESDLAs from \citet{Ranjan2020} for reference that is used for this study. In continuation, we further searched for and analysed the following neutral gas species: Oxygen (\OI), Argon (\ArI), Nitrogen (\NI), Chlorine (\ClI), Sodium (\NaI), singly ionised calcium (\CaII), Sulphur (\SII), Magnesium (\MgII), Phosphorous (\PII), Titanium (\TiII), Manganese (\MnII), Nickel (\NiII), and four higher ionisation species -- Carbon (\CIV), Silicon (\SiIV), Nitrogen (\NV), and Oxygen (\OVI). The results of our analysis (column density, rest-frame equivalent width, and \deltav\ estimates) for 11 ESDLAs are provided in Table~\ref{Column_density_table_1} and Table~\ref{Column_density_table_2}. Table~\ref{Column_density_table_1} also provides additional information about the seven ESDLAs studied in detail in \citet{Telikova2022article}. We provide important details for each species in the following subsections. \\

In addition to these, we also use the relevant ESDLA data from literature.\ They include the following: \citet{Noterdaeme2007}, towards QSO HE0027$-$1836, with log $N$(\ArI) = 14.42$\pm$0.02, log $N$(\HI) = 21.75$\pm$0.1, and log $N$(\HH) = 17.3$\pm$0.07]; \citet{Noterdaeme+12}, towards QSO J1135$-$0010, with $w$(\MgII$\lambda$2796) = 3.6\AA, and log $N$(\HI) = 22.1$\pm$0.05; \citet{Balashev2017}, towards QSO J0843$+$0221, with log $N$(\ClI) = $13.63^{+0.20}_{-0.05}$, log $N$(\HI) = 21.82$\pm$0.11, and log $N$(\HH) = 21.21$\pm$0.02; and \citet{Kulkarni2015}, towards QSO J0230$-$0334, with log $N$(\HI) = 21.74$\pm$0.1, log $N$(\CIV) = 14.6$\pm$0.05, and log $N$(\SiIV) = 13.89$\pm$0.04 as well as towards QSO Q0743$+$1421, with log $N$(\HI) = 21.9$\pm$0.1, log $N$(\CIV) = 14.41$\pm$0.07, and log $N$(\SiIV) = 13.84$\pm$0.06.

\begin{table*}[]
\setlength\tabcolsep{2.5pt}
\begin{tabular}{cccccccccc}
\hline
Quasar & z$_{QSO}$ & z$_{abs}$ & N(\HI) & N(\HH) & E(B-V) & [M/H] & \deltav & \deltav(\CIV) & N(\SiII) \\ 
\hline
\hline
SDSS J001743.8+130739.8 & 2.594 & 2.326 & 21.62$\pm$0.03 & $<$18.3 & 0.11 & -1.5$\pm$0.09 & 120 & 430 & 16.01$\pm$0.13 \\
SDSS J002503.0+114547.8 & 2.961 & 2.304 & 21.92$\pm$0.09 & $\sim$20 & 0.19 & -0.53$\pm$0.11 & 240 & - & 16.67 $-$ 17.18\tablefootmark{a} \\
SDSS J114347.2+142021.6 & 2.583 & 2.323 & 21.64$\pm$0.06 & 18.3$\pm$0.1 & 0.08 & -0.8$\pm$0.06 & 130 & - & 16.30$\pm$0.03 \\
SDSS J125855.4+121250.2 & 3.055 & 2.444 & 21.9$\pm$0.03 & $<$18.3 & 0.02 & -1.43$\pm$0.04 & 100 & 270 & 16.10$\pm$0.04 \\
SDSS J134910.4+044819.9 & 3.353 & 2.482 & 21.8$\pm$0.01 & $<$18.1 & 0.03 & -1.35$\pm$0.06 & 60 & - & 16.47 $-$ 17.24\tablefootmark{a} \\
SDSS J141120.5+122935.9 & 2.713 & 2.545 & 21.83$\pm$0.03 & $<$15.9 & 0.03 & -1.59$\pm$0.08 & 50 & 220 & 16.07$\pm$0.18 \\
SDSS J151349.5+035211.6 & 2.68 & 2.464 & 21.83$\pm$0.01 & 21.31$\pm$0.01 & 0.13 & -0.84$\pm$0.23 & 90 & 200 & 16.92$\pm$0.24 \\
SDSS J214043.0$-$032139.2 & 2.479 & 2.339 & 22.41$\pm$0.03 & 20.13$\pm$0.07 & 0.04 & -1.52$\pm$0.08 & 70 & - & 16.04 $-$ 17.31\tablefootmark{a} \\
SDSS J223251.0+124225.3 & 2.299 & 2.23 & 21.75$\pm$0.03 & 18.56$\pm$0.02 & 0.004 & -1.48$\pm$0.05 & 75 & 320 & 15.69 $-$ 17.23\tablefootmark{a} \\
SDSS J224621.1+132821.3 & 2.514 & 2.215 & 21.73$\pm$0.03 & $<$16.3 & $<$0.004 & -1.84$\pm$0.1 & 65 & 230 & 15.63$\pm$0.21 \\
SDSS J232207.3+003349.0 & 2.693 & 2.477 & 21.58$\pm$0.03 & $<$16.0 & $<$0.004 & -1.71$\pm$0.13 & 40 & 140 & 15.47$\pm$0.06 \\ 
\hline
\end{tabular}
\caption{Quasar name, redshift of the quasar (z$_{QSO}$), redshift of the ESDLA (z$_{abs}$), column density (in logscale) of \HI\ and \HH, dust content measured in the ESDLA (E(B-V)), metallicity relative to solar (in [M/H], where M represents the most undepleted neutral gas species), {\deltav}, {\deltav(\CIV)}, and column density (in logscale) of \SiII for ESDLAs. \tablefoot{{\deltav} is derived from unsaturated transitions of singly ionised species, such as \NiII\, and \ZnII. All data are taken from \citet{Ranjan2020} (and references therein), except for {\deltav(\CIV)} and N(\SiII), which is calculated in this work. \tablefoottext{a}{Taking into account the effects of saturation and blends, we provide revised estimates of N(\SiII) here. See section  \ref{siii_saturation_subsection} for discussion.}}}
\label{Column_density_table_3}
\end{table*}

\begin{table*}[]
\setlength\tabcolsep{2.5pt}
\centering
\begin{tabular}{cccccccccc}
\hline
{QSO}        & ${N(\OI)}$             & ${N(\ArI)}$           & ${N(\ClI)}$           & ${N(\SiIV)}$        & ${N(\CIV)}$         & {$w(\CaII\lambda$3934)}   & {$w(\MgII\lambda$2796)} & {\delv(\MgII)}\tablefootmark{b} \\
\hline
\hline
J0017+1307 & \textgreater{}17.0                 & 14.72$\pm$0.19   & \textless{}12.55 & 13.38$\pm$0.23 & 14.02$\pm$0.22 & \textless{}0.57 & 1.88$\pm$0.01 & 190             \\

J0025+1145 & \textgreater{}18.6                 & -                & 13.5 $-$ 15.5   & \textgreater{}14.19 & \textgreater{}14.71 & 0.7$\pm$0.02    & 4.36$\pm$0.01 & 390             \\

J1143+1420 & \textgreater{}18  & \textless{}15.2 & \textless{}12.68 & \textgreater{}14.02 & \textgreater{}14.63 & \textless{}1.09 & 3.23$\pm$0.02 & 450             \\

J1258+1212 & \textgreater{}17      & -                & \textless{}12.91 & 13.57$\pm$0.08 & 14.28$\pm$0.03 & \textless{}0.23 & 2.52$\pm$0.02 & 190             \\

J1349+0448 & \textgreater{}16.6 & -                & -                & -              & \textgreater{}14.38 & 0.43$\pm$0.14   & 1.80$\pm$0.06 & 160             \\

J1411+1229 & \textgreater{}17.1    & $\sim$14.4       & \textless{}12.66 & 13.45$\pm$0.05 & 13.47$\pm$0.1  & 0.78$\pm$0.07   & 0.95$\pm$0.07 & 110             \\

J1513+0352 & \textgreater{}17     & -                & $14.63\,\pm\,0.74$   & 13.65$\pm$0.32 & 14.12$\pm$0.32 & 0.15$\pm$0.05   & 2.89$\pm$0.04 & 340             \\
J2140-0321 & 18.01$\pm$0.18\tablefootmark{a}    & \textless{}15.7 & $13.37\,\pm\,0.07$\tablefootmark{a}   & \textgreater{}14.06 & \textgreater{}14.63 & 0.26$\pm$0.03   & 1.19$\pm$0.02 & 150             \\

J2232+1242 & \textgreater{}16.7 & 14.43$\pm$0.14   & \textless{}12.72 & 13.4$\pm$0.02  & 13.8$\pm$0.04  & 0.34$\pm$0.04   & 2.31$\pm$0.01 & 200             \\
J2246+1328 & \textgreater{}16.8    & 14.15$\pm$0.32   & \textless{}12.85 & 13.48$\pm$0.08 & 14.28$\pm$0.19 & 0.60$\pm$0.07   & 0.77$\pm$0.01 & 100             \\
J2322+0033 & \textgreater{}15.7 & 14.46$\pm$0.21   & \textless{}12.72 & 13.36$\pm$0.05 & 13.84$\pm$0.04 & \textless{}4.24 & 1.68$\pm$0.06 & 150  \\

\hline
\hline

J0024$-$0725 & \textgreater{}17.6 & $\rm \sim\,14.07$   & \textless{}13.25 & - & - & 0.29$\pm$0.07 & 1.38$\pm$0.20 & 110  \\  
J1238$+$1620 & - & $\rm \sim\,14.69$   & \textless{}13.44 & - & - & - & 3.94$\pm$0.07 & 290  \\   
J1353$+$0956 & \textgreater{}16.7 & \textless{}14.46   & \textless{}12.61 & - & - & \textless{}0.05 & 1.47$\pm$0.18 & 200  \\   
J1418$+$0718 & \textgreater{}16.7 & \textless{}14.41   & \textless{}12.22 & - & - & - & $<0.97$ & 90  \\   
J2205$+$1021 & - & $\rm \sim\,15.25$   & \textless{}13.63 & - & - & 0.31$\pm$0.08 & 2.63$\pm$0.07 & 210  \\   
J2351$-$0639 & - & -   & \textless{}13.19 & - & - & - & 1.47$\pm$0.17 & 130  \\   
J2359$+$1354 & - & -   & \textless{}14.13 & - & - & 0.79$\pm$0.11 & 5.63$\pm$0.02 & 380  \\

\hline
\end{tabular}
\caption{Column density (in logscale) of gas species, \OI, \ArI, \ClI, \SiIV,\ and \CIV,\, the absorber rest-frame equivalent width ($w$ in \AA) of '\CaII$\lambda$3934' and '\MgII$\lambda$2796' lines, and the velocity spread, \delv\,\tablefootmark{b}(in \kms) of the '\MgII$\lambda$2796' line for ESDLAs.} 
\tablefoot{Values starting with $\sim$ represent detections deemed tentative due to either strong blends with sky lines, low signal-to-noise, and/or contamination from the \lya\, forest. The estimates for seven additional ESDLAs presented in the bottom of the table demarcated by the horizontal lines are from \citet{Telikova2022article}.
\tablefoottext{a}{Robust estimate for exceptionally high value of N(\OI) measured by comparing X-shooter spectra with high resolution UVES spectra \citep[studied in][]{Noterdaeme2015a}. The UVES spectra were also used to measure N(\ClI).}\\
\tablefoottext{b}{\delv\, was obtained here for saturated '\MgII$\lambda$2796' profiles, and it is defined as the velocity separation between the two extreme pixels where the optical depth $\tau\,<$0.1 \citep[see][for detailed discussion]{Zou2018}. We note that this is different from the standard \deltav\, measurement of unsaturated lines. }}
\label{Column_density_table_1}
\end{table*}

\subsection{Neutral Oxygen}
\label{sect:OI}

Neutral oxygen absorption is present in all ESDLAs in our sample. We primarily used the '\OI$\lambda$1302' transition to estimate the \OI\, column density. We also used the '\OI$\lambda$1039' line wherever possible, but with extreme caution as this transition is inside the \lya\, forest and can also be blended with \HH\, lines. Since our study is related to gas associated with galaxies, there is a chance that \OI\, column densities are high and hence, most of the '\OI$\lambda$1302' lines are probably visibly or intrinsically saturated (a situation where the convolved absorption profile appears to be unsaturated, but the true unconvolved profile is saturated). Since our spectra are of medium resolution, the convolved absorption profile may appear unsaturated in many cases, when the true unconvolved profile is indeed saturated.  In such cases, there is a degeneracy between the $b$-value and column density, and in the absence of additional information, the column density estimate might not be robust. \\

Therefore, we used the '\OI$\lambda$1302' profile taking component information (redshift and $b$-value) from other singly ionised species (such as \FeII\, and \ZnII) to report a lower limit for the \OI\, column density in all ESDLAs, except for the z$_{abs}$=2.339 system towards QSO J2140$-$0321. We obtained a log~$N$(\OI)=18.01$\pm$0.18 for this ESDLA. \citet{Noterdaeme2015a} have also published the results for this ESDLA using spectra obtained with the high resolution Ultraviolet and Visual Echelle (VLT-UVES, R$\rm \sim$48000) spectrograph. We further cross-checked our measured values for column densities with \citet{Noterdaeme2015a}. Our estimates match with theirs (log~$N$(\OI)=17.9$\pm$0.2) within the measured uncertainties. However, we note that this robust estimate from single X-shooter spectra may not be representative of the sample. The column density estimates for \OI\, are listed in Table.~\ref{Column_density_table_1}. \\

We also searched for the presence of absorption from the fine-structure transition of \OI. In only one case, that is for ESDLA towards QSO J2140$-$0321, are we able to confirm the detection of \OI$^{*}$ absorption that is well separated from the \SiII$\lambda$1304 line and measure log~N(\OI$^{*}$)=13.82$\pm$0.10.  This is consistent with that measured by \citet{Noterdaeme2015b} using the UVES spectrum. 
There is a $2\sigma$ detection in the case of the ESDLA towards J1411+1228. For the ESDLA towards QSO J2232$+$1242, the \OI$^{*}\lambda$1304 absorption is blended with the \SiII$\lambda$1304 line (log N(\OI$^{*}$)=13.46$\pm$0.19). In the remaining cases, we do not have a clear detection at more than the 3$\sigma$ level.  The N(\OI$^{*}$) upper limits ($X\sigma$ level) estimated from the non-detection of the $\lambda$=1304 line for all other ESDLAs are listed in Table.~\ref{Column_density_table_2}.

\subsection{Neutral Argon}

\begin{figure}
\centering
   \includegraphics[width=1.0\hsize]{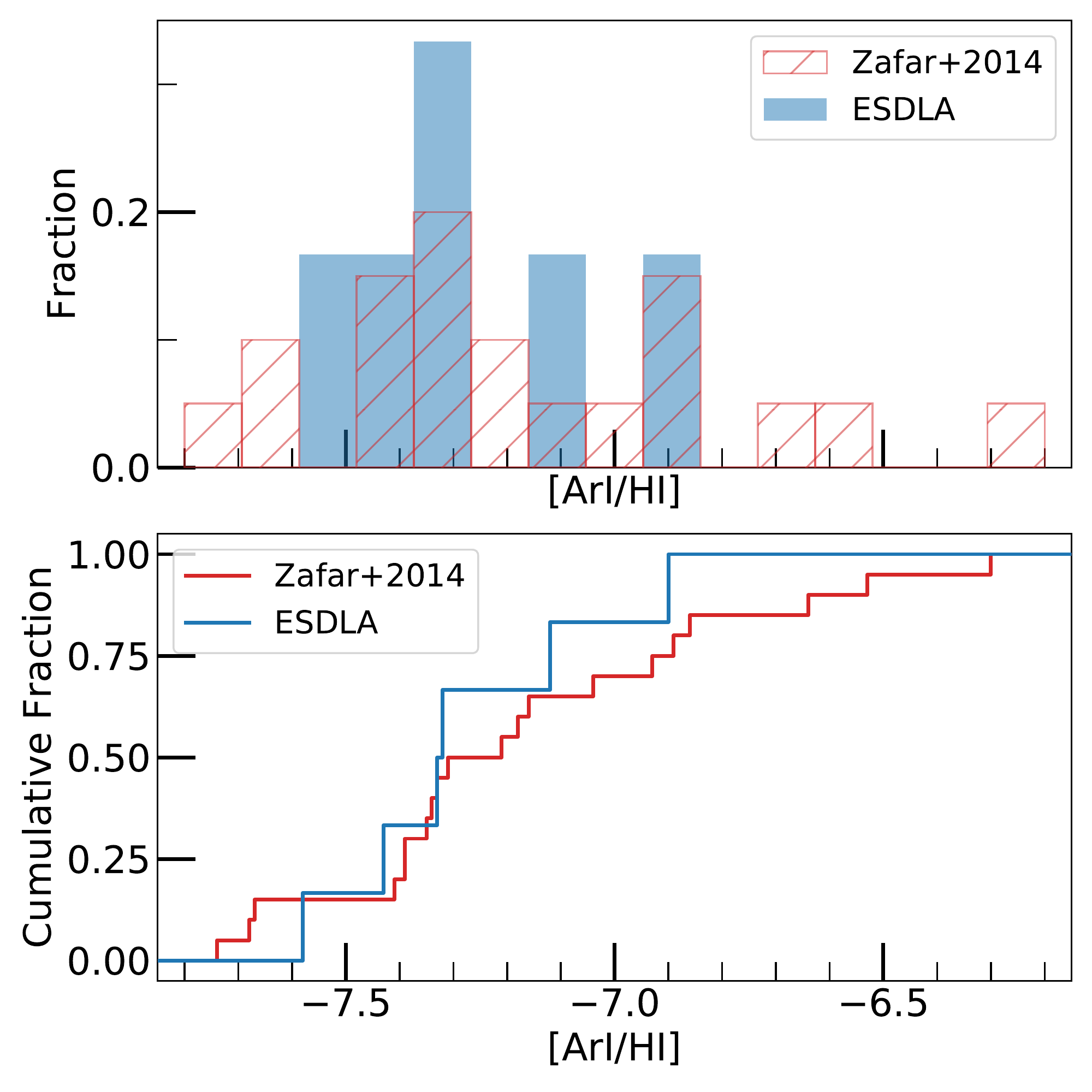}
    \caption{ Distributions of the column density ratio, $N$(\ArI)/$N$(\HI),
    in ESDLAs (blue curves, this work) and the general DLA population \citep[red curves,][]{Zafar2014}.  The histogram  and cumulative distributions from the two samples are compared in the top and bottom panels, respectively. } 
    \label{argon_distribution_fig_1}
\end{figure}

Neutral argon is difficult to probe in high-$z$ clouds especially because the twin transitions -- '\ArI\, $\lambda$1066' and '\ArI\, $\lambda$1048' are located within the \lya\, forest. We fitted these transitions together taking any possible contamination from intervening \lya\, absorption  into account and using the same component structure as other low ionisation species such as \FeII\, and \ZnII. We estimated the column density of \ArI\, in four of our ESDLAs, towards the QSOs J0017$+$1307, J2232$+$1242, J2246$+$1328, and J2322$+$0033. For the ESDLA towards QSO J1411$+$1229, we could only fit the '\ArI $\lambda$1066' line and hence, we consider this detection as tentative. For the system towards QSO J2140$-$0321, the \ArI\, transitions are blended strongly with the \lya\, forest. Hence, we used our X-shooter spectra in combination with the high-resolution UVES spectra to obtain a tentative upper limit estimate on the column density. We intend to use these tentative estimates in our discussion as the component structure (redshift and $b$-value) used is consistent with that of lines from other low-ionisation species (such as \FeII\ and \ZnII). Apart from this, we have a system towards QSO J1143$+$1420, for which the \ArI\, transition is too weak to be considered a confirmed detection. For this system, we obtained a 3 $\sigma$ upper limit on the column density. The column density for all detections and upper limits are listed in Table. ~\ref{Column_density_table_1}. In Fig. ~\ref{argon_distribution_fig_1}, we compare our distribution of the $N$(\ArI)/$N$(\HI) ratio in ESDLAs with that of the general DLA population studied by \citet{Zafar2014}. Based on the Kolmogorov--Smirnov test (K-S test) p-value of 0.9 between ESDLAs from this work and DLAs from \citet{Zafar2014}, we can argue that the distribution of a neutral Argon abundance in DLAs and ESDLAs are indistinguishable. 

\subsection{Neutral Chlorine}

In the presence of even small amounts of \HH, chlorine becomes principally neutral as Cl$^{+}$ reacts rapidly with \HH, giving HCl$^{+}$ that recombines quickly into H and Cl \citep[see e.g.][]{Jura1974}. \citet{Balashev2015} also noted a strong association of neutral chlorine with \HH\, gas in high-$z$ absorbers. \citet{Ranjan2020} found diffuse \HH\, signatures in their rest-frame Lyman-Werner (L-W) band absorption in about $\sim$ 50\% of the ESDLAs. Although, the \lya\, forest at high redshift is dense enough to create strong blends with the L-W band and create confusion. In addition, there are cases (such as the ESDLA towards QSO J0025$+$1145), where the L-W band signatures are washed away due to the presence of a Lyman-limit system. Hence, tracers such as \CI\, and \ClI\, can be used as indirect evidence for the presence of \HH\, gas. Thus, we would like to study the presence of \ClI\, in our sample. \\

We primarily looked for \ClI\, transitions at $\lambda = 1347\AA$ and $\lambda = 1363\AA$ \citep[atomic data taken from][]{Schectman1993} and the transitions at $\lambda = 1335\AA$, $\lambda = 1379\AA$, and $\lambda = 1389\AA$ \citep[atomic data taken from][and references therein]{Welty2020}. Other \ClI\, transitions located in the \lya\ forest \citep[such as \ClI\, transitions at $\lambda = 1088\AA$, $\lambda = 1188\AA$, and $\lambda = 1084\AA$, reported in][and references therein]{Oliveira_Hebrard_2006} have a high probability of being blended with other absorption lines. We note the non-detection of '\ClI$\lambda$1363' in all our spectra. We also note that the transition '\ClI$\lambda$1335' is heavily blended with the saturated '\CII$\lambda$1334' and '\CII$\rm ^{*}\lambda$1335' transitions. Using the other three transitions (at $\lambda = 1347\AA$, $\lambda = 1379\AA$, and $\lambda = 1389\AA$), we fitted the \ClI\, lines with the component structure (redshift and $b$-value) tied with \CI, as neutral carbon is also found in the presence of \HH\, gas. We note that for two ESDLAs towards QSO J0025$+$1145 and J1513$+$0352, the \ClI\, $\lambda = 1347\AA$ profile has contamination from other absorbers in proximity. For ESDLAs towards QSO J1513$+$0352, the other two \ClI\, transitions are clear and the contamination near $\lambda = 1347\AA$ is identified. Hence, we get a robust estimate on $N$(\ClI). For ESDLAs towards QSO J0025$+$1145, the other transitions are also contaminated and the $N$(\ClI) is highly uncertain. We report the $N$(\ClI) lower limit from the $\lambda = 1347\AA$ transition and the upper limit from the $\lambda = 1389\AA$ transition. We also report the robust measurement of $N$(\ClI) in ESDLA towards QSO J2140$-$0321 by combining our medium resolution X-shooter spectra and high-resolution UVES spectra \citep[initially studied in][]{Noterdaeme2015b}. All measurements and upper limits are reported in Table. ~\ref{Column_density_table_1}. \\        

We detected \ClI\ in all systems where \HH\ was previously detected. In the new ESDLA sample studied in \citet{Telikova2022article}, they detected \HH\, towards J2205$+$1021 and J2359$+$1354. We note that \ClI\, is not detected in these systems. Although, the $N$(\HH) in these two systems are much lower than the other ESDLAs with conformed \ClI\, detection. Since $N$(\HH)-$N$(\ClI) are correlated, the non-detection could just indicate the the \ClI\, detection limit with an X-shooter wavelength range and resolution. The detection of \HH\, towards J0025$+$1145 was declared as tentative in \citet{Ranjan2020} despite a high inferred column density. The reason being the presence of a Lyman-limit system towards QSO J0025$+$1145 which implies that we could detect only one L-W band of \HH\ in that system. The presence of neutral chlorine in this system confirms our previous claim. We report 3-$\rm \sigma$ upper limits on the \ClI\ column density for the rest of the ESDLAs, except for ESDLA towards QSO J1349$+$0448. For this system, all \ClI\, transitions are strongly blended.

\subsection{Neutral Nitrogen}

Nitrogen is produced in different stages of hydrogen and helium burning shells in stars. There is large uncertainty in the contribution of nitrogen production from different stages in various types of stars such as low, intermediate, or massive stars \citep[see][]{Meynet_Maeder_2002}. The study of nitrogen abundance in neutral gas clouds gives important insight into resolving this uncertainty \citep[see e.g.][]{Petitjean_2008}. Hence, we looked for neutral Nitrogen (\NI) in our ESDLA sample as well. We note that \NI\, is another challenging species to detect as its transitions fall within the \lya\, forest. However, due to multiple transitions (centred around rest frame $\lambda = 1134\AA$ and $\lambda = 1200\AA$), we were able to detect neutral Nitrogen in ESDLAs. In our sample, we report the detection of neutral nitrogen in four ESDLAs towards QSOs J0017$+$1307, J1411$+$1229, J2140$-$0321, and J2232$+$1242. For the rest of the ESDLAs, the \NI\, profile was either too weak or heavily contaminated with forest lines to be detected robustly in our medium resolution study. The measured column densities and 3-$\rm \sigma$ upper limits are reported in Table. ~\ref{Column_density_table_2}.

\subsection{\MgII}

The \MgII\, absorption signature has been a common feature of gas associated with galaxies both in the local \citep[see e.g.][]{Bergeron1991, Steidel1992, Feltre2018} and distant \citep[see e.g.][]{Bouche2004, Rao2005, Bouche2007} universe. We detect \MgII\, absorption in all of our ESDLAs. We note that \MgII\, is quite abundant in strong \HI\, absorbers (DLAs). The primary transitions, '\MgII$\lambda$2796' and '\MgII$\lambda$2803', are strongly saturated in most DLAs. Hence we report the equivalent width of the mentioned \MgII\, transitions as previously done in the literature \citep[see e.g.][]{Zou2018, Matejek2013}. We compare our distribution of '\MgII$\lambda$2796' equivalent widths in ESDLAs (median - 2.1\AA) with that of other high-$z$ absorption selected samples associated with galaxies, such as the DLA population \citep[taken from \MgII-selected absorbers studied in][median - 1.72\AA]{Matejek2013}, the \CI-selected high metallicity absorber sample \citep[see,][median - 2.89\AA]{Zou2018}, and the \CaII-selected dusty absorber sample \citep[see,][median - 2.27\AA]{Wild_and_Hewett2005}. \\

\begin{figure}
\centering
   \includegraphics[width=1.0\hsize]{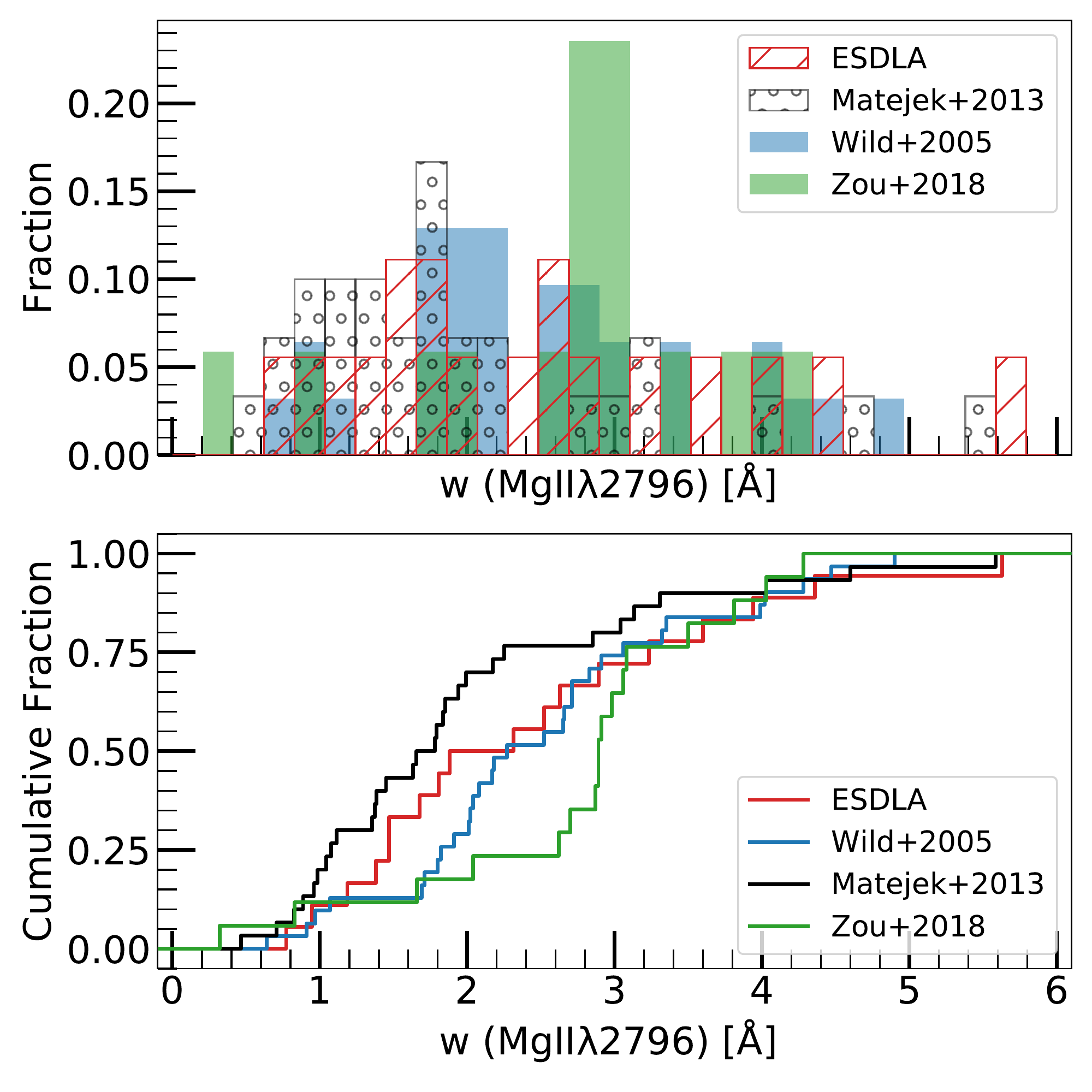} 
    \caption{{\sl $w$('\MgII\,$\lambda$2796') in different samples. Top panel}: Distribution of the '\MgII\,$\lambda$2796'
    equivalent width ($w$) in the ESDLAs (red hashed histogram, this work), \CI-selected absorbers \citep[green filled histogram, from][]{Zou2018}, \CaII-selected absorbers \citep[blue filled histogram, from][]{Wild_and_Hewett2005}, and the general DLA population \citep[grey dotted histogram, from][]{Matejek2013}. {\sl Bottom panel}: Cumulative distributions of the same samples using the same colours.}
    \label{mg_ii_distribution_fig_1}
\end{figure}

\begin{figure}
\centering
   \includegraphics[width=1.0\hsize]{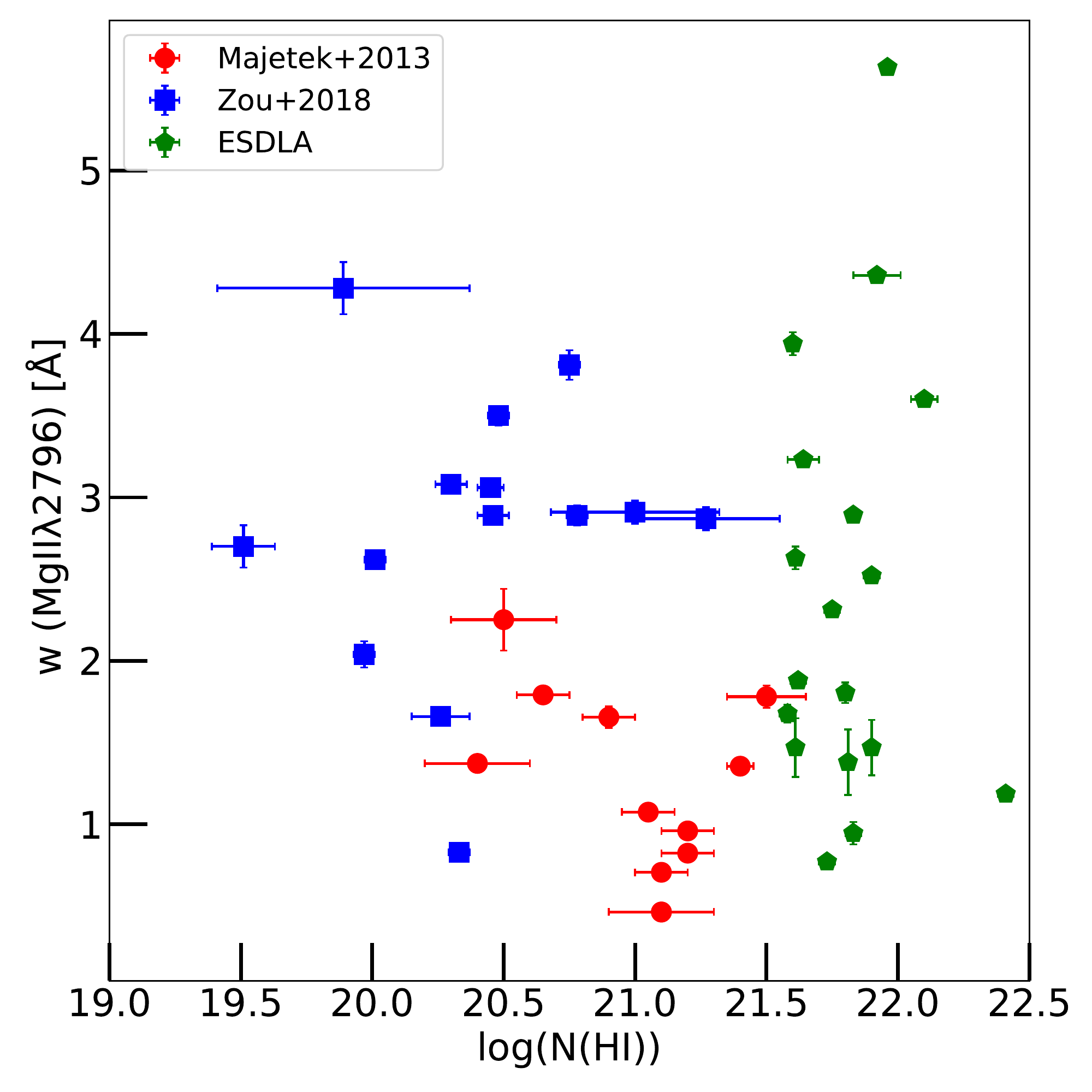} 
    \caption{log $N$(\HI) [atoms cm$\rm ^{-2}$] vs. $w$(\MgII\,$\lambda$2796) [$\rm \AA$] for ESDLAs (green points, this work), \CI-selected absorbers \citep[blue points, from][]{Zou2018}, and the \MgII-selected population of DLAs \citep[red points, from][]{Matejek2013}.}
    \label{nhi_vs_wmg_ii_fig}
\end{figure}

Fig.~\ref{mg_ii_distribution_fig_1} shows the distributions of $w$(\MgII$\lambda$2796) in the above-mentioned samples. The K-S test reveals that the \MgII\, equivalent width distribution of ESDLAs. ESDLAs are indistinguishable from that of the \MgII-selected DLA population (p-value=0.35). ESDLAs also tend to have $w$(\MgII$\lambda$2796) similar to the \CaII-selected sample as testified by the medians of the samples. The K-S test of ESDLAs with the \CaII\, sample (p-value=0.37) also indicates that the sample distributions are indistinguishable. We performed a K-S test with the \CI-selected sample as well and found that the samples are consistent with being drawn from the same parent population (p-value=0.13). In addition, we show the different samples mentioned above in $N$(\HI) -- $w(\MgII\lambda$2796) plane in Fig. ~\ref{nhi_vs_wmg_ii_fig}. We note that the $N$(\HI) for ESDLAs and $w(\MgII\lambda$2796) for the \citet{Matejek2013} sample are restricted by their respective selection technique. Although, we note that the $w(\MgII\lambda$2796) for the sample of ESDLAs is quite varied and no significant relation can be drawn between the two quantities shown in the figure. \\

In addition to the \MgII\, equivalent width, we also report the velocity spread of \MgII\, absorption using \delv\,(\MgII$\lambda$2796) similar to \citet{Zou2018} in Table~\ref{Column_density_table_1}. In their paper, \citet{Zou2018} define \delv\, as the velocity separation between the two extreme pixels where the optical depth $\tau\,<$ 0.1. We note that the velocity spread of ESDLAs is different (K-S test p-value=0.02) and statistically smaller (median=150 \kms) than that of the \CI-selected absorbers (median=390 \kms) \footnote{The velocity spread of '\MgII$\lambda$2796' in \CI-selected absorbers was obtained with private communication from the authors of \citet{Zou2018}.}.

\subsection{\NaI\  and \CaII} 

With an ionisation potential of 5.1~eV, neutral Sodium (\NaI) serves as a tracer for cold neutral gas \citep[see][]{Crawford1992}. We did not find any significant trace of \NaI\, in our sample. The stronger transitions on \NaI\, lie in the far end of the X-shooter near-infrared (NIR) arm (rest wavelengths 5891\AA\ and 5897\AA), where the signal-to-noise ratio (S/N) is very poor in our observations. Hence, we note that the non-detection in our observations is not necessarily an indication of a low abundance of \NaI. In Table \ref{Column_density_table_2}, we report the 3$\rm \sigma$ upper limit on the column density of \NaI\, in our sample. \\

A \CaII\, absorption signature has been associated with dusty DLAs at high redshift \citep[see e.g.][]{Wild2006, Nestor2008}.  We found signatures of \CaII\, absorption in seven ESDLAs. For weak transitions, we calculated the equivalent width ($w$) of the absorption profile. In cases where the spectra have a low S/N, we checked whether the value for the equivalent width is greater than a 3 $\sigma$ uncertainty and we report it as detection if this is true. If not, we used the Voigt profile model to estimate a 3$\sigma$ upper limit on the column density. The NIR spectrum has significant sky residuals hindering a robust estimate of the equivalent width of some transitions such as '\CaII$\lambda$3934' and '\CaII$\lambda$3969'. For systems towards QSO J1513$+$0352 and J2246$+$1328, the respective transitions, '\CaII$\lambda$3934' and '\CaII$\lambda$3969', are contaminated. Hence, we declare them to be tentative in our list. In Table \ref{Column_density_table_2}, we report the column density of \CaII\, in our sample. Additionally, in Table ~\ref{Column_density_table_1}, we report the equivalent width of the '\CaII$\lambda$3934' transition as this has been previously used in the literature. \\

\begin{figure*}[!t]
 \centering
 \addtolength{\tabcolsep}{-3pt}
 \begin{tabular}{cc}
    \includegraphics[trim=10 10 10 0,clip,width=0.48\hsize]{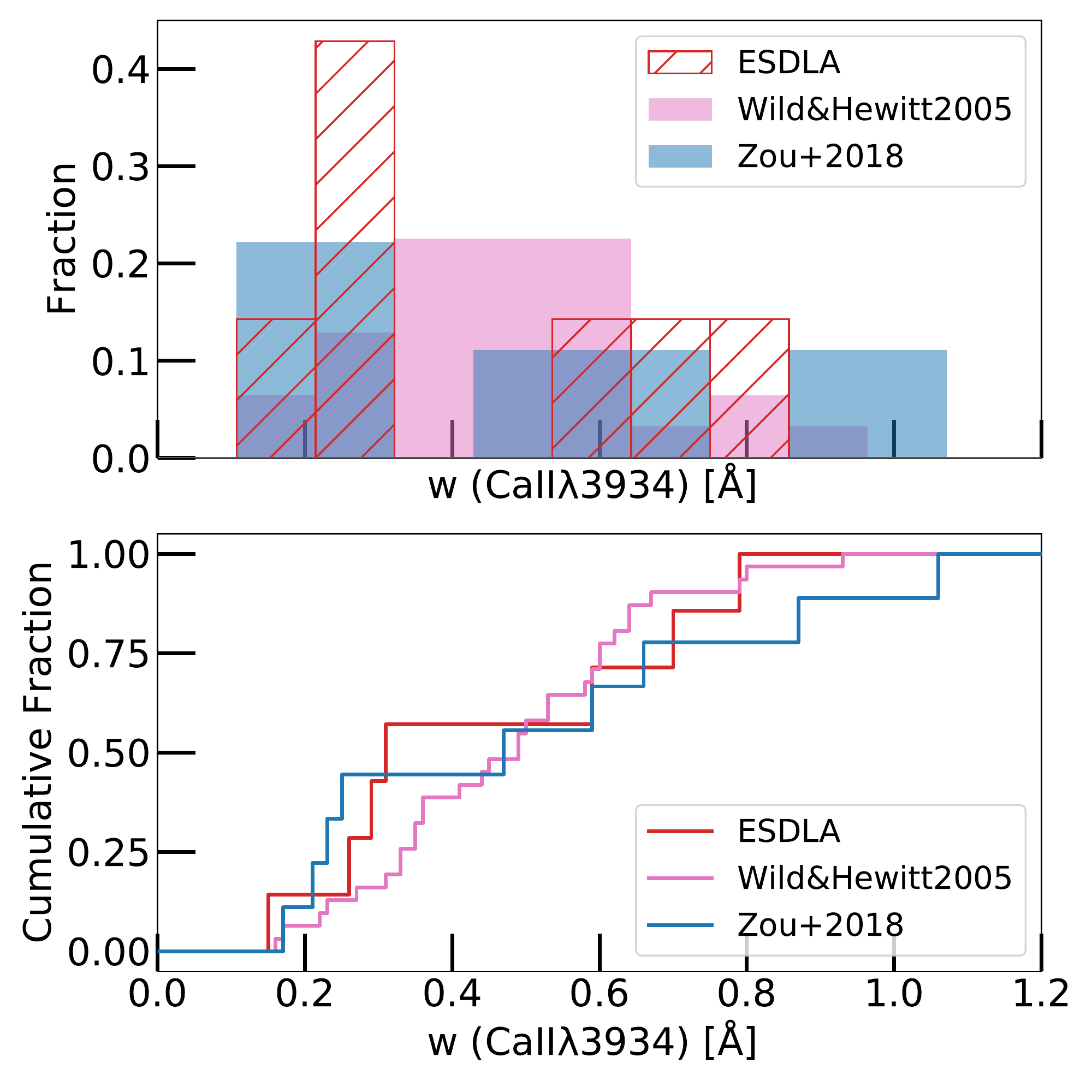} & 
    \includegraphics[trim=10 10 10 0,clip,width=0.48\hsize]{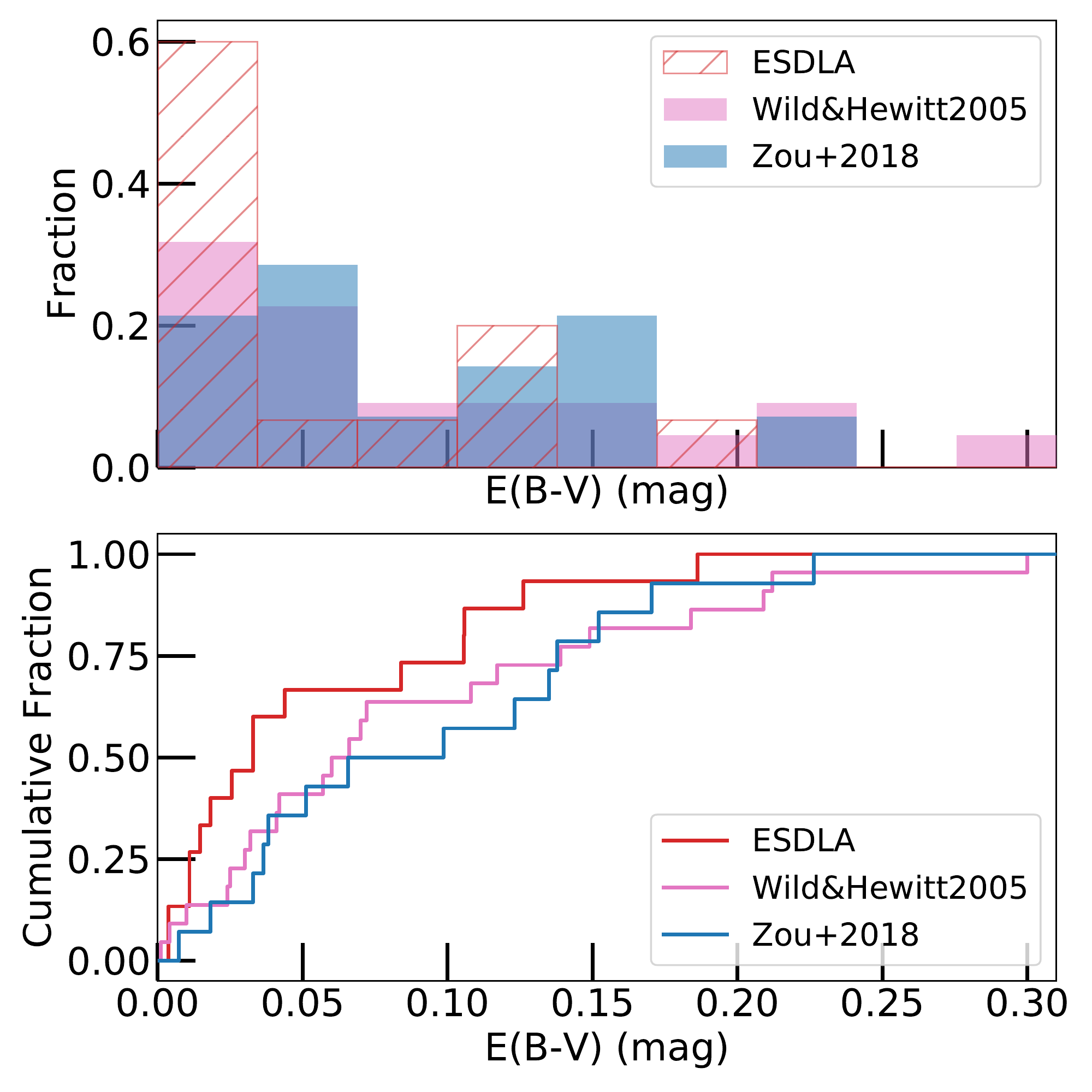} \\
 \end{tabular}
  \addtolength{\tabcolsep}{+3pt}
  \caption{{\sl Distribution of $w$(\CaII\,$\lambda$3934) and dust in different samples. Left panels}: $w$(\CaII\,$\lambda$3934) distribution from different samples of DLAs (top). ESDLAs are plotted as red lines, while blue and pink shades represent the \CI-selected \citep{Zou2018} and \CaII-selected \citep{Wild_and_Hewett2005} samples, respectively. Bottom panel: Cumulative distributions for the same samples in the same colours. 
  {\sl Right panels}: Similar plot for the measured colour-excess, E(B-V), quantifying dust extinction.}
  \label{caii_3934_eq_width_distribution}
\end{figure*}

Fig. ~\ref{caii_3934_eq_width_distribution} shows the distribution of $w$(\CaII$\lambda$3934) (left) and E(B-V) (right) for ESDLAs in our sample, CI-selected absorbers from \citet{Zou2018}, and the \CaII-selected sample from \citet{Wild_and_Hewett2005}. The E(B-V) for ESDLAs were obtained from \citet{Ranjan2020}. K-S tests between ESDLAs and CI-selected absorbers for $w$(\CaII$\lambda$3934) (p-value=0.79) and E(B-V) (p-value=0.18) and with the \CaII-selected absorber sample, with a p-value=0.31 for $w$(\CaII$\lambda$3934) and a p-value=0.41 for E(B-V) indicate that the dust content and \CaII\, abundance for all three samples are indistinguishable.

\subsection{\SII, \NiII, \MnII, \TiII,\, and \PII}

In this subsection, we further report the detection of five other singly ionised species, \SII, \NiII, \MnII, \TiII,\, and \PII\, in our sample. These low ionisation species are commonly detected in DLAs. For all five species, the redshift and $b$-values are tied to other low ionisation species such as \FeII\, and \SiII. We note that \NiII\,  is detected in all ESDLAs, and \MnII\, is also detected in all ESDLAs, except towards QSO J1513+0352. For this ESDLA, the two \MnII\, transitions, $\lambda = 1162\AA$ and $\lambda = 1199\AA$, are contaminated by the \lya\, forest. The remaining three \MnII\, transitions, $\lambda = 2576\AA$, $\lambda = 2594\AA$, and $\lambda = 2606\AA$, are strongly blended with the sky residuals in the VIS arm spectra of X-shooter. \\

Single ionised sulphur (\SII) is a good tracer of gas phase metallicity (similar to \ZnII) as sulphur is hardly depleted onto dust. However, the relevant \SII\, transitions, $\lambda = 1250\AA$, $\lambda = 1253\AA$, and $\lambda = 1259\AA$, are often contaminated by the \lya\, forest lines. Due to this, the robust measurement of the column density is not possible in all cases. For six of our ESDLAs, the \SII\, profiles were free of \lya\, contamination. Out of these, the non-blended profile of one system was saturated (towards QSO J0025$+$1145) and hence we could only estimate the lower limit on $N$(\SII). For another ESDLA system, towards QSO J1258$+$1212, we found one slightly blended transition, '\SII$\lambda$1250', and using that line, we give a tentative estimate on the column density. For the remaining four ESDLAs, towards QSOs J2140$-$0321, J2232$+$1242, J2246$+$1328, and J2322$+$0033, we obtained a robust estimate on the column density. All the estimations are given in Table ~\ref{Column_density_table_2}. \\

We further looked for the \TiII\, transitions at $\lambda = 3384\AA$, $\lambda = 3242\AA$, $\lambda = 3230\AA$, $\lambda = 3073\AA$, $\lambda = 3067\AA$, and $\lambda = 1910\AA$. For all ESDLA rest frames in our study, all mentioned transitions (except $\lambda = 1910\AA$) lie in the NIR arm region of the X-shooter spectrograph (with the exception of the ESDLA towards QSO J2246$+$1328 for which the $\lambda = 3073\AA$ and $\lambda = 3067\AA$ transitions also lie in VIS arm region). There is strong contamination from sky lines in the NIR region. In addition, the spectra in the NIR region have significantly lower S/Ns and they are more strongly blended with the sky lines relative to the VIS and UVB band. Using different combinations of the above-mentioned transitions, we obtained a robust estimate of the \TiII\, column density. For systems where none of the NIR lines can be used for the fit, we report a tentative value for the lower limit using the $\lambda = 1910\AA$ transition. The column density estimates and limits for \TiII\, are reported in Table. ~\ref{Column_density_table_2}. \\

For most ESDLAs, the \PII\, transitions, $\lambda = 961\AA$, $\lambda = 963\AA$ and $\lambda = 1152\AA$, are strongly contaminated by the \lya\, forest and $\lambda = 1301\AA$ and $\lambda = 1532\AA$ are often weak and not detected. We used a combination of multiple transitions mentioned above to obtain a robust column density estimate of \PII\, in ESDLAs towards QSO J1411$+$1229, J2140$-$0321, J2232$+$1242, and J2246$+$1328. In the appendix, the metal line plots (see section \ref{abs_figs}) for all ESDLAs show the relevant transitions used for fitting \SII, \NiII, \MnII, \TiII,\, and \PII. \\

\subsection{Dealing with saturation effects for \SiII\label{siii_saturation_subsection}} 

Absorption lines from \SiII\ are frequently detected in high-z DLAs and they are a good tracer of neutral \HI\ gas. While \SiII\ has a detectable absorption spread over a wide range of rest wavelengths, only a few transitions are outside the Lyman-$\alpha$ forest. For estimating the total column density of \SiII,\, we tied the $b$-value and redshift of individual \SiII,\, components to \FeII\, and fitted the observed absorption lines in the spectra with a combination of many \SiII\, transitions with a varying oscillator strength, with namely $\lambda = 989, 1020, 1190, 1193, 1260, 1304, 1526, 1808,$ and $2335~\AA$ taking care of possible blending. The total \SiII\ column densities from this analysis are listed in Table~\ref{Column_density_table_3}. However, for some systems, such as ESDLAs toward QSO J0025$+$1145, J1349$+$0448, J2140$-$0321, and J2232$+$1242, it seems that the absorption profiles of most \SiII\, transitions are saturated. In these cases, in Table~\ref{Column_density_table_3}, we give the allowed range for the \SiII\ column density by getting a conservative upper limit on $N$(\SiII) using the line at  $\lambda = 2335\AA$ with the weakest oscillator strength.

\subsection{Warm and hot gas tracers\label{warm_hot_gas_sections}}

In addition to cold gas, a DLA line of sight also reveals ionised warm gas ($T\sim\,10^{4}$~K, traced by \SiIV\, and \CIV) and hot gas (traced by \NV\, and \OVI), residing in the warm and hot ionised medium as well as the halo of the associated galaxy. \citet{Fox2007} searched for warm and hot gas signatures in DLAs using VLT/UVES data. We searched for the same thing in our medium resolution ESDLA sample. \\

All the ESDLAs in our sample show absorption lines of warm gas tracers (\SiIV\, and \CIV). We performed a multi-component Voigt profile fitting by simultaneously tying the $b$-value and redshift of all detected warm and hot gas species. Contrary to the plethora of lines for singly ionised species such as \FeII,\, there are fewer detected lines for warm gas species. Hence, we have to be careful when dealing with intrinsic saturation (see section \ref{sect:OI} for details). To deal with this issue, we created a mock Voigt absorption profile models for the transitions with a relatively lower oscillator strength -- '\CIV$\lambda$1550' and '\SiIV$\lambda$1402'. To create this mock, we needed to estimate a certain minimal $b$-value  for the fitting. The $b$-value is dependant on the thermal properties of the transition ($b_{\rm thermal}\,=\sqrt{2kT/m}$, where $k$=Boltzmann's constant, $T$=temperature of the gas, and $m$=atomic mass) as well as the turbulence in the medium. Since we do not have any way to distinguish between the two, for a conservative lower limit, we assumed that the gas is non-turbulent and hence, the lower limit on $b$-value was obtained just from the thermal component. Taking $T\sim$1000\,K, we got $b$-value$\rm \sim$3.7 \kms\, for \CIV\, and $\rm \sim$2.4 \kms\, for \SiIV. Using these as lower limit estimates for $b$-value, we fitted our \CIV\, and \SiIV\, profiles. We note that the $b$-value for our individual fitted components is always $>$25 \kms. Assuming that there are no hidden saturated components with a $b$-value between 2.4 \kms and 25 \kms, we note that the "\CIV$\lambda$1550" and "\SiIV$\lambda$1402" profiles have an optical depth, $\rm \tau_{0}\sim\,1$ at log$N$(\CIV)$\sim$14.3 and log$N$(\SiIV)$\sim$13.7. Any column density measurement higher than these limits might indicate saturation of the line and hence, the corresponding column density estimates are reported as lower limits\footnote{We note that high spectral resolution (R$\sim$40000) observations would be required to check the presence of any hidden components with a $b$-value between $\sim$2.5 to 25 \kms. We still declare our results as robust assuming that the warm gas $b$-value would likely have a turbulent component of b$_{\rm turb}\sim$20 \kms\, or higher.}. Using this information, we report the total column density estimates of \CIV\, and \SiIV\, lines in Table.~\ref{Column_density_table_1}. The column density for \SiIV\, and \CIV\, in our sample ranges from 13.3 to $>$13.7 (in log scale) for \SiIV\, and 13.5 to $>$14.3 (in log scale) for \CIV,\, respectively. For the ESDLA towards QSO J1349$+$0448, the \SiIV\, transitions are heavily contaminated by the \lya\, forest lines and hence, the corresponding $N$(\SiIV) could not be estimated. \\

\begin{figure}
\centering
   \includegraphics[width=1\hsize]{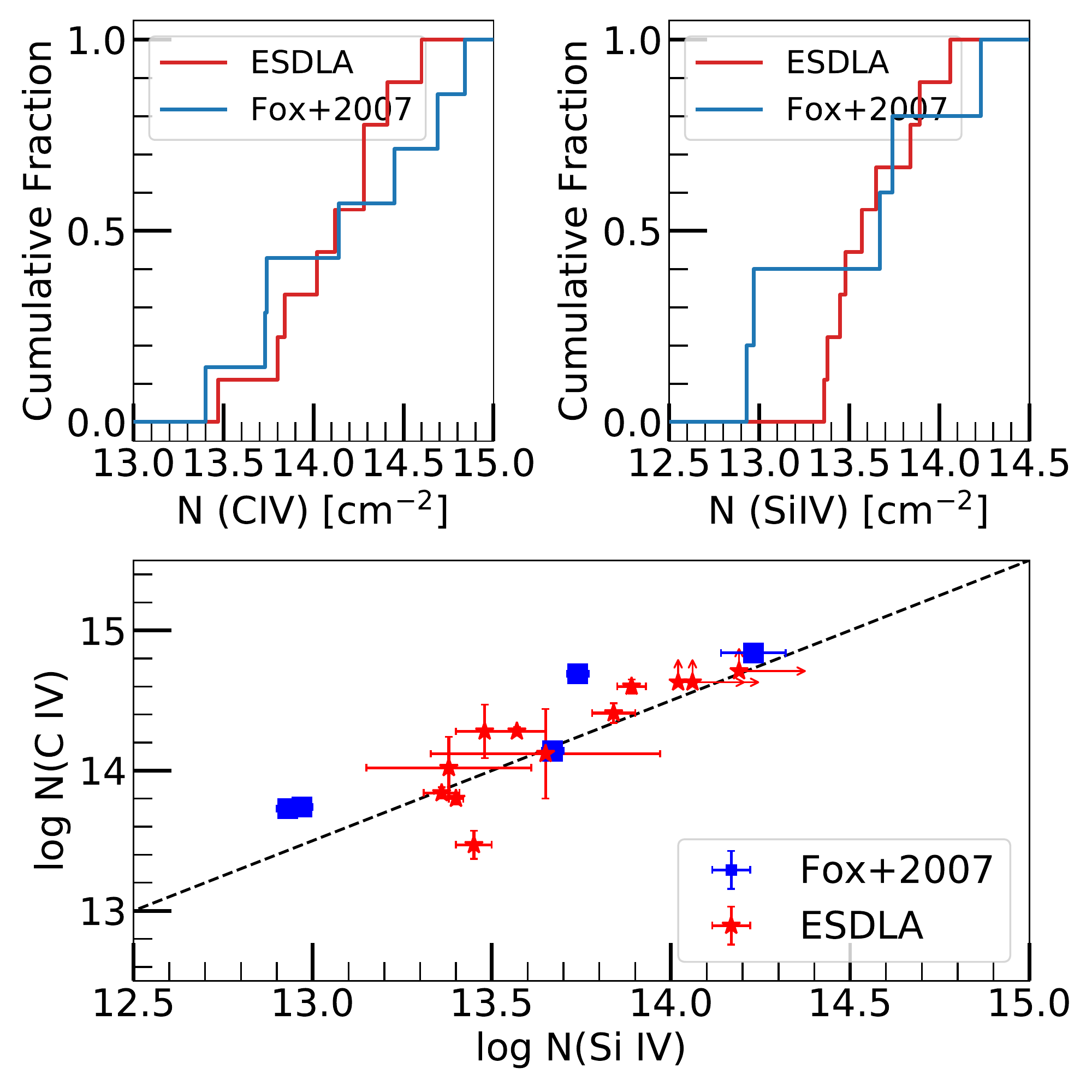}
    \caption{{\sl $N$(\CIV) distribution. Top Left panel}: log $N$(\CIV) cumulative distributions for the general DLA population \citep[blue line, from][]{Fox2007} and ESDLAs (red line). 
    {\sl Top Right panel}: Same for the log $N$(\SiIV) distributions. 
    {\sl Bottom panel}: log $N$(C~{\sc iv}) vs. log $N$(Si~{\sc iv}) for ESDLAs (red points) and DLAs \citep[blue points, taken from][]{Fox2007}. The straight line indicates log($N$(\CIV))=log($N$(\SiIV))+0.5.}
    \label{n_civ_vs_n_siiv_fig_1_new}
\end{figure}

The bottom panel of Fig.~\ref{n_civ_vs_n_siiv_fig_1_new} shows that the column densities of both of these species, \SiIV\, and \CIV, for the general DLA population and ESDLAs are correlated. In the top subplots of the figure, we show the cumulative distributions of \CIV\, and \SiIV\, column densities for the general DLA population and ESDLAs. The corresponding high p-values obtained for the K-S tests between the samples [0.74 for $N$(\CIV) and 0.56 for $N$(\SiIV)] indicate that they originate from the same distribution. This is expected, as both ESDLA and normal DLA sightlines will also sample warm gas associated with the host galaxy (from their circumgalactic medium, or CGM) despite sampling different H~{\sc i} regions. We note that these warm gas clouds (traced by \CIV\, and \SiIV) can also originate from a combined effect of many ionised bubbles present within the interstellar medium (ISM) of the associated galaxy. Yet, this seems unlikely because the cross-section of ionised bubbles in the ISM gas is small (a volume filling factor of $\sim$15-20\% found in studies in the local universe, see \citet{Berkhuijsen2006}). There could also be a contribution of warm gas from within the galactic disk originating from a multi-phase ISM model \citep[as described by][]{Ferriere2001, Cox2005}. More recently, \citet{Werk2019} have shown that the size of the warm ionised medium in the Milky Way can be greater than 1 kpc in size. Given that the typical size of ESDLA galaxies is around $\sim$2.5 kpc \citep[see discussions in][]{Guimaraes2012, Ranjan+2018, Ranjan2020}, we cannot ignore the contribution of warm gas from within the galactic disk. However, it is also imperative from geometry that any sightline passing through the galaxy also has to cross its corresponding CGM. This idea is further supported by the fact that warm gas is detected in all DLAs (including ESDLAs). Hence, the warm gas signature seen in ESDLAs can be from a mixed contribution from both the CGM and the ISM of the associated galaxy. \\

Due to the \lya\, forest confusion, we could not estimate the column density of any hot gas tracer (\NV\, and \OVI\, transitions) 
in our sample, except towards QSO J1349$+$0448, in which we detected the '\NV$\lambda$1238' transition (fixing the position and $b$-value from \CIV). However, we declare this detection as tentative (with log $N$(\NV)$\sim$14.16) as the second transition of the doublet ('\NV\,$\lambda$1242') is contaminated by the \lya\, forest. We do show the normalised spectra of \NV\, and \OVI\, transitions for some other ESDLA systems (see section \ref{abs_figs}), yet they are heavily contaminated with the \lya\, forest lines.

\section{Discussion \label{discussion}}

\subsection{Comparing dust-corrected abundances}
In absorption line literature, the metallicity of the neutral gas in a system is given as the ratio of the column density of the least dust depleted element, X, to the total hydrogen column density. Volatile elements, such as Zinc or Sulphur, are commonly used in the literature \citep[see e.g.][]{Kulkarni2002}, since they are usually accessible in the spectrum and their lines are unsaturated. We note that \ZnII\, was detected in all ESDLAs in our sample and we used Zn (assuming $N$(Zn)=$N$(\ZnII) in neutral gas clouds) to calculate the metallicity of ESDLAs in \citet{Ranjan2020}\footnote{We note that the uncertainty in $N$(\ZnII) along QSO J2246$+$1328 was reported incorrectly in Paper I, leading to relatively large errors in metallicity and depletion measurements. The revised value for the \ZnII\, column density along J2246$+$1328 is log $N$(\ZnII)=12.52$\pm$0.1. We further revised the value of metallicity as [Zn/H]=-1.84$\pm$0.1 and the depletion as [Fe/Zn]=-0.14$\pm$0.13.}. \\

\begin{figure}
\centering
   \includegraphics[width=1.0\hsize]{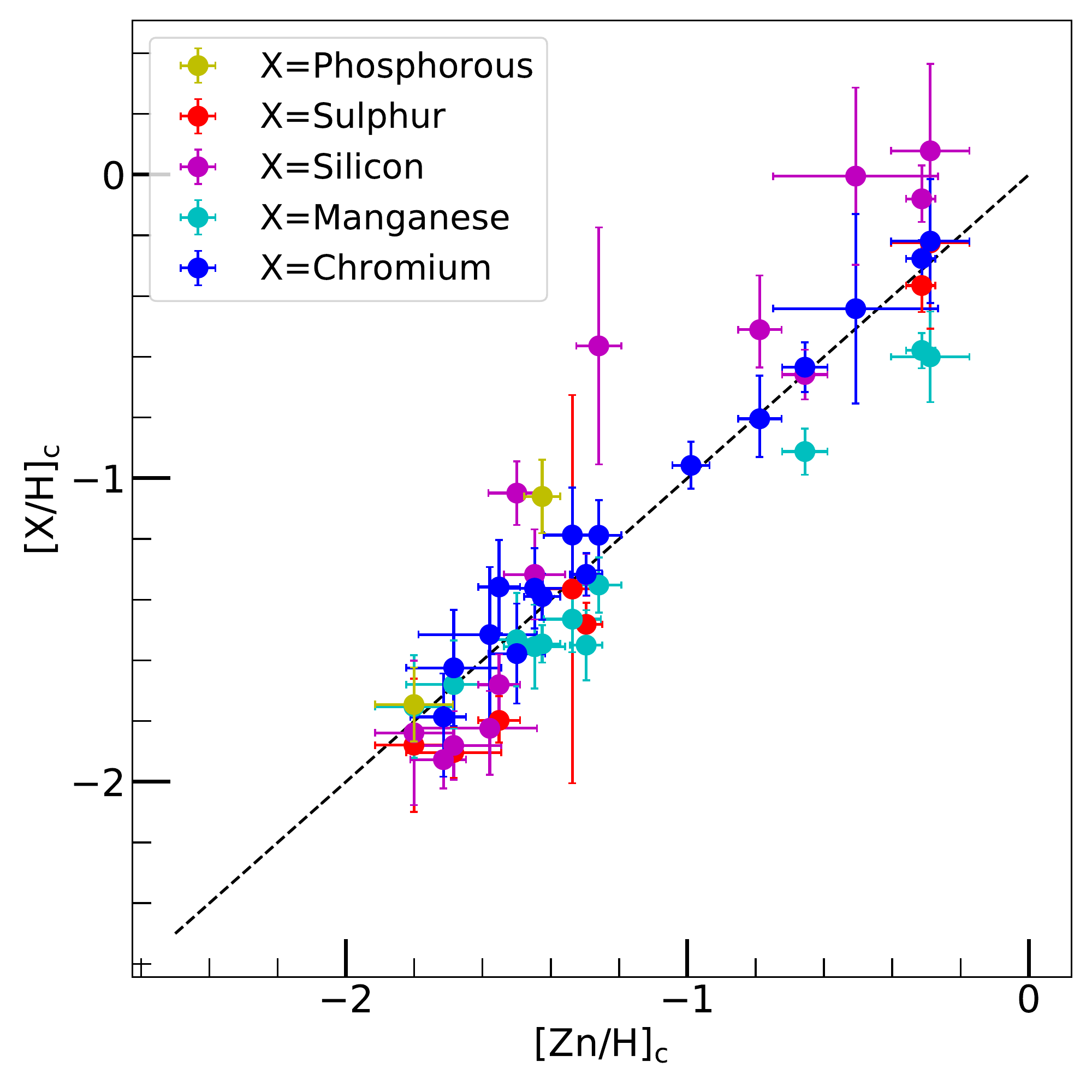}
    \caption{Depletion-corrected metallicity obtained from different species ([X/H]$\rm _c$) as a function of [Zn/H]$\rm _c$, where X is a common, singly ionised species found in DLAs as described in the legend. The dashed black line describes [X/H]$\rm _c$=[Zn/H]$\rm _c$. The depletion corrections were performed using the equations and information from \citet{DeCia2016}}
    \label{metallicity_comparison_fig_1}
\end{figure}

Since we obtained the column density of many other neutral gas species in our ESDLA systems, we could further check the dust-corrected metallicity for all ESDLAs. We did this in an attempt to help understand the influence of dust depletion on individual neutral gas tracer species. For this comparison, we corrected the metallicity (say, e.g. [X/H], where X=Phosphorous, Sulphur, Silicon, Manganese, and Chromium) for depletion as per the method described in \citet{DeCia2016} to obtain a depletion-corrected metallicity ([X/H]$\rm _c$) for all ESDLA systems with a robust estimate of X (uncertainty in log$N$(X)$<$0.5). In Fig.~\ref{metallicity_comparison_fig_1}, we compare the depletion-corrected metallicity obtained using Zinc, with metallicities measured based on other singly ionised atomic gas species commonly reported in high-$z$ absorption line studies, such as \PII, \SII, \SiII, \MnII,\, and \CrII. In the same figure, we also show a straight line indicating [X/H]$\rm _c$ = [Zn/H]$\rm _c$. We conclude that the dust correction using the method described in \citet{DeCia2016} is robust and that the metallicities calculated using different species are consistent within $\sim$3-$\rm \sigma$ uncertainty in all cases. We also note that after dust corrections, the metallicity, [X/H]$\rm _c$ of ESDLAs range from $\sim$ -1.8 to -0.2.

\subsection{Neutral Argon in ESDLAs \label{neutral_argon_discussion_subsection}}

\begin{figure}
    \centering
    \includegraphics[width=\hsize]{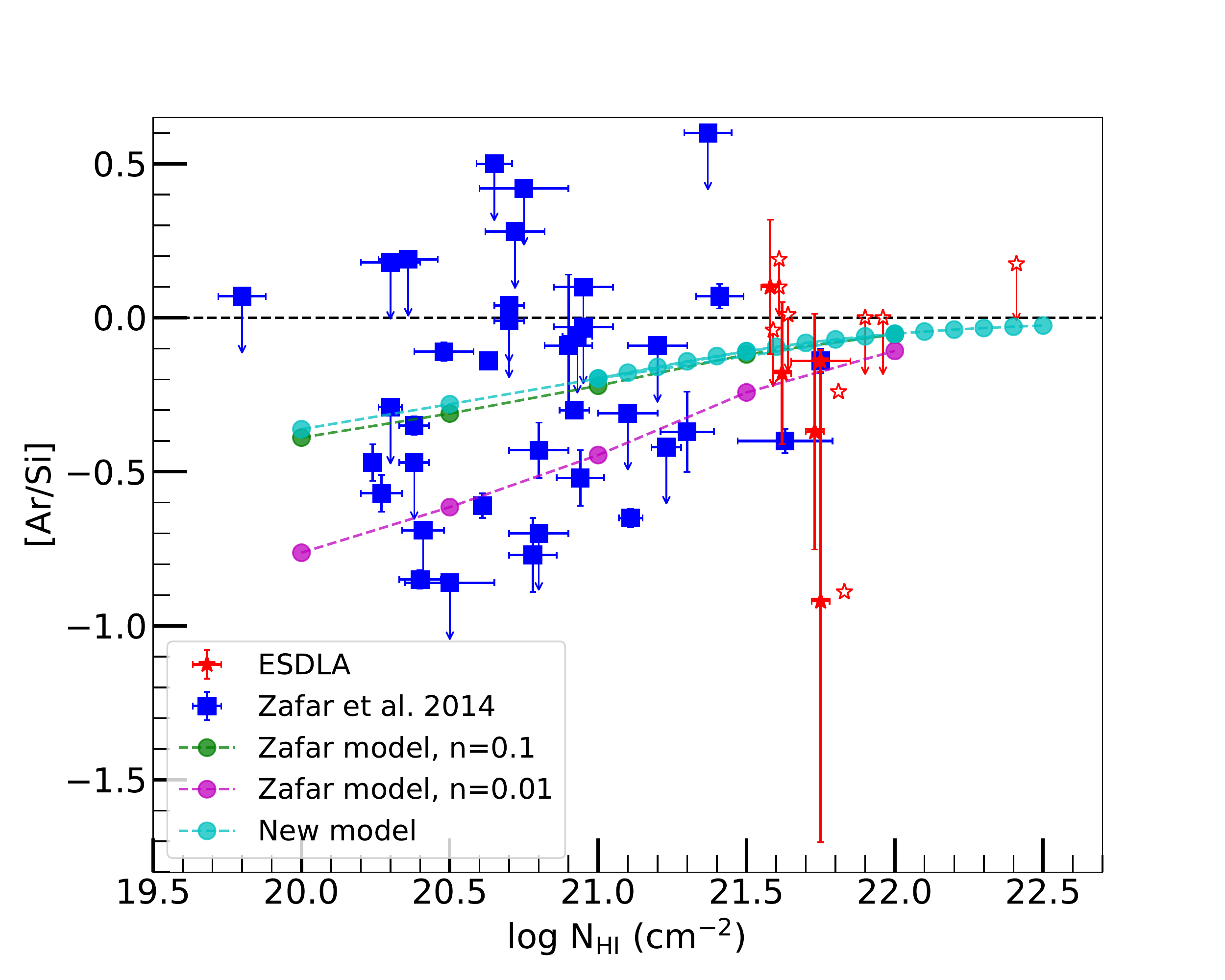}
    \caption{Ratio of neutral argon \ArI\, to \SiII\, column density (relative to solar), [Ar/Si], plotted as a function of the \HI\ column density. Red stars indicate measurements for ESDLAs from this work, with empty stars indicating tentative detections of \ArI\, and down arrows representing upper limits. Blue squares are measurements for DLAs from \citet{Zafar2014}. The green dots represent a series of photoionisation models run by \citet{Zafar2014} for gas with a solar abundance ratio, low density ($n_{\rm H}$=0.1 atoms cm$^{-3}$), and Haardt \& Madau (HM) extragalactic background radiation \citep[see][]{Haardt_Madau_2012} at $z=2.5$. The purple dots also represent similar models, but with a lower density ($n_{\rm H}$=0.01 atoms cm$^{-3}$). The cyan points are a series of photo-ionisation models from this work with an updated Khaire \& Srianand extragalactic background radiation field \citep[see][]{Khaire_Srianand_2019} and the cosmic microwave background (CMB) at redshift, $z=2.5$, solar abundance ratios, log$N$(\HI) ranging from 20 to 22.5, a metallicity of [Zn/H]$\sim$0.05 solar (average ESDLA metallicity), and a density, $n_{\rm H}$=0.1 ~cm$^{-3}$. The photoionisation model results were obtained using CLOUDY. }
    \label{Ar_over_Si_vs_n_HI}
\end{figure}

The first ionisation potential of Argon (15.76~eV) is higher than the \HI\, ionisation threshold (13.6~eV) and hence, Argon largely should remain neutral in \HI\, clouds. However, \citet{Sofia_Jenkins_1998} showed that the photo-ionisation to recombination rates of \ArI\, is one order of magnitude higher than \HI,\, indicating that \ArI\, is quite sensitive to high energy ionising photons. Hence, in the presence of UV-background photons, \ArI\, might become deficient relative to other $\rm \alpha$-capture elements, such as Silicon and Sulphur. Indeed, \citet{Jenkins2013} show that \ArI\, is deficient in diffuse \HI\, gas in the Milky Way. \citet{Zafar2014} note a similar deficiency in their sample of DLAs observed at high redshift (2.0 $\rm \leq\,z_{\rm abs}\leq$ 4.2). They conclude that, given the typical DLA metallicity, the deficiency is caused by extragalactic UV photons that ionise neutral Argon in the absence of \HI\, self-shielding and that such a deficiency should not exist in the presence of high-$N$(\HI) gas clouds originating from within the associated galaxy. Since ESDLAs are high-$N$(\HI) gas clouds with a similar metallicity as DLAs that likely reside within their associated galaxy, our sample is ideal to test this conclusion. Following up on the work of \citet{Zafar2014} (and references therein), we plotted Fig.~\ref{Ar_over_Si_vs_n_HI} showing the ratio of \ArI\, over \SiII\, relative to solar ([Ar/Si]), as a function of the \HI\, column density.\\

\citet{Zafar2014} combined modelling with observations to exclude the possibility that \ArI\, is deficient due to dust depletion or nucleosynthesis and they conclude that the deficiency is due to ionizing photons originating from the extragalactic background at that redshift. Their photo-ionisation models are for gas with solar abundance ratios, low density ($n_{\rm H}$=0.1 atoms cm$^{-3}$), and Haardt \& Madau (HM) extragalactic background radiation \citep[see][]{Haardt_Madau_2012} at $z=2.5$. We performed photo-ionisation modelling using the spectral synthesis code CLOUDY \citep[as described by][]{Ferland2017, Shaw2005, Shaw2020} using these parameters. We show the models in Fig. ~\ref{Ar_over_Si_vs_n_HI} as green dots. \citet{Zafar2014} also have a separate set of models with similar parameters and twice the radiation intensity to include the effects of high energy ionising photons. For our analysis, instead of doubling the radiation intensity which would not be physical, we lowered the density of our cloud models (to $n_{\rm H}$=0.01 atoms cm$^{-3}$) instead (shown in the figure with purple dots). The lowering of the density is a better assumption also because, if the ESDLAs are made up of multiple gas clouds, we are unsure about the mean particle density in the clouds, even if we know the total column density. We performed additional modelling with an updated background radiation field \citep[from][]{Khaire_Srianand_2019}. We assumed a plane parallel gas cloud under constant pressure irradiated from both sides. The gas cloud extends up to a given $N$(\HI). The radiation field consists of cosmic microwave background (CMB) and Khaire and Srianand \citep[see][]{Khaire_Srianand_2019} metagalactic radiation at $z=$2.5 along with diffuse radiation due to in situ star formation. This is similar to the photo-dissociation region (PDR) modelling used to interpret observations of high-$z$ DLAs in the literature \citep[see e.g.][]{Shaw2016, Rawlins2018}. We used the updated parameters relative to \citet{Zafar2014} to verify that we obtained consistent results. For the new models, we used solar abundance ratios and log$N$(\HI), ranging from 20 to 22.5. We used a metallicity of [Zn/H]$\sim$-1.3 (average ESDLA metallicity) and a density ranging from $n_{\rm H}$=0.1 to $n_{\rm H}$=100~cm$^{-3}$. Models with $n_{\rm H}\geq$1 do not show any under-abundance in the range of $N$(\HI) mentioned. Hence, we do not show these systems in the plot for clarity. We do, however, show the series of models with $n_{\rm H}$=0.1 (as cyan dots), which seems to be similar to \citet{Zafar2014} for the high $N$(\HI) regime. \\  

Comparing observations with modelling, we note that for a robust $N$(\ArI) measurement, neutral Argon is almost as abundant as singly ionised Silicon in ESDLAs and hence, consistent with photoionisation models, there is no pronounced under-abundance of \ArI\, compared to \SiII. Two other ESDLA systems with tentative \ArI\, detections, towards QSOs J0024$-$0725 and J1411$+$1229, show an under-abundance in \ArI\, (compared to \SiII). For these systems, the column density was estimated using only one transition due to strong \lya\, forest blends in the other \ArI\, transition. Hence, we expect these column density estimates to be quite uncertain. Since, we have no way to robustly quantify the uncertainty (hence the measurements are shown as open stars in the figure), we refrained from interpreting the trend arising from these tentative detections.\\

Our ESDLA sample result indicates that the Argon-Silicon column density ratio seems to be increasing as predicted by the photoionisation models at high column densities. The average [Ar/Si] in ESDLAs is -0.21$\rm \pm$0.1 and that of the photoionisation models is $\sim$-0.05 for $n_{\rm H}$=0.1 ~cm$^{-3}$ and $\sim$-0.11 for $n_{\rm H}$=0.01 to $n_{\rm H}$=100~cm$^{-3}$. We conclude that the mean deficiency of Argon as compared to Silicon (other $\alpha$-capture element) as seen by \citet{Zafar2014} in DLAs is non-existent in ESDLAs (within a 2-$\rm \sigma$ uncertainty). There is a residual deficiency of $\sim$0.2 dex in the Argon-Silicon column density ratio in ESDLAs, but as shown by photoionisation models, this can be attributed to low metallicity and/or a number density of the clouds. We conclude that \ArI\, is likely ionised in low $N$(\HI) (DLAs). This ionisation is likely driven by UV-background photons which cannot penetrate high $N$(\HI) self-shielding ESDLAs. Hence, the under-abundance of \ArI\, relative to \SiII\, is not seen in ESDLAs.

\subsection{Neutral Chlorine in \HH\, bearing gas}

\begin{figure}
\centering
   \includegraphics[width=\hsize]{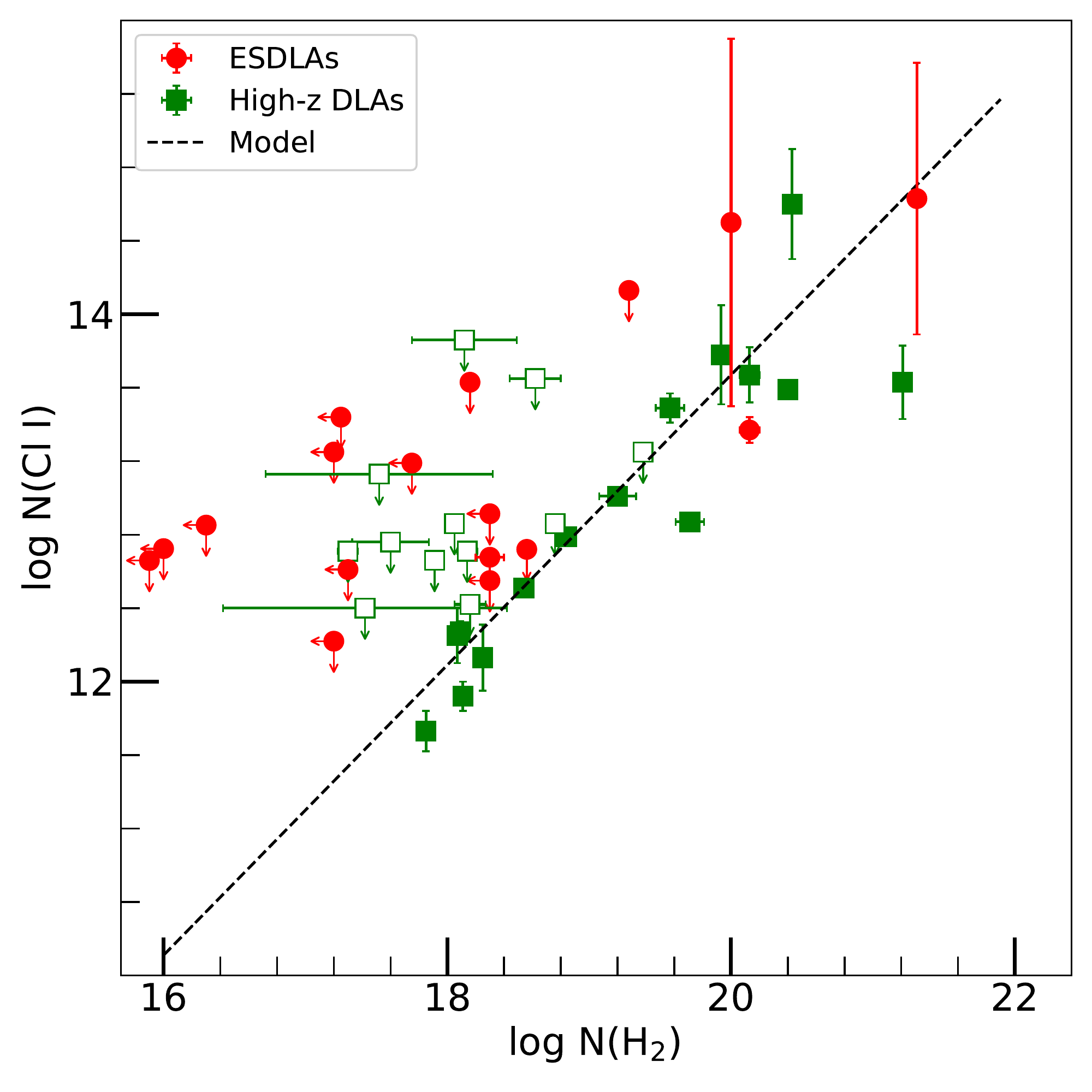} 
    \caption{Neutral chlorine column density (log $N$(\ClI)) vs. molecular hydrogen column density (log $N$(\HH)) in different gas clouds observed in the local universe and high-$z$ as labelled. Most ESDLA points are from this work. Some ESDLA and all high-$z$ DLA points are taken from studies described in the text.}
    \label{neutral_chlorine_vs_h2}
\end{figure}

The production of neutral chlorine (\ClI) is very efficient in the presence of molecular gas \citep[see][]{Jura1974}. Hence, neutral chlorine (\ClI) is an important tracer of molecular hydrogen \citep[see e.g.][]{Balashev2015}. It has been shown in the Milky Way sightlines that there is a direct correlation between $N$(\HH) and $N$(\ClI) \citep[see e.g.][]{Jura1974, Sonnentrucker2006, Moomey2012} (See Figure.~\ref{neutral_chlorine_vs_h2}). Using the data points from the literature, \citet{Wallstrom2019} give a relation between $N$(\HH) and $N$(\ClI), holding for up to two orders of magnitude (18 $<$ log($N$(\HH) $<$ 20), as given below: 

\begin{equation}
    \rm log [N({\ClI})] = (0.79 \pm 0.06)\,\times\,[log(N({\HH}))] - (2.13 \pm 1.15)
.\end{equation}

This correlation has also been shown to be true for other \HH\, bearing high-$z$ DLAs \citep[see e.g.][]{Balashev2015}. The high-$z$ observations available in \citet{Balashev2015} are limited to log $N$(\HH)$<$20.2. In our sample, we have an ESDLA system towards QSO J1513$+$0352 with the highest log $N$(\HH) (=21.31$\pm$0.01) amongst QSO-DLAs. Even for such a high $N$(\HH), the neutral chlorine measured in this study follows the above-mentioned relation between $N$(\ClI) and $N$(\HH). The other two $N$(\ClI) measurements in ESDLAs towards QSO J0025$+$1145 and J2140$-$0321 also show consistency with the relation. Further, \citet{Noterdaeme17} also reported another DLA (towards QSO J0000$+$0048) with log $N$(\HH)=20.43$\pm$0.02 and log $N$(\ClI)=14.6$\pm$0.3. Studies of \CI-selected metal-rich DLAs \citep[see e.g.][]{Zou2018, Noterdaeme2018} report a system towards QSO J0917$+$0154 with a high log $N$(\HH)=20.11$\pm$0.06 with no significant detection of neutral chlorine \citep[as discussed in a private conversation with the authors of][]{Zou2018}. \citet{Balashev2017} reported an ESDLA towards QSO J0843$+$0221 with log $N$(\HH)=21.21$\pm$0.02 and log $N$(\ClI)=13.21$\pm$0.18. We found that the system, J0843$+$0221, slightly deviates from the $N$(\ClI)$-N$(\HH) relation. The environment of the associated GRB-DLAs (gas probed from the host galaxy associated with a $\rm \gamma$-ray burst) have also been shown to be similar to ESDLAs \citep[see][]{Ranjan2020}. In GRB-DLA studies, \citet{heintz2019b} reported a GRB-DLA (towards QSO 181020A) with log $N$(\HH)~=~20.4$\pm$0.04. We report the neutral chlorine in this GRB-DLA to be log $N$(\ClI)=14.6$\pm$0.3 \citep[information received from private conversation with the authors of][]{heintz2019b}. Fig.~\ref{neutral_chlorine_vs_h2} shows all the above discussed data points along with the mentioned $N$(\ClI)$-N$(\HH) relation as a straight line. We conclude that the $N$(\HH)$-N$(\ClI) relation remains consistent up to another two orders of magnitude, log($N$(\HH))$<$22 for all high-$z$ \HH\, bearing clouds.

\subsection{\MgII\, in different DLA sub-samples}

Strong \MgII\, absorption ($w$(\MgII$\lambda$2796)$>1$\AA) is seen in ESDLAs similar to \CI-selected absorbers. Such strong \MgII\, absorption (high $w$(\MgII) $>$1.0\AA) signatures have been linked with starburst related feedback processes in low mass galactic halos \citep[see][]{Prochter2006}. Although further studies such as \citet{Bouche2012} note that strong \MgII\, absorption might not always represent strong star-formation activity. \citet{Zou2018} further conjecture that strong \MgII\, absorption in their sample along with a large velocity spread might be a consequence of either an interaction or star-formation activity in the associated galaxy. To continue this discussion further, we compared the N(\HI), metallicity, velocity spread, \delv(\MgII$\lambda$2796), and equivalent width, $w(\MgII\lambda$2796), for \MgII-selected DLAs from \citet[][]{Matejek2013}, \CI-selected absorbers from \citet[][]{Zou2018}, and ESDLAs from this work. We attempted to understand different properties of these sub-samples that are shown to be associated with galaxies by comparing their corresponding \MgII\, profiles. \\       

\begin{figure*}[!t]
 \centering
 \addtolength{\tabcolsep}{-3pt}
 \begin{tabular}{cc}
    \includegraphics[trim=0 0 0 0,clip,width=0.48\hsize]{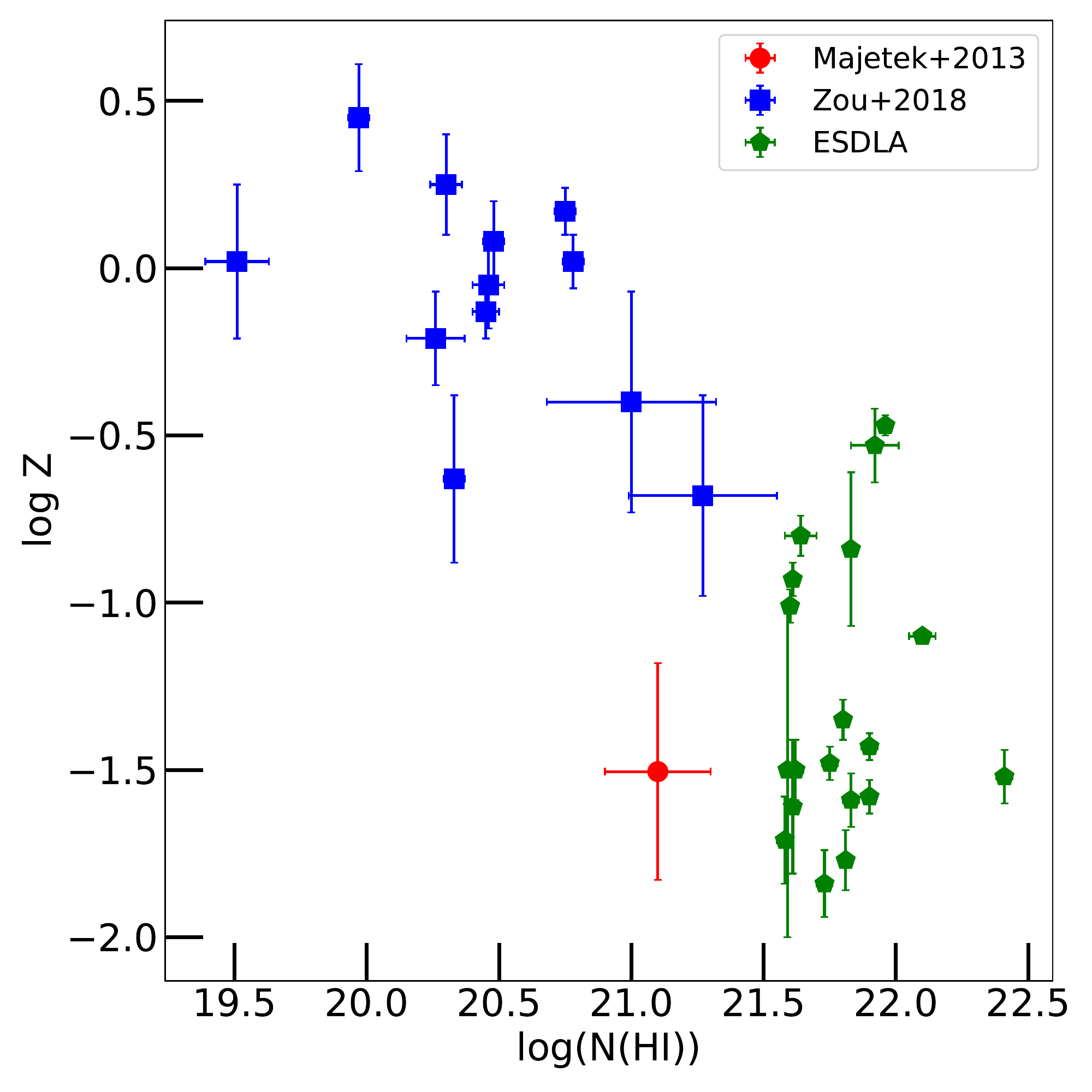} & 
    \includegraphics[trim=0 0 0 0,clip,width=0.48\hsize]{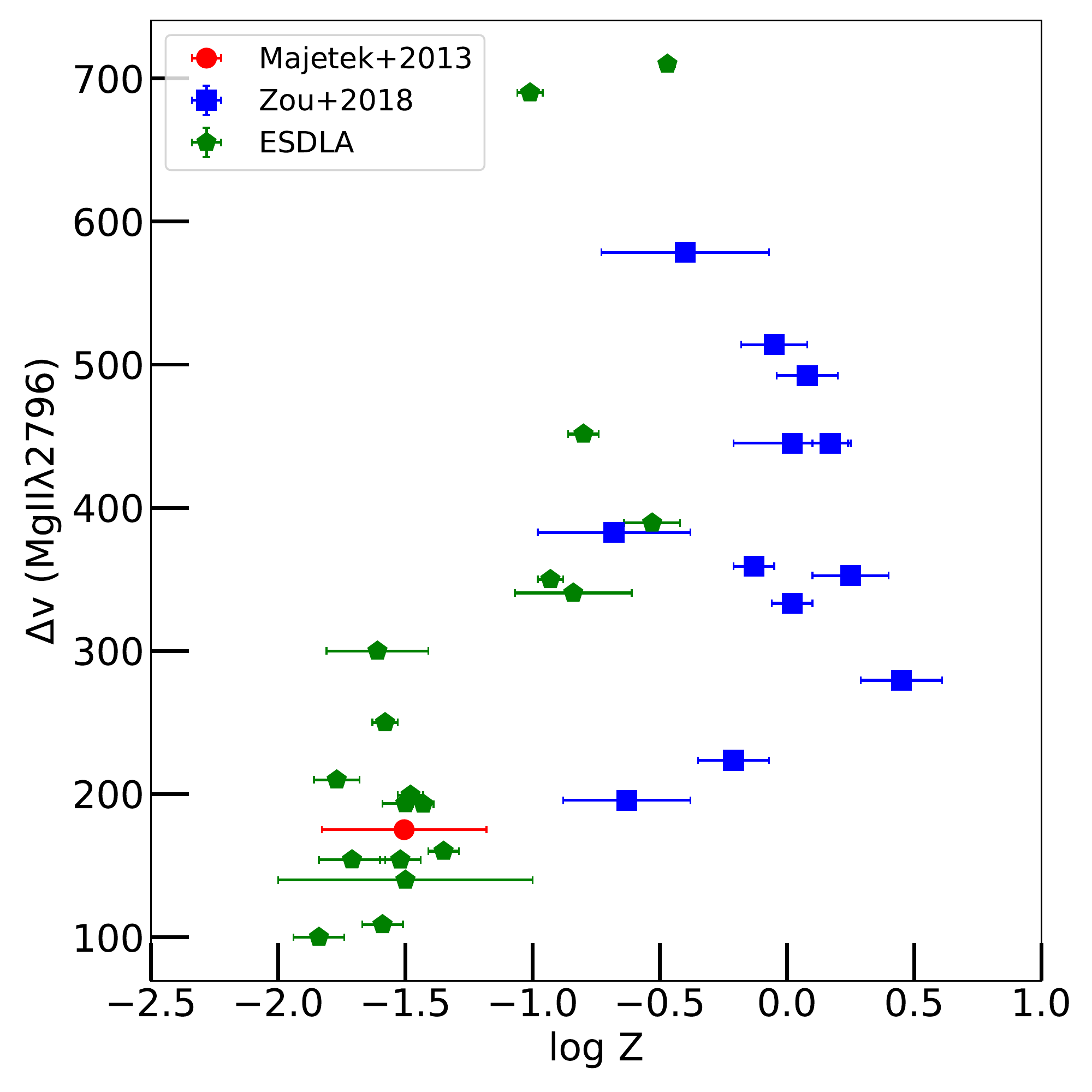} \\
    \includegraphics[trim=0 0 0 0,clip,width=0.48\hsize]{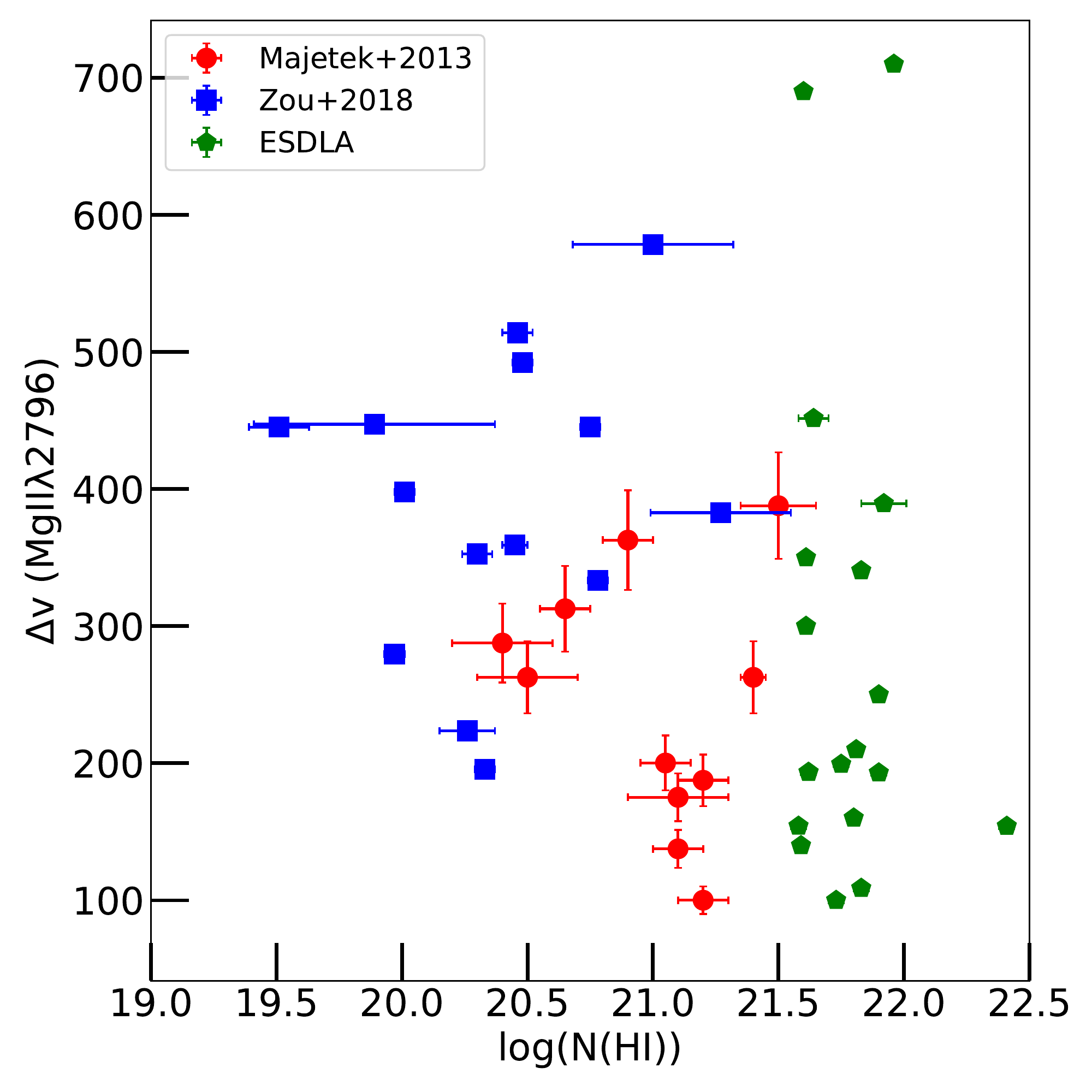} & 
    \includegraphics[trim=0 0 0 0,clip,width=0.48\hsize]{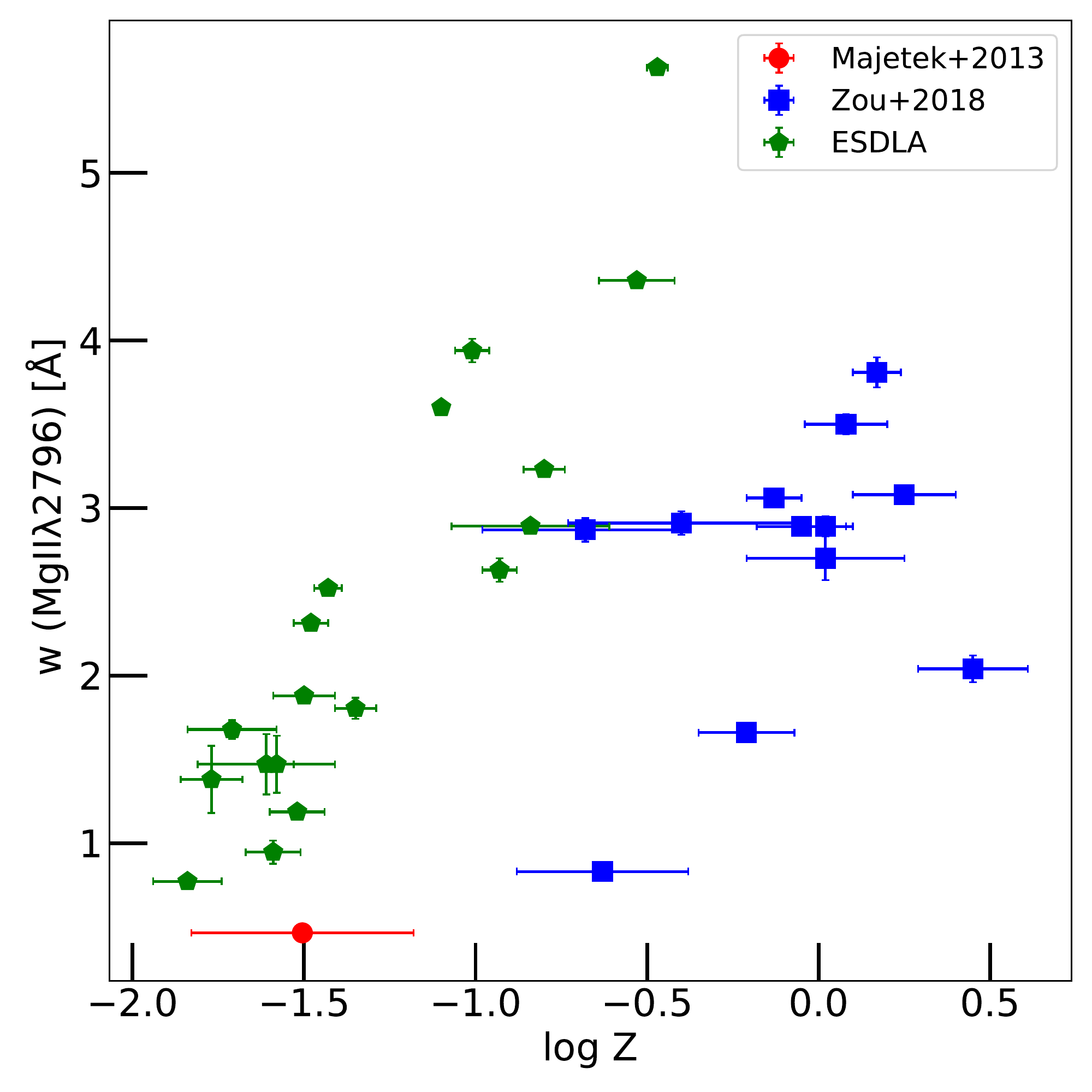} \\
 \end{tabular}
  \addtolength{\tabcolsep}{+3pt}
  \caption{{\sl Comparison between properties of \MgII\, profiles in different samples. Top left panel}: log $N$(\HI) [atoms cm$\rm ^{-2}$] vs. metallicity (log $Z$=[X/H]). {\sl Top right panel}: \delv(\MgII\,$\lambda$2796) [km s$\rm ^{-1}$] vs. metallicity (log $Z$). {\sl Bottom left panel}: Equivalent width $w$(\MgII\,$\lambda$2796) [$\rm \AA$] vs. log $N$(\HI) [atoms cm$\rm ^{-2}$]. {\sl Bottom right panel}: Equivalent width $w$(\MgII\,$\lambda$2796) [$\rm \AA$] vs. metallicity (log $Z$) for ESDLAs (green pentagons, this work), \CI-selected absorbers \citep[blue points, from][]{Zou2018}, and the \MgII-selected DLAs \citep[red circles, from][]{Matejek2013}.}
  \label{mgiivel_nhi}
\end{figure*}

To explore these, Fig.~\ref{mgiivel_nhi} shows the plots between $N$(\HI), metallicity (log $Z$), \delv(\MgII\,$\lambda$2796), and $w(\MgII$ $\lambda$2796). We found that ESDLA metallicity strongly correlates with \delv(\MgII\,$\lambda$2796, Pearson correlation coefficient, $r$=0.79) and $w(\MgII$ $\lambda$2796, $r$=0.91). We also see a correlation between $N$(\HI) of \CI-selected DLAs and \delv(\MgII\,$\lambda$2796, $r$=0.58). We note that while both \CI-selected absorbers and ESDLAs have a high \MgII\, equivalent width (compared with general DLAs), the associated kinematics of \MgII\ indicate that these two sets may be sampling different populations of galaxies. \\   

We note that while \CI-selected DLAs were targeted to have a high \CI\, content and were found to have a high metallicity and ESDLAs to be of high $N$(\HI) by definition, they do indeed probe low $N$(\HI) and low metallicity gas consecutively. Although, the absence of metal-rich high $N$(\HI) gas might just represent the limitation of our selection criteria. Metal- and dust-rich systems will be missed in the optical colour-excess selection based criteria \citep[see][]{Richards2002} for observing QSOs as used by SDSS. Future surveys such as DESI that implement a more robust selection combining optical and NIR colours \citep[see][]{Yeche2020} might help understand whether such systems were just missed in our observations or are indeed rarer to find. \\

\begin{figure}
\centering
   \includegraphics[width=1.0\hsize]{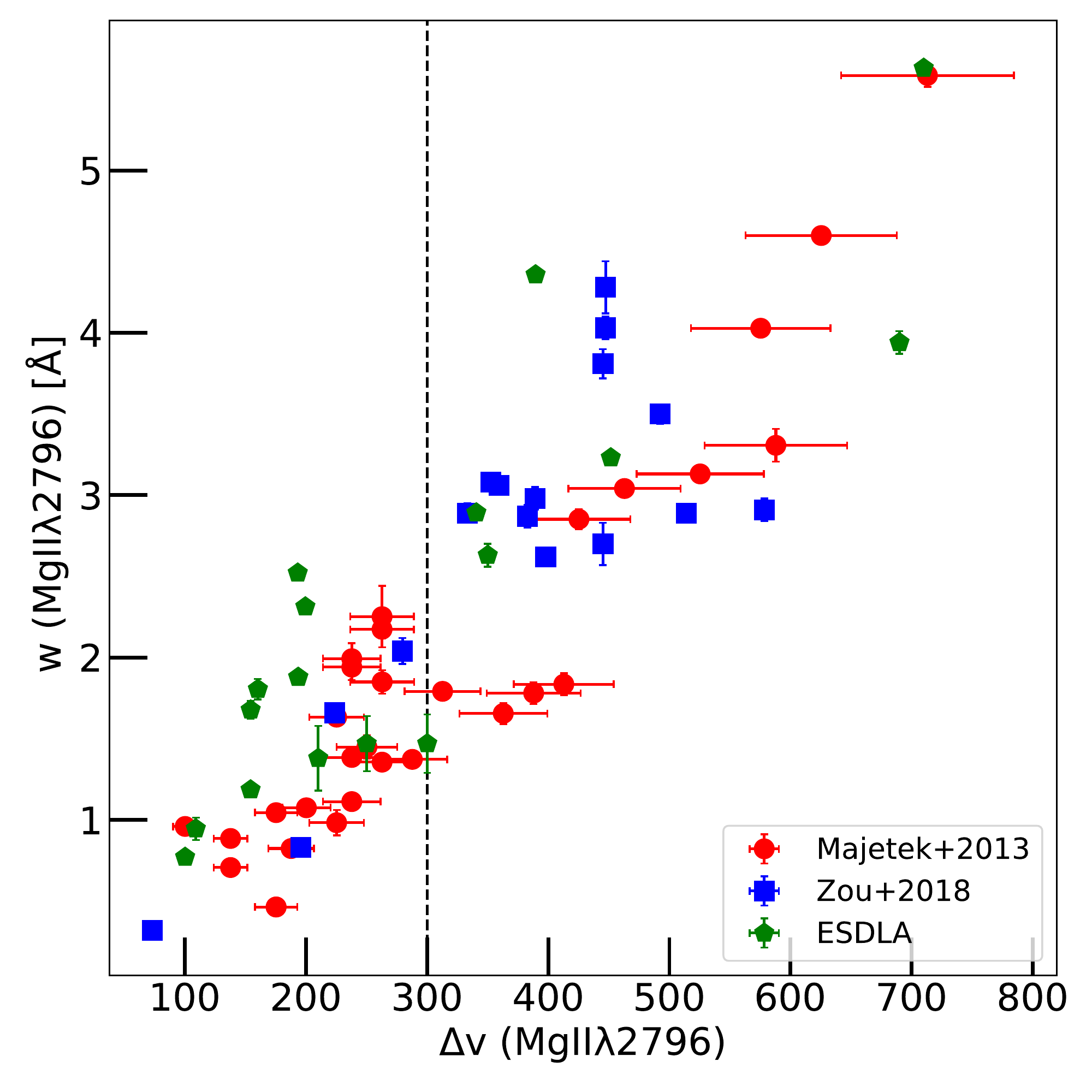}
    \caption{Velocity spread \delv\, vs. the equivalent width ($w$) of the '\MgII$\lambda$2796' absorption line for C~{\sc i}-selected absorbers \citep[taken from][]{Zou2018}), \MgII-selected DLAs \citep[taken from][]{Matejek2013}, and ESDLAs (from this work). The dashed black vertical line represents \delv\,(\MgII$\lambda$2796)=300 \kms.  
   }
    \label{delta_v90_vs_w_mgii_2796}
\end{figure}

In kinematics-metallicity space, we show that ESDLAs and \CI-selected absorbers are clearly distinct entities, while their $w$(\MgII$\lambda$2796) distribution is similar. The C~{\sc i}-selected sample shows high kinematical extension (measured as a \delv\, value for their '\MgII$\lambda$2796' absorption profile) with all but four (out of 17) systems having '\MgII$\lambda$2796' \delv\,$>$ 300 \kms\footnote{We note that the \delv\, mentioned here is measured for saturated or intrinsically saturated lines for both \CI-selected absorbers and ESDLAs and hence cannot be compared with the standard \textbf{\deltav} measurements of unsaturated lines \citep[as shown in][]{Ledoux2006}.}. Fig.~\ref{delta_v90_vs_w_mgii_2796} plots the \delv\, against the equivalent width ($w$) of '\MgII$\lambda$2796' line. We can see that in comparison to \CI-selected absorbers, ESDLAs have a relatively lower kinematical extension (only six out of 18 systems have \delv\,$>$ 300 \kms). In addition to having a high \delv\, value, \citet{Zou2018} also show that some of their DLAs have subsystems in which the \MgII\, and \CI\, absorption profiles are separated by more than $\sim$200 \kms, termed as gas with {disturbed} kinematics in their work. In this work, \MgII\  and \CI\, profiles \citep[see][]{Ranjan+2018, Ranjan2020} in ESDLAs do not show such {disturbed} kinematics as opposed to what is seen in \CI-selected clouds. This is representative of different origins for the ESDLA gas clouds compared to \CI-selected absorbers. \\

As discussed in literature, the connection of strong \MgII\, absorbers ($w$(\MgII$\lambda$2796)$>$1\AA) with star-formation activity is rather ambiguous. In addition, the current observations using X-shooter and other high-resolution ground-based spectrographs do not detect the galaxy stellar continuum in any of the samples discussed above. While emission lines have been detected in ESDLAs that indicate a modest instantaneous star-formation rate \citep[see][]{Ranjan2020}, they are not enough to form a general consensus about the galaxy morphology.  \citet{Rafelski2016} note that the star-formation rate (SFR) deficiency in DLAs are a consequence of poor conversion efficiency of \HI\, to \HH\, in low metallicity environments. While we do note that ESDLAs are metal-poor compared to \CI-selected absorbers, the presence of \HH\, is high in ESDLAs (more than half of the ESDLAs have confirmed diffuse \HH\, detection), similar to the \CI-selected sample. Although, ESDLAs are dust-poor (A$_V\,\sim$0.1) and due to our limited line of sight study, there is no way to confirm as to how much of this diffuse \HH\, gas indeed cools down to form fully molecular regions that can further form stars. Using the analysis above, we conclude that, while \CI-selected absorbers and ESDLAs indeed probe different environments, no strong conclusion can be drawn by this about the nature of star formation in their associated galaxy.

\subsection{Trend of \CaII\, with dust}

The presence of calcium (as \CaII) in high-$z$ DLAs and sub-DLAs have been linked to dusty gas systems from within the halo of their associated galaxy. Calcium gets depleted onto dust and hence, the \CaII\, column density does not necessarily scale linearly with dust. The ionisation potential of \CaII\, is 11.87~eV and is lower than \HI\, (13.6~eV) and hence, \CaII\, is not the main ionisation state of calcium in \HI\, gas. \citet{Nestor2008} indicated that the systems with $w$(\CaII)$>$0.25\AA\, should be DLAs. We note that, among robust detections, all our ESDLAs fulfil this criterion except the ESDLA towards QSO J1513$+$0352. We note that this is the only ESDLA in our sample with \lya\, emission detected in proximity as well as the presence of higher rotational levels of \HH\, \citep[see][]{Ranjan+2018}, indicating a possibly enhanced radiation environment as compared to other ESDLAs. Such enhanced radiation might ionise \CaII\, significantly to show the observed under-abundance. However, further investigation of the effects of radiation on \CaII\, requires detailed modelling and is beyond the scope of discussion for this article.    \\
 
Studies in the literature \citep[such as][]{Zych2009} also suggest that absorbers with $w$(\CaII$\lambda$3934)$\gtrsim$0.7\AA\, should probe diffuse \HH\, gas clouds from within their associated galactic disks. However, \CaII\, is easily depleted onto dust and hence, it does not simply correlate with $N$(\HH) in diffuse gas clouds with a moderate dust content (0.2$<$A$_{V}<$1). The ESDLAs toward QSOs J0025$+$1145, J1411$+$1229, and J2359$+$1354 fulfil this criterion. We found traces of \HH\, in ESDLAs towards QSO J0025$+$1145 (with A$_V$=0.51) in \citet{Ranjan2020} and J2359$+$1354 (with A$_V$=0.29) \citep{Telikova2022article}. In J0025$+$1145, we also found '[\OIII]$\lambda$5007' and H-$\alpha$ emission in very close proximity ($\rho=$1.9$\pm$0.1 kpc) to the absorber line of sight, indicating that the gas originates from within the star-forming disk of the associated galaxy. However, we did not find a significant trace of \HH\, in the ESDLA system towards QSO J1411$+$1229. Since the survival of \HH\, is associated with dust, the extremely low dust content (A$_V$=0.09) as compared to the other two cases can be the reason for the absence of \HH. \\

In the local universe, \citet{Nestor2008} initially showed that $w$(\CaII$\lambda$3934) might not increase linearly with $N$(\HI). However, later studies from the local universe \citep[such as][and references therein]{Murga2015} indicate an increasing trend of $N$(\HI) and E(B-V) with increasing $w$(\CaII$\lambda$3934). Yet the trend disappears once saturation effects become significant (at $N$(\HI)$\sim$ 5$\rm \times\,10^{20}\,cm^{-2}$ and E(B-V) $\sim$ 0.08 mag). At high-$z$, \citet{Wild_and_Hewett2005} were the first to search the SDSS catalogue for \CaII-bearing absorbers and they reported a trend of increasing dust content with an equivalent width of \CaII. Although they also mention that $\sim$40\% of the \CaII\, absorbers would be missed due to the optical selection criteria of SDSS. \citet{Nestor2008} further noted that no trend was found between the \CaII\, equivalent width and metallicity or degree of depletion. They also noted that the strength of \CaII\, lines is determined by a combination of particle density, background UV photons, and the dust content of the gas clouds. \citet{Zou2018} also state that the relation between $w$(\CaII$\lambda$3934) and dust reddening E(B-V) is ambiguous for high-$z$ observations. \\

\begin{figure}
\centering
   \includegraphics[width=1.0\hsize]{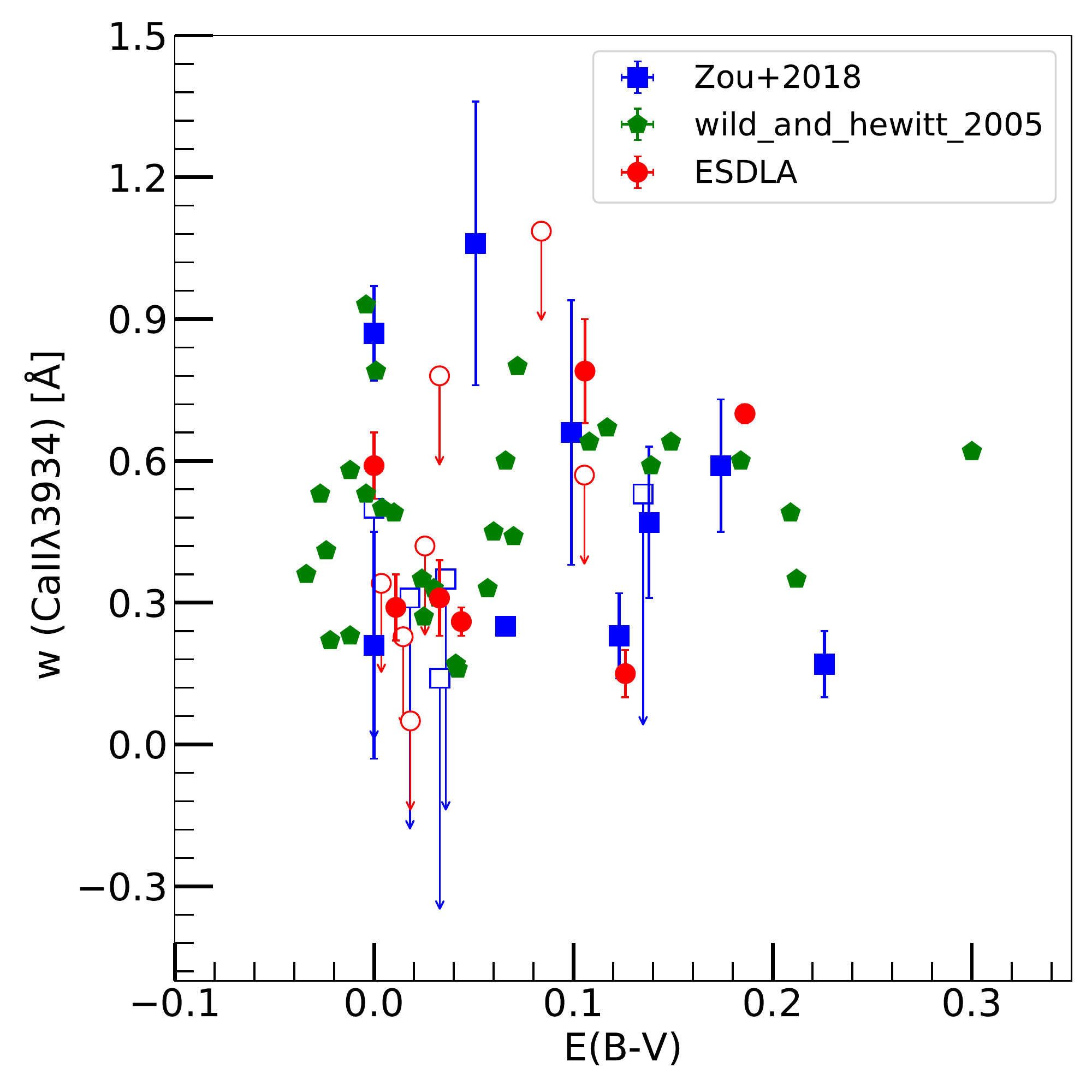}
    \caption{Equivalent width, $w$(\CaII$\lambda$3934), as a function of the colour excess E(B-V). The red circles, blue squares, and green pentagons represent the ESDLAs, as well as the \CI-selected \citep{Zou2018} and \CaII-selected \citep{Wild_and_Hewett2005} sample, respectively. The $w$(\CaII$\lambda$3934) upper limits are represented by a hollow shape and a down arrow. }
    \label{caii_distribution_fig_1}
\end{figure}

Given that our ESDLAs sample metal- and dust-poor environments that are, as of yet, often \HH\, bearing (at least for 50\% of the cases), we intend to check whether we see any similar trend as mentioned above. Fig.~\ref{caii_distribution_fig_1} shows the plot of $w$(\CaII$\lambda$3934) versus E(B-V) for ESDLAs in our sample, the C~{\sc i}-selected absorbers from \citet{Zou2018}, and the \CaII-selected sample from \citet{Wild_and_Hewett2005}. The figure shows that $w$(\CaII$\lambda$3934) does not correlate with E(B-V) globally or even in the individual high-$z$ samples. Hence, we cannot conclude that there is any strong trend of $w$(\CaII$\lambda$3934) with dust extinction in high-$z$ absorbers. We note that this lack of a trend cannot be necessarily attributed to a low dust content (E(B-V)$\rm \lesssim$0.3) in QSO absorber samples. The correlation between $w$(\CaII$\lambda$3934) and E(B-V) is seen in previous studies for samples with E(B-V)$<$0.08 and $w$(\CaII$\lambda$3934)$<$0.35. However, for all high-$z$ absorbers seen here, there seems to be no significant correlation between $w$(\CaII$\lambda$3934) and E(B-V). We also checked for correlation in sub-samples with E(B-V)$<$0.08 and $w$(\CaII$\lambda$3934)$<$0.35 and found no significant correlation to report. Probing {dustier} sightlines (E(B-V)$\rm \gtrsim$0.3), which might be possible in future deeper surveys such as DESI \citep[see][]{Yeche2020} and WEAVE-QSO \citep[see][]{Pieri2016}, will help understand if there is any strong trend of $w$(\CaII$\lambda$3934) with dust in absorbers with (E(B-V)$\rm \gtrsim$0.3.

\subsection{Kinematics\label{kinematics_section}}

\begin{figure}
\centering
   \includegraphics[width=1.0\hsize]{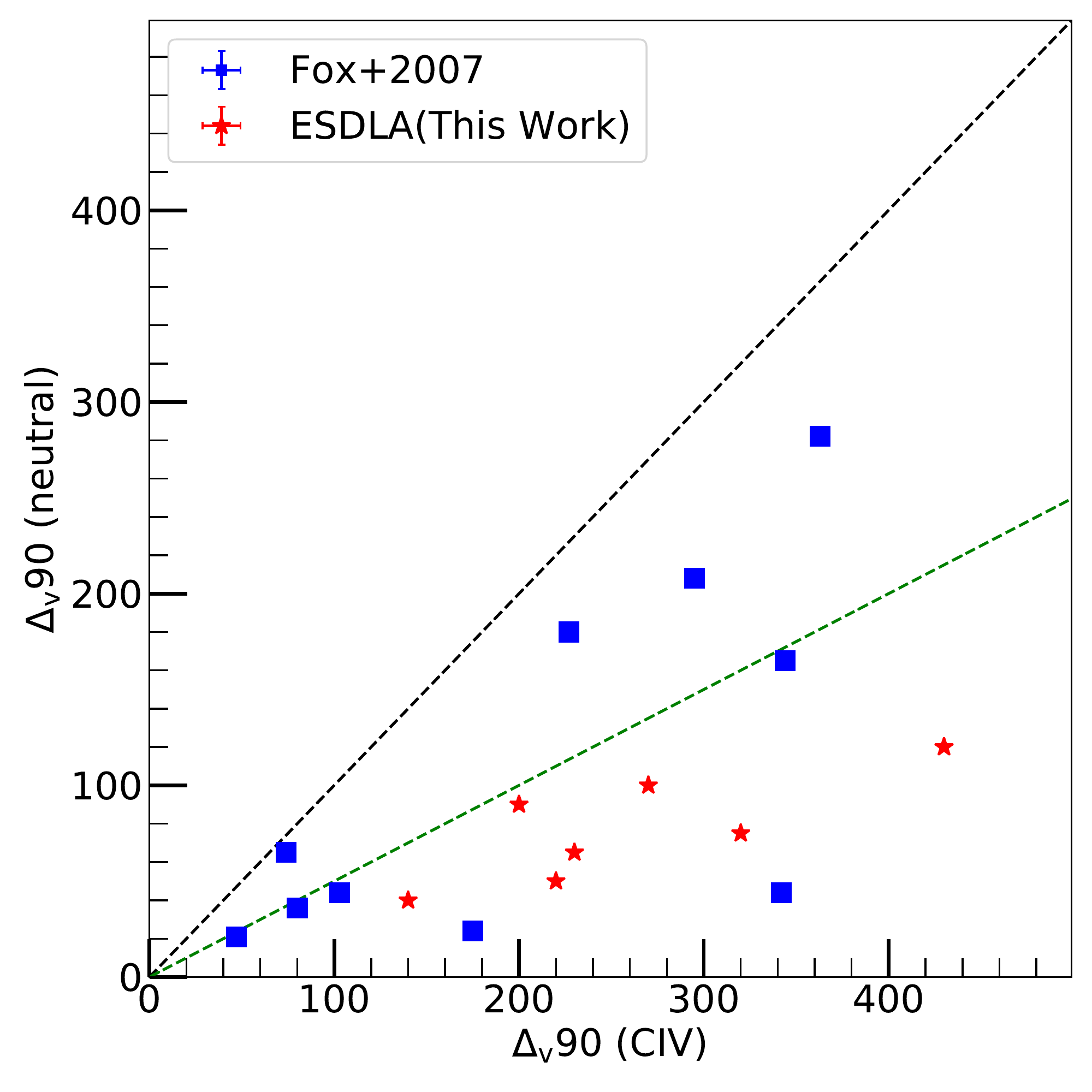}
    \caption{Comparison between \deltav\, estimates of warm gas (\CIV) (from this work) and neutral gas as obtained from unsaturated transitions of singly ionised species such as \ZnII\, and \FeII\, in different ESDLAs \citep[taken from][]{Ranjan2020} and DLAs \citep[taken from][]{Fox2007}. The dashed black and green lines indicate \deltav(\CIV) = \deltav(neutral) and \deltav(\CIV)=2$\times$ \deltav(neutral), respectively. }
    \label{n_civ_vs_n_siiv_fig_2}
\end{figure}

ESDLAs provide a unique opportunity to study the nature of gas, not only from within the star-forming disk of an associated faint galaxy \citep[see][]{Ranjan2020}, but also study the warm and hot medium associated with it. Since ESDLAs do probe gas from within the galactic disk, they are likely to probe warm and hot gas both from within the disk (see discussion in Section. ~\ref{warm_hot_gas_sections}) and the outskirts of the associated galaxy. Warm gas is likely to have relatively disturbed kinematics as compared to their neutral gas counterparts. \citet{Fox2007} showed that the mean \deltav(\CIV) is approximately twice that of the mean \deltav(neutral). In our ESDLA sample, we make a similar comparison of the standard \textbf{\deltav} measurements of singly ionised unsaturated lines (taken from \citet{Ranjan2020} and shown in Table.~\ref{Column_density_table_3}) with the \deltav\, of unsaturated \CIV\, profiles. We see that the mean \deltav(\CIV)(=260 \kms) is $\sim$ 3 times the mean \deltav(neutral)(=80 \kms) for unsaturated lines indicating less disturbed \HI\, gas in ESDLAs compared to DLAs. Fig.~\ref{n_civ_vs_n_siiv_fig_2} shows the distribution of kinematics (\deltav) of unsaturated \CIV\, and neutral gas. We show that the distribution has a large spread, but the \deltav\, of \CIV\, gas is larger than neutral gas for all the absorbers. We also note that this is especially evident for ESDLAs as the neutral gas is much less disturbed. While the distribution of \deltav\, of \CIV\, gas is quite similar for DLAs and ESDLAs (as evident in Figures. ~\ref{n_civ_vs_n_siiv_fig_1_new} and \ref{n_civ_vs_n_siiv_fig_2}), the \deltav\, of neutral gas in ESDLAs is smaller than that of DLAs.  Figure ~\ref{n_civ_vs_n_siiv_fig_2} further shows that the \deltav\, of \CIV\, profiles is always more than double than that of \deltav\, of neutral gas in ESDLAs, while the same is not true for DLAs. In DLAs, the \deltav\, of neutral gas can be higher in many cases. In addition, the average metallicity of the DLAs of  \citet{Fox2007} is -1.68$\rm \pm$0.08, which is also lower than that of the ESDLAs (-1.30$\rm \pm$0.05). Hence, this trend in \deltav\, of neutral gas cannot be attributed to the standard \deltav-metallicity relation. Using the above argument, we conclude that while the \deltav\, of \CIV\, in DLAs and ESDLAs are not distinguishable, \HI\, gas is statistically less disturbed for ESDLAs (with a relatively smaller \deltav) and hence, indicative of different origins of the \HI\, gas in DLAs and ESDLAs.\\

\section{Summary \label{Conclusion}}

We study a sample of extremely strong damped \lya\, absorber systems (ESDLAs) observed in medium and higher spectral resolution spectra towards quasars. In \citet{Ranjan2020}, we reported the column density and kinematics of \HI\, gas, diffuse \HH\, and some associated species in addition to reporting the emission signatures from nearby star-forming regions. The impact parameter of these absorbers relative to the centroid of the emission indicates that ESDLAs originate from the star-forming disk of their associated galaxies. In this paper, we extend our analysis towards the multi-phase nature of ESDLAs. We measured the column density and equivalent width and kinematics (\deltav) of low ionisation species (such as \OI, \ArI, \ClI, \NI, \NaI, \MgII, \CaII, \SII, \NiII, \MnII,\, \TiII,\, and \PII). We found tracers of all of these species in different systems, except for \NaI. We also looked for signatures of warm and hot gas tracers (\CIV, \SiIV, \NV,\ and \OVI) associated with these ESDLAs and identified all these species in different ESDLAs, except for \OVI. Since \OVI\ transitions are heavily contaminated with \lya\ forest, it was difficult to confirm their detection. Although, the presence of \OVI\, in ESDLAs cannot be ruled out. \\

We further compare the distribution of column densities, equivalent widths, and kinematics of the above-mentioned species with other absorber samples, which are known to probe gas associated with galaxies. Specifically, we compare our ESDLAs with the \MgII-selected DLA population \citep[studied by][]{Matejek2013}, dusty gas clouds probed using \CaII\, H-K absorption bands \citep[studied by][]{Wild_and_Hewett2005}, metal-rich \CI-selected absorbers \citep[studied by][]{Zou2018}, and \CIV\, absorption \citep[studied in the DLA sample by][]{Fox2007}. ESDLAs and DLAs have the same column density distribution for their warm gas tracers, \CIV\, and \SiIV. The robust estimate of \NV\, and \OVI\, column densities are not possible in our medium resolution study due to the contamination by \lya\, forest absorption lines. \\

We estimate dust-corrected metallicity measured using different neutral gas species such as \PII, \SII, \SiII, \MnII,\, and \CrII\, and compare it with the standard \ZnII-based measurements, which is common in DLA literature. We find that, after applying dust correction as prescribed by \citet{DeCia2016}, the measured metallicities are consistent for all mentioned species in all ESDLAs within a 3-$\rm \sigma$ uncertainty. \\

By comparing the neutral Argon (\ArI) in ESDLAs and DLAs, we conclude that the distribution of [Ar/H] in DLAs is similar to our sample of ESDLAs. We also note that the under-abundance of \ArI\, compared to $\rm \alpha$-elements as found in DLAs \citep[see][]{Zafar2014} do not extend to our ESDLA sub-sample. \citet{Zafar2014} claim that as opposed to dust depletion, or nucleosynthesis effects, the mentioned under-abundance is caused by background UV photons ionising the \ArI\, in DLAs. They also conclude that, in the presence of self-shielding high $N$(\HI) gas, \ArI\, should be able to survive. As our sample has high $N$(\HI), we tested this hypothesis. Comparing our observations and with updated CLOUDY photo-ionisation models, we conclude that ESDLAs do not show any under-abundance of \ArI\, relative to another $\alpha$-capture element, Silicon relative to the photoionisation models. Hence, the ionisation seen in DLAs is likely driven by UV-background photons that cannot penetrate the high $N$(\HI) in ESDLAs. We further note that a residual under-abundance of \ArI\, relative to \SiII\, of $\sim$0.2 dex in ESDLAs can be attributed to the low metallicity of ESDLAs. \\    

Compared to the \MgII-selected DLA population, we detect a large equivalent width, $w$(\MgII$\lambda$2796), for some ESDLAs similar to some metal-rich \CI-selected absorbers. Although, we note that ESDLAs and \CI-selected samples can have a rather different velocity spread (\delv) of '\MgII$\lambda$2796'. In the combined sample of \MgII-selected DLAs, ESDLAs, and \CI-selected absorbers, we find a correlation between metallicity and velocity spread. The dust extinction, E(B-V), and $w$(\CaII$\lambda$3934) distribution in ESDLAs is quite similar to that of \CI- and \CaII-selected absorbers.       \\

We study the relationship between $N$(\HH) and $N$(\ClI), previously studied in the literature \citep[see e.g.][]{Balashev2015}. By extending studies in literature (with ESDLAs and other translucent clouds with log($N$(\HH))$>$20), we note that the $N$(\ClI)-$N$(\HH) correlation shown in the literature is consistent even for high $N$(\HH) translucent clouds (up to log $N$(\HH)$<$22). We conclude that the \HH\, production seems as favourable in gas clouds with a low metallicity and high $N$(\HI) environments (such as ESDLAs) as they are in metal-rich, relatively lower $N$(\HI) environments \citep[such as \CI-selected systems studied by][and reference therein]{Zou2018}. Additionally, we report a strong \CaII\, presence ($w$(\CaII)$>$0.3\AA) in most of our ESDLAs. By comparing the dust extinction measurement with the \CaII\, measurement, we also conclude that $w$(\CaII$\lambda$3934) does not correlate with the dust content in different high-$z$ absorbers. Although, the dust content in all absorbers discussed above is rather limited (E(B-V)$\rm \lesssim$0.3). We note that this might be indicative of the SDSS optical colour-excess selection technique \citep[see][]{Richards2002} being biased towards dust-poor absorbers. Future large-sky surveys such as DESI that use a more robust selection technique combining optical colour with  NIR band colour from the Wide-field Infrared Survey Explorer (WISE) \citep[see e.g.][]{Yeche2020} will likely probe dustier (A$\rm _V\,\sim\,1$), possibly metal-rich, high $N$(\HH) gas environments towards QSOs. These samples could be then be used for robust comparison between high $N$(\HH) metal-rich systems and ESDLAs. \\ 

The column density and kinematics of \CIV\, and \SiIV\, associated with ESDLAs are similar to those found in DLAs. This is expected as both DLA and ESDLA sightlines will sample warm gas in the CGM and the warm neutral medium associated with their host galaxy despite sampling different \HI\, regions. Interestingly, the \deltav\, of a warm ionised medium (traced by unsaturated \CIV\, lines) is always more than double that of \HI\, gas (traced by an unsaturated, singly ionised species such as \FeII\, and \ZnII) in ESDLAs. This trend is not often seen in DLAs. We also show that the \deltav\, of the \HI\, region in DLAs are higher than in ESDLAs, despite having a lower average metallicity as compared to ESDLAs. This indicates that the \HI\, regions sampled by DLAs might be different to the \HI\, regions sampled by ESDLAs. \\   

ESDLAs provide a unique way to study gas within a metal-poor, general galaxy population at a high redshift and hence, warrant follow-up studies of their physical properties and kinematics. We believe that future studies of ESDLAs in higher spectral resolution (R$\sim$40000) will help model the physical conditions for each individual ESDLA. In addition, the NIR-optical-based QSO selection technique in large sky surveys such as WEAVE-QSO and DESI will help identify if dustier (A$\rm _V\, \sim$1) ESDLA environments also exist.

\begin{acknowledgements}

PPJ thanks allegorical Camille Noûs (Laboratoire Cogitamus) for inappreciable and often unnoticed discussions, advice and support. G.S. acknowledges WOS-A grant from Department of Science and Technology (SR/WOS-A/PM-9/2017). YKS acknowledges support from the National Research Foundation of Korea (NRF) grant funded by the Ministry of Science and ICT (NRF-2019R1C1C1010279). SB and KT are supported by RSF grant 18-12-00301. We thank the authors of \citet{Telikova2022article}, \citet{Telikova2020proceedings}, \citet{Zou2018}, and \citet{heintz2019b} for providing us useful information about their sample. We thank Pasquier Noterdaeme and J.-K.~Krogager for help with the observations and comments on early versions of the manuscript. AR also thanks Evelyne Roueff for her important correction in text.

\end{acknowledgements}

\bibliographystyle{aa}
\bibliography{bibliography.bib}

\begin{appendix}

\section{Absorption line properties and figures}

\subsection{Details on abundance and kinematics of neutral gas ions and other fine structure transitions}

\begin{sidewaystable*}[]
\begin{tabular}{lllllllllllll}
\hline
{QSO}        & {$N$(\NI)}            & {$N$(\OI$^{*}$)}                            & {$N$(\CaII)} & {$N$(\SII)} & {$N$(\NaI)}          & {$w$(\CaII$\lambda$3969)}     & {$w$(\MgII$\lambda$2803)}   & $N$(\PII) & $N$(\TiII) & $N$(\MnII) & $N$(\NiII) \\
\hline
\hline
J0017+1307 & 16.53$\pm$0.23   & \textless{}13.30                             & 13.22$\pm$0.33 & B & \textless{}13.53 & 0.34$\pm$0.04   & 2.22$\pm$0.03 & - & {$\sim$13} & 13.1$\pm$0.1 & {14.1$\pm$0.1} \\
J0025+1145 & -                & \textless{}15.26                             & 12.94$\pm$0.02 & $>$16.4   & - & 0.36$\pm$0.02   & 4.2$\pm$0.01  & - & {13.31$\rm \pm$0.04} & {14.00$\pm$0.01} & {14.88$\pm$0.02} \\
J1143+1420 & \textless{}15.7  & \textless{}19.16                & \textless{}13.19 & B & \textless{}13.65 & \textless{}0.68 & 4.08$\pm$0.08 & - & {$\sim$13.68} & {13.59$\pm$0.01} & {14.53$\pm$0.02} \\
J1258+1212 & \textless{}16.25 & \textless{}14.12                             & -     & $\sim$15.40     & \textless{}11.96      & 0.72$\pm$0.07   & 1.90$\pm$0.02 & - & 13.1$\pm$0.2 & {13.1$\pm$0.1} & {14.0$\pm$0.1} \\
J1349+0448 & -                & \textless{}13.87   & -  & B           & \textless{}13.48   & 0.92$\pm$0.13   & 2.13$\pm$0.07 & - & {$\sim$13.3} & {13.35$\pm$0.04} & {14.4$\pm$0.1} \\
J1411+1229 & 15.52$\pm$0.11   & $\rm \sim$13.42                               & - & B           & \textless{}15.03     & 1.23$\pm$0.09   & 0.53$\pm$0.05 & {13.9$\pm$0.7} & {$\sim$12.9} & {13.2$\pm$0.1} & {14.2$\pm$0.1} \\
J1513+0352 & -                & \textless{}14.70                             & $\sim$12.3 & B     & \textless{}12.79  & \textless{}0.13 & 2.84$\pm$0.04 & - & {$\sim$12.8} & - & 14.1$\pm$0.1 \\
J2140$-$0321 & 15.75$\pm$0.11   & 13.82$\pm$0.1        & 12.71$\pm$0.1 & 15.61$\pm$0.07  & \textless{}12.83  & 0.13$\pm$0.03   & 1.31$\pm$0.01 & {15.5$\pm$0.7} & 13.2$\pm$0.1 & 13.51$\pm$0.02 & 14.35$\pm$0.06 \\
J2232+1242 & 15.5$\pm$0.11    & $\rm \sim$13.46                               & \textless{}12.57 & 15.49$\pm$0.01 & \textless{}12.76 & 0.18$\pm$0.03   & 1.98$\pm$0.01 & {14.1$\pm$0.1} & 13.1$\pm$0.1 & 13.23$\pm$0.01 & 14.26$\pm$0.02 \\
J2246+1328 & \textless{}13.87 & \textless{}13.22                             & $\sim$12.85 & 15.16$\pm$0.27  & \textless{}13.45    & \textless{}0.38 & 0.72$\pm$0.01 & {13.4$\pm$0.1} & {\textless{}12.7} & 12.96$\pm$0.05 & 13.7$\pm$0.1 \\
J2322+0033 & \textless{}15.24 & \textless{}13.59   & - & 14.63$\pm$0.07           & \textless{}11.39     & \textless{}0.70 & 1.50$\pm$0.04 & - & {$\sim$13.2} & 12.93$\pm$0.05 & 13.83$\pm$0.05 \\
\hline
\end{tabular}
\caption{Column density (in logscale) of low ionisation gas species, \NI, \OI$^{*}$, \CaII, \SII, \NaI\, \PII, \TiII, \MnII, \NiII,\, and equivalent width (w) measurement of "\CaII$\lambda$3934" and "\MgII$\lambda$2803" for the ESDLA sample.}
\tablefoot{Symbol `B' means that all the transition lines for that species are blended and hence the column density cannot be estimated.}
\label{Column_density_table_2}
\end{sidewaystable*}

\subsection{Absorption line figures\label{abs_figs} }

\begin{figure*}
    \centering
    \includegraphics[trim=0 0 0 0,clip,width=1.1\hsize]{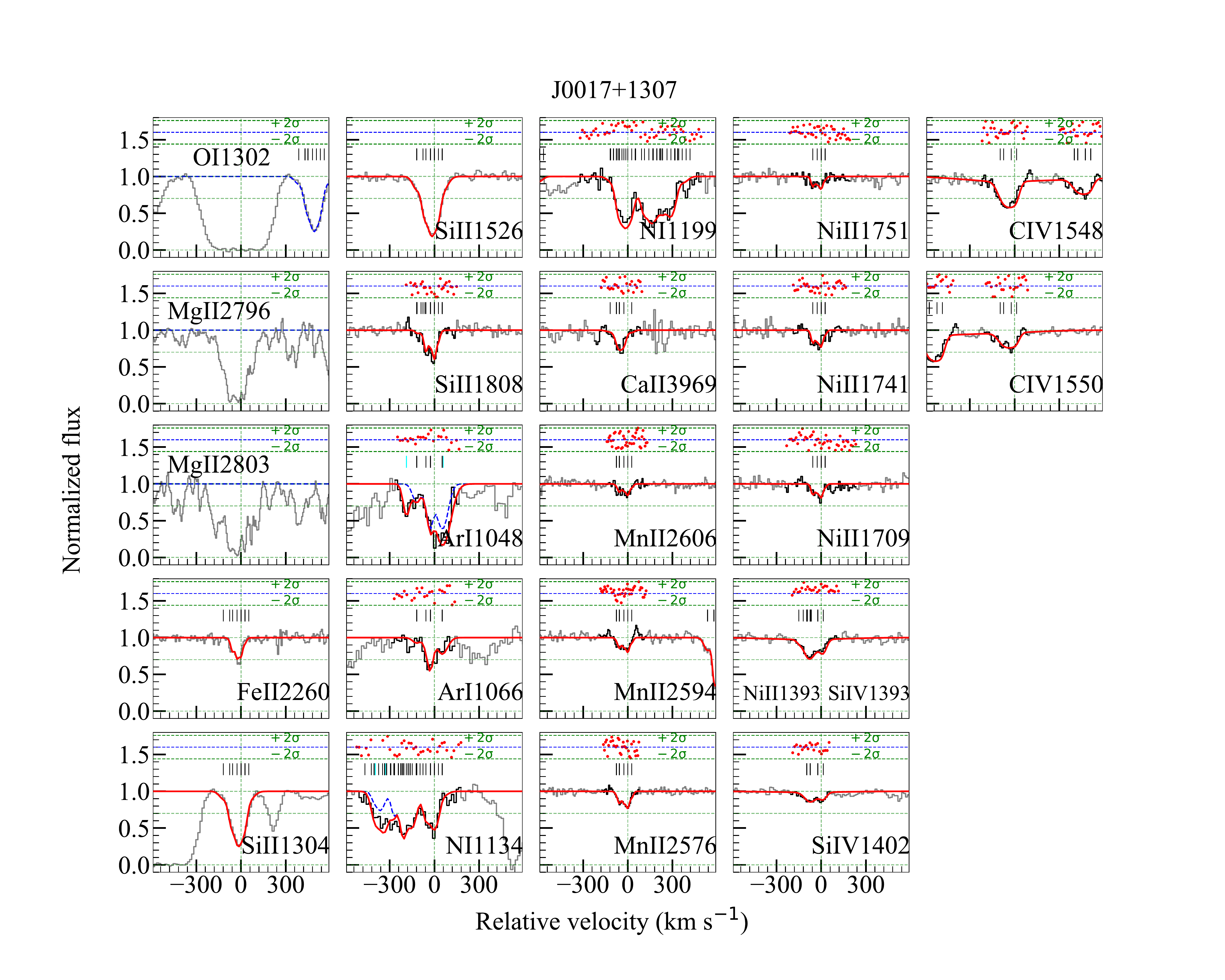}
    \caption{Different ionisation metal lines (\OI, \MgII, \FeII, \SiII, \ArI, \NI, \CaII, \MnII, \NiII, \SiIV,\, and \CIV) associated with the $\zabs=2.326$ ESDLA system towards QSO SDSS J0017$+$1307. The normalised X-shooter spectrum is shown in grey, the continuum is shown as a dashed green line, with the total best-fit multi-component Voigt profile over-plotted in red. The highlighted part of the spectrum (in black) shows the data that were used to constrain the fit. The residuals are shown above each line in areas used for the fit in units of the standard deviation ($\pm\,2\,\sigma$) from the error spectrum. In some of the plots where there is contamination either from the \lya\, forest or from other transitions from different absorber systems, apart from the total modelled profile in red, the modelled profile for the labelled transition is also shown as a dashed blue line. Short vertical marks (or ticks) in black show the location of the different velocity components. The position of blends with the \lya\, forest and other absorption systems unrelated to the ESDLA are shown with cyan ticks. There is a velocity shift between the UVB and the VIS arm spectra for some systems \citep[see][for more details]{Ranjan2020}. The velocity shifted positions, if any, are indicated with magenta ticks. The sub-component redshifts and $b$-values of neutral and singly ionised gas species, such as \ArI, \NI,\, and \CaII,\, are tied together. The sub-component redshifts and $b$-values for \SiIV\, and \CIV\, profiles are also tied together. The saturated absorption profiles for \OI\, and \MgII\, are shown for reference. The subplot with \FeII\, is taken from \citet{Ranjan2020} and is shown here just for reference.}
    \label{J0017met}
\end{figure*}{}

\begin{figure*}
    \centering
    \includegraphics[trim=0 0 0 0,clip,width=1.1\hsize]{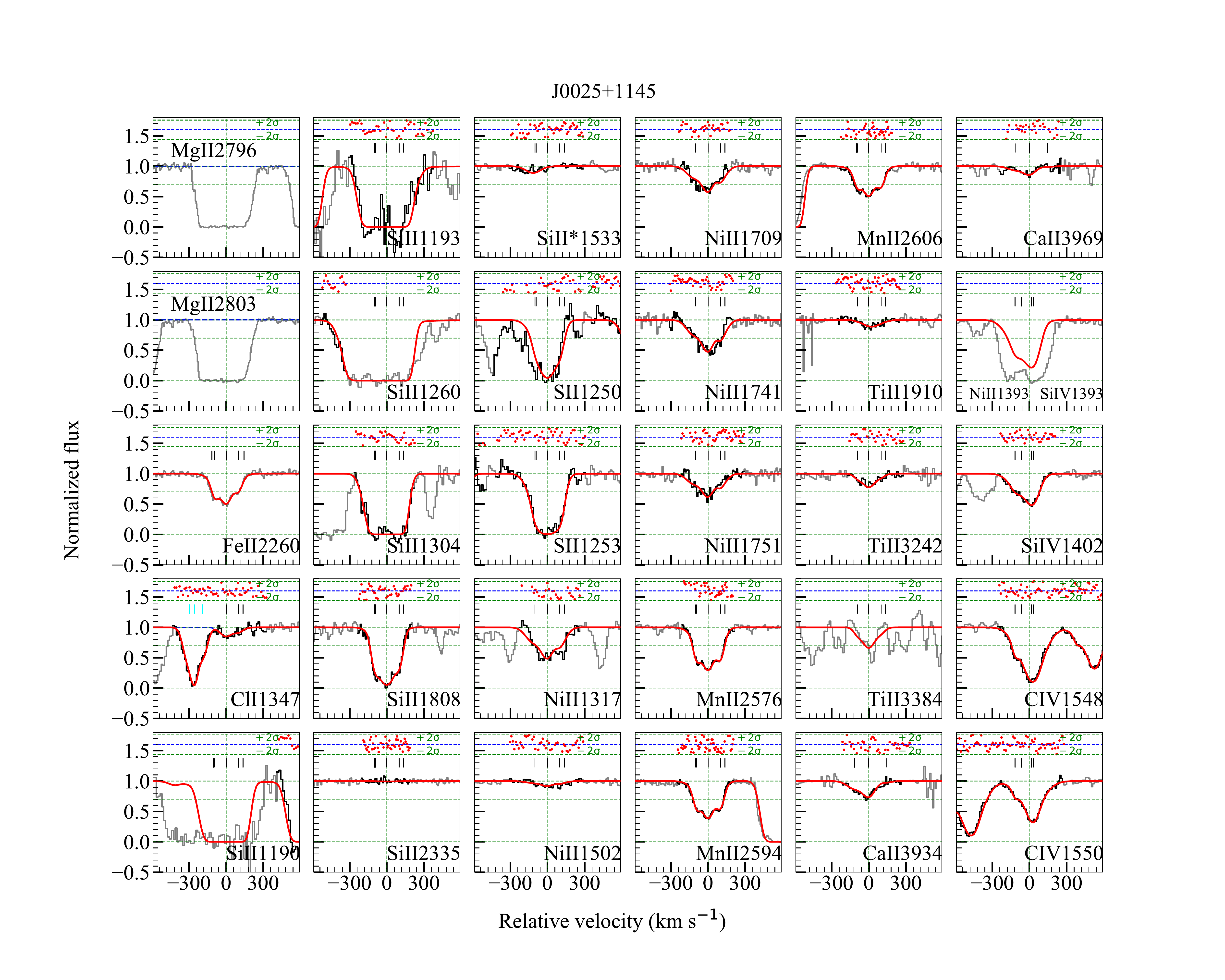}
    \caption{Different ionisation metal lines (\OI, \MgII, \ClI, \FeII, \SiII, \SII, \NiII, \MnII, \TiII, \CaII, \SiIV,\, and \CIV) associated with the $\zabs=2.304$ ESDLA system towards QSO SDSS J0025$+$1145. The legends as well as the assumption on sub-component redshifts and $b$-values are the same as in Fig.~\ref{J0017met}.}
    \label{J0025met}
\end{figure*}{}

\begin{figure*}
    \centering
    \includegraphics[trim=0 0 0 0,clip,width=1.1\hsize]{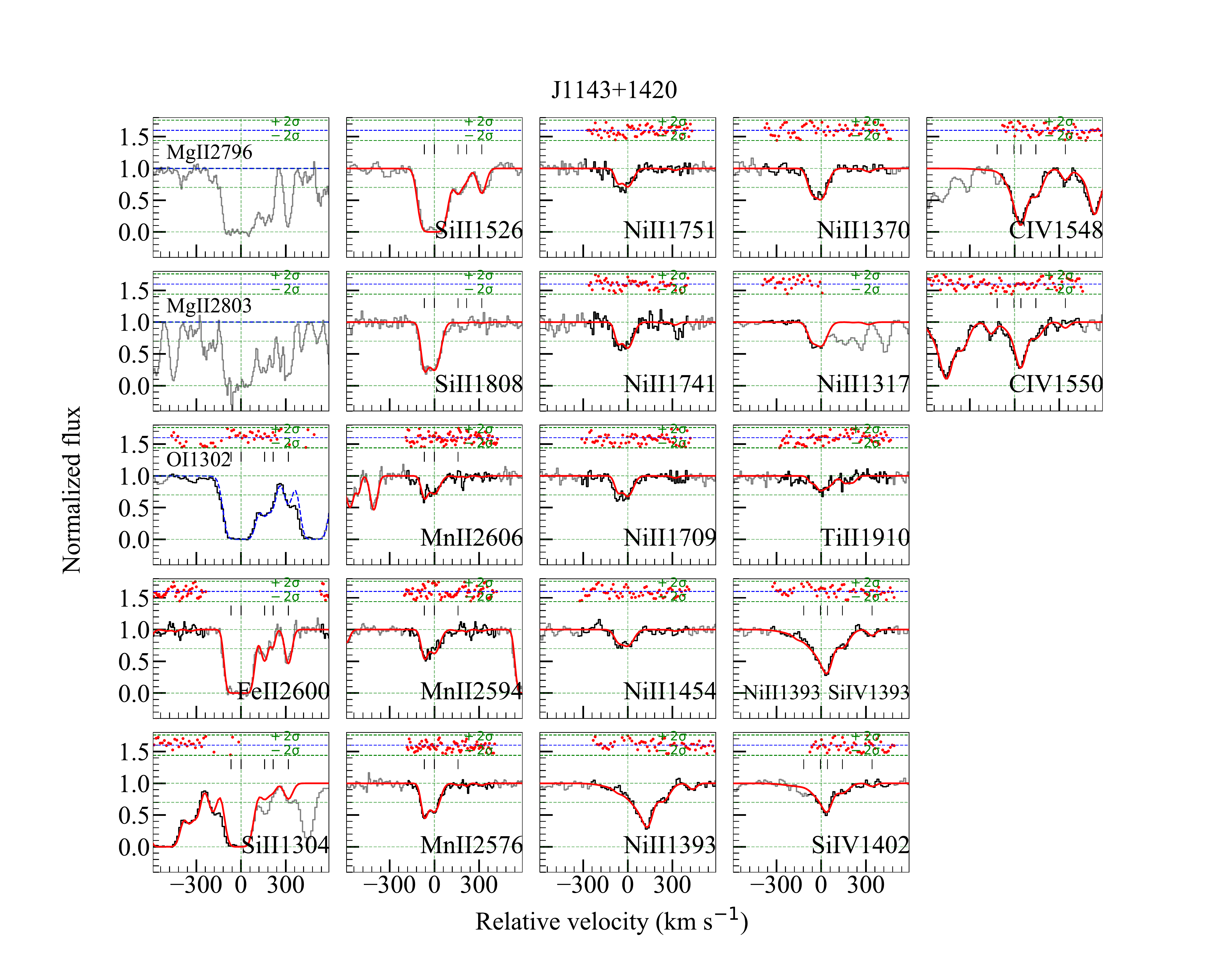}
    \caption{Different ionisation metal lines (\MgII, \OI, \FeII, \SiII, \MnII, \NiII, \TiII, \SiIV,\, and \CIV) associated with the $\zabs=2.323$ ESDLA system towards QSO SDSS J1143$+$1420. The legends as well as the assumption on sub-component redshifts and $b$-values are the same as in Fig.~\ref{J0017met}.}
    \label{J1143met}
\end{figure*}{}

\begin{figure*}
    \centering
    \includegraphics[trim=0 0 0 0,clip,width=1.1\hsize]{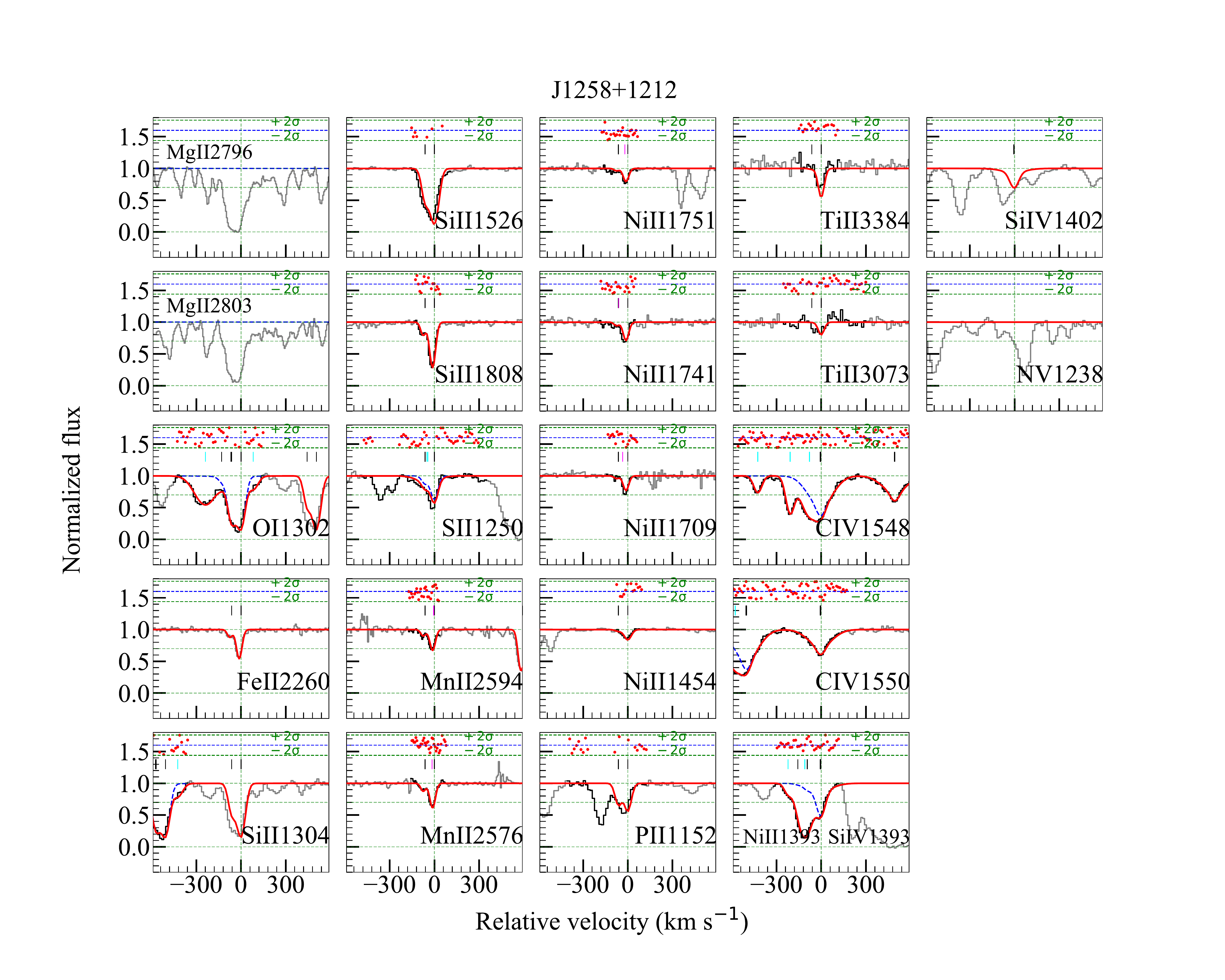}
    \caption{Different ionisation metal lines (\MgII, \OI, \FeII, \SiII, \SII, \MnII, \NiII, \PII, \TiII, \CIV, \SiIV,\, and \NV) associated with the $\zabs=2.444$ ESDLA system towards QSO SDSS J1258$+$1212. We note that \NV\, transitions are heavily contaminated with the \lya\, forest and hence we could not find any robust solution for the column density tied with \CIV. Hence, we only show the spectrum of "\NV$\lambda\,1238$" here without the fit. The legends as well as the assumption on sub-component redshifts and $b$-values are the same as in Fig.~\ref{J0017met}.}
    \label{J1258met}
\end{figure*}{}

\begin{figure*}
    \centering
    \includegraphics[trim=0 0 0 0,clip,width=1.1\hsize]{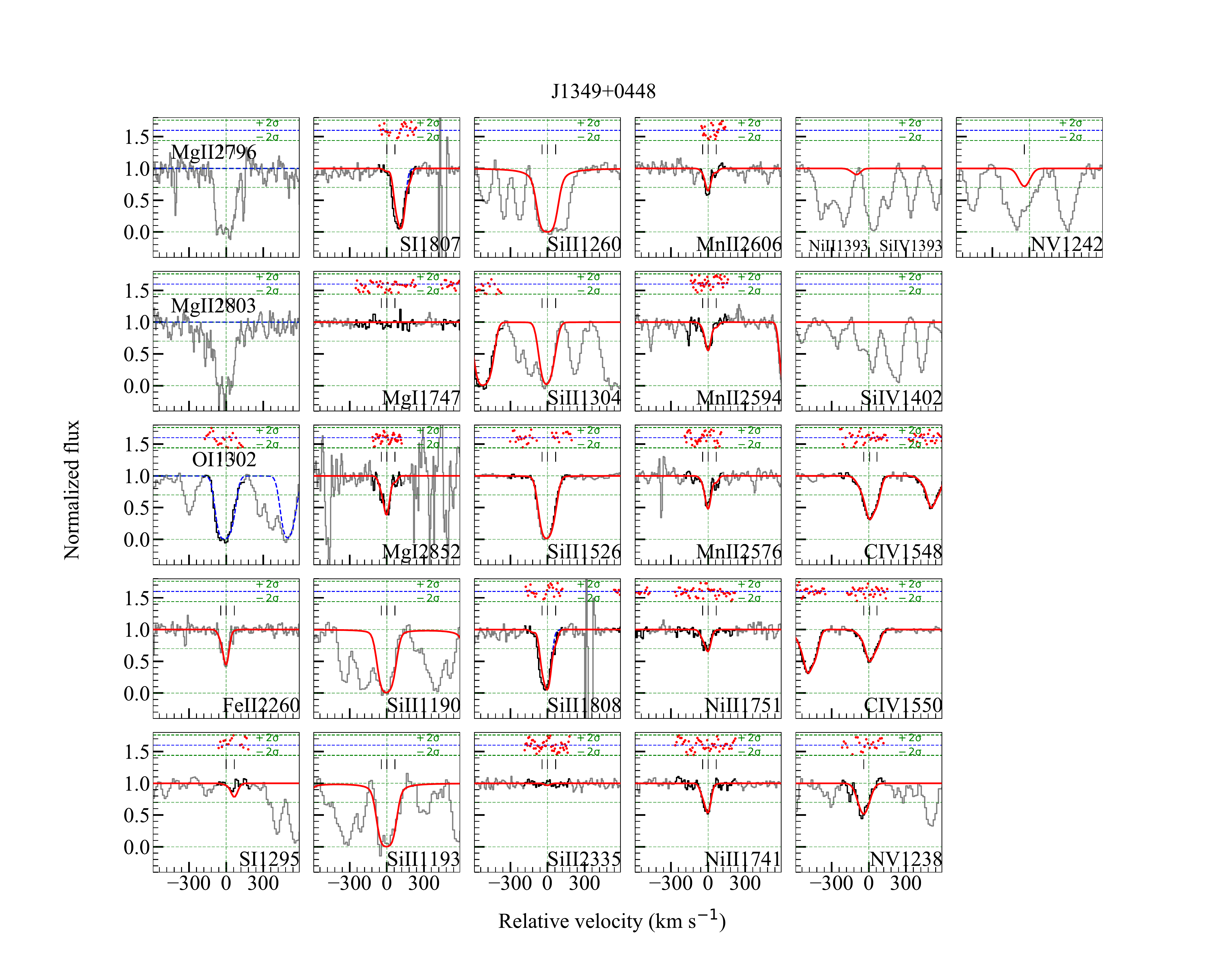}
    \caption{Different ionisation metal lines (\MgII, \OI, \FeII, \SiII, \MnII, \NiII, \SiIV, \CIV,\, and \NV) associated with the $\zabs=2.482$ ESDLA system towards QSO SDSS J1349$+$0448. The sub-component redshifts and $b$-values of \CIV\, and \NV\, are also tied together. '\NV$\lambda$1242' is contaminated and hence the transition was not used for fitting. The \SiIV\, profiles are strongly blended and we could not find any robust solution for the column density tied with \CIV. Here, we just show the unfitted spectrum for reference. The cyan tick in the '\SiII$\lambda$1526' subplot represents a blend with an \MgII\, absorption system at z$\rm _{abs}$ = 0.8998. The legends as well as the assumption on sub-component redshifts and $b$-values are the same as in Fig.~\ref{J0017met}.}
    \label{J1349met}
\end{figure*}{}

\begin{figure*}
    \centering
    \includegraphics[trim=0 0 0 0,clip,width=1.1\hsize]{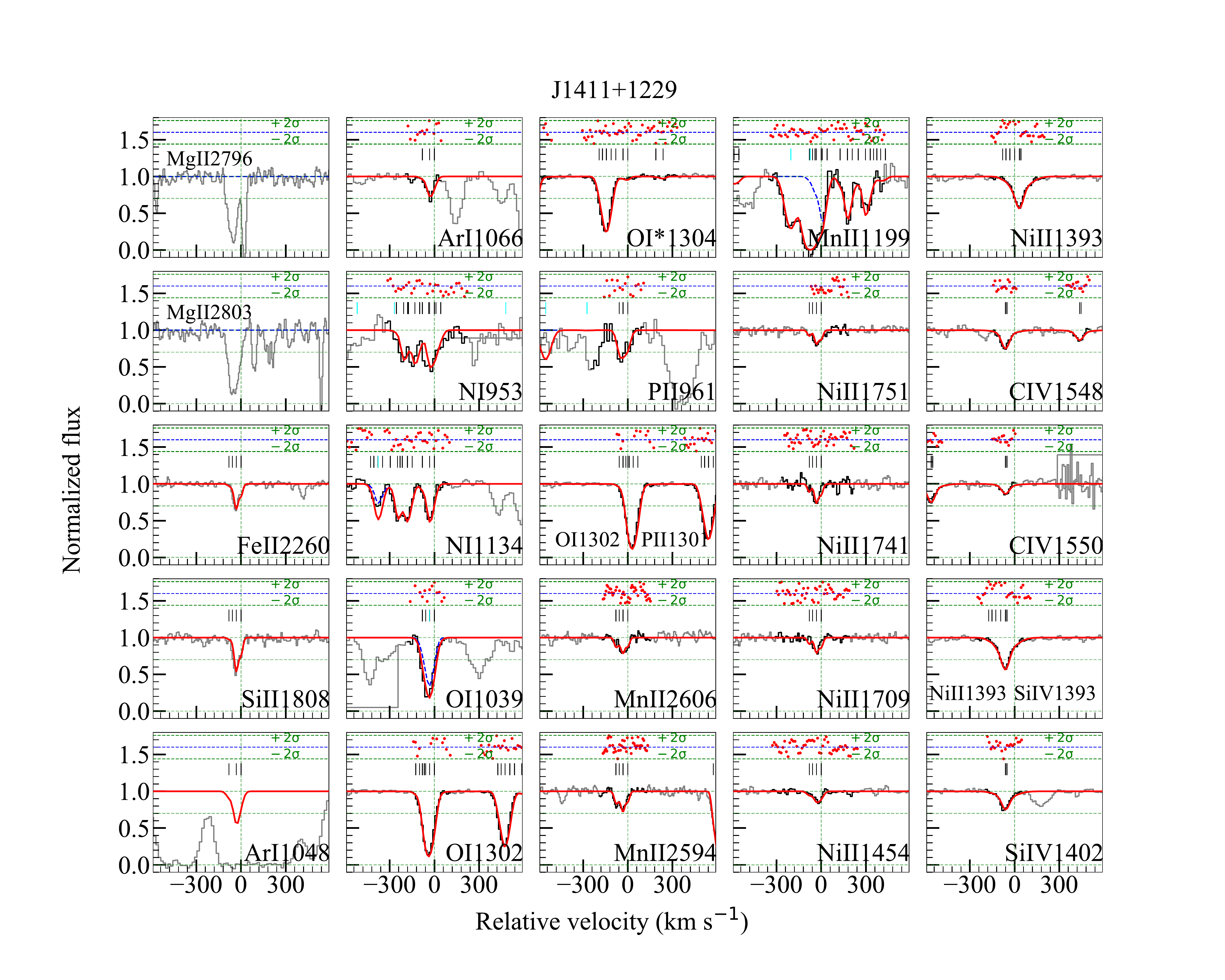}
    \caption{Different ionisation metal lines (\MgII, \FeII, \SiII, \ArI, \NI, \OI, \OI$^{*}$, \PII, \MnII, \NiII, \CIV,\, and \SiIV) associated with the $\zabs=2.545$ ESDLA system towards QSO SDSS J1411$+$1229. The subplot with '\ArI$\lambda$1048' shows the mock \ArI\, profile on top of the spectrum, heavily contaminated with the \lya\, forest and it was not used for fitting. The legends as well as the assumption on sub-component redshifts and $b$-values are the same as in Fig.~\ref{J0017met}.}
    \label{J1411met}
\end{figure*}{}

\begin{figure*}
    \centering
    \includegraphics[trim=0 0 0 0,clip,width=1.1\hsize]{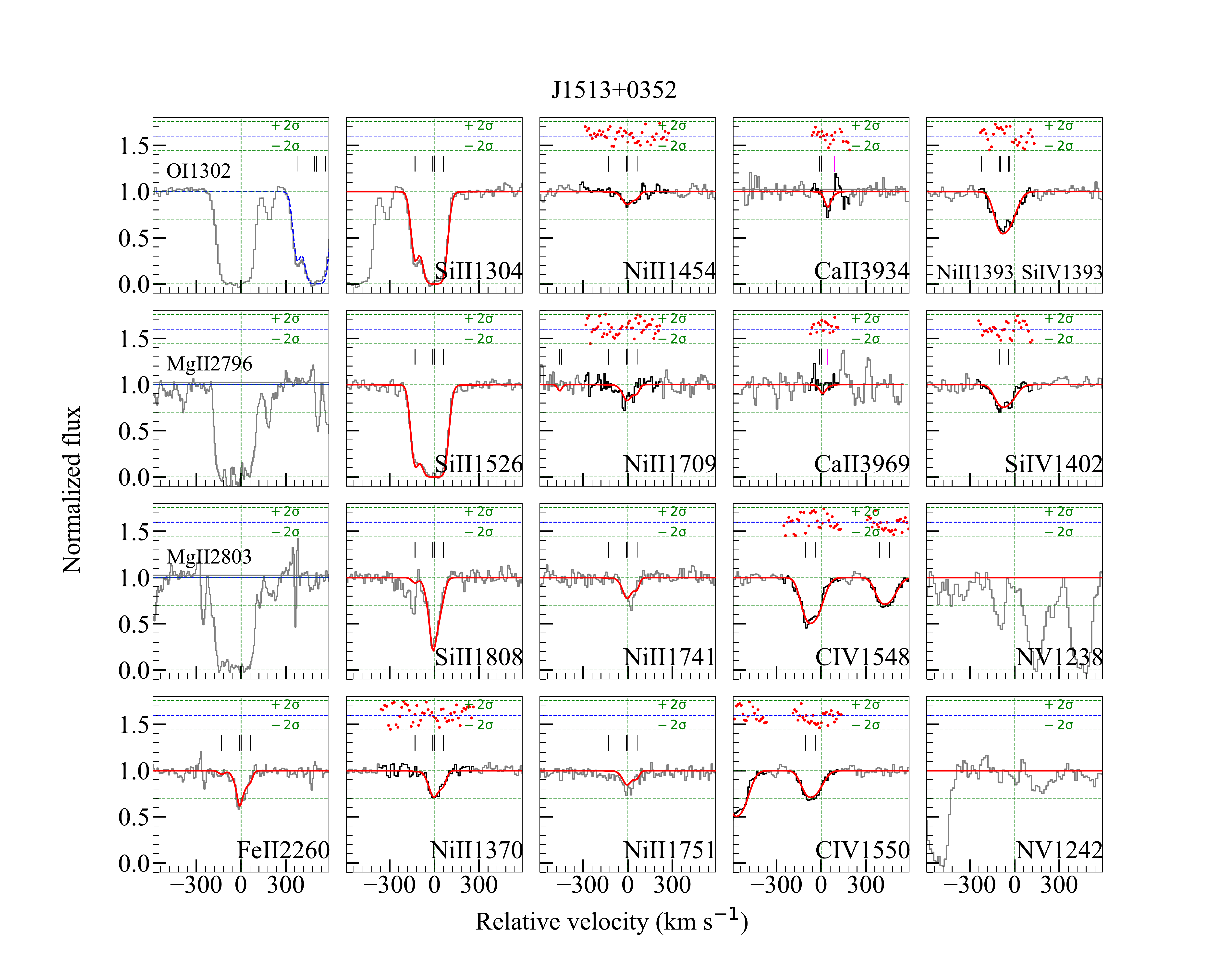}
    \caption{Different ionisation metal lines (\OI, \MgII, \FeII, \SiII, \NiII, \CaII,  \CIV, \SiIV,\, and \NV) associated with the $\zabs=2.464$ ESDLA system towards QSO SDSS J1513$+$0352. The \NV\, transitions are heavily contaminated with the \lya\, forest and hence we could not find any robust solution for the column density tied with \CIV. Hence, we only show the \NV\, absorption profiles here without the fit. The legends as well as the assumption on sub-component redshifts and $b$-values are the same as in Fig.~\ref{J0017met}.}
    \label{1513met}
\end{figure*}{}

\begin{figure*}
    \centering
    \includegraphics[trim=0 0 0 0,clip,width=1.1\hsize]{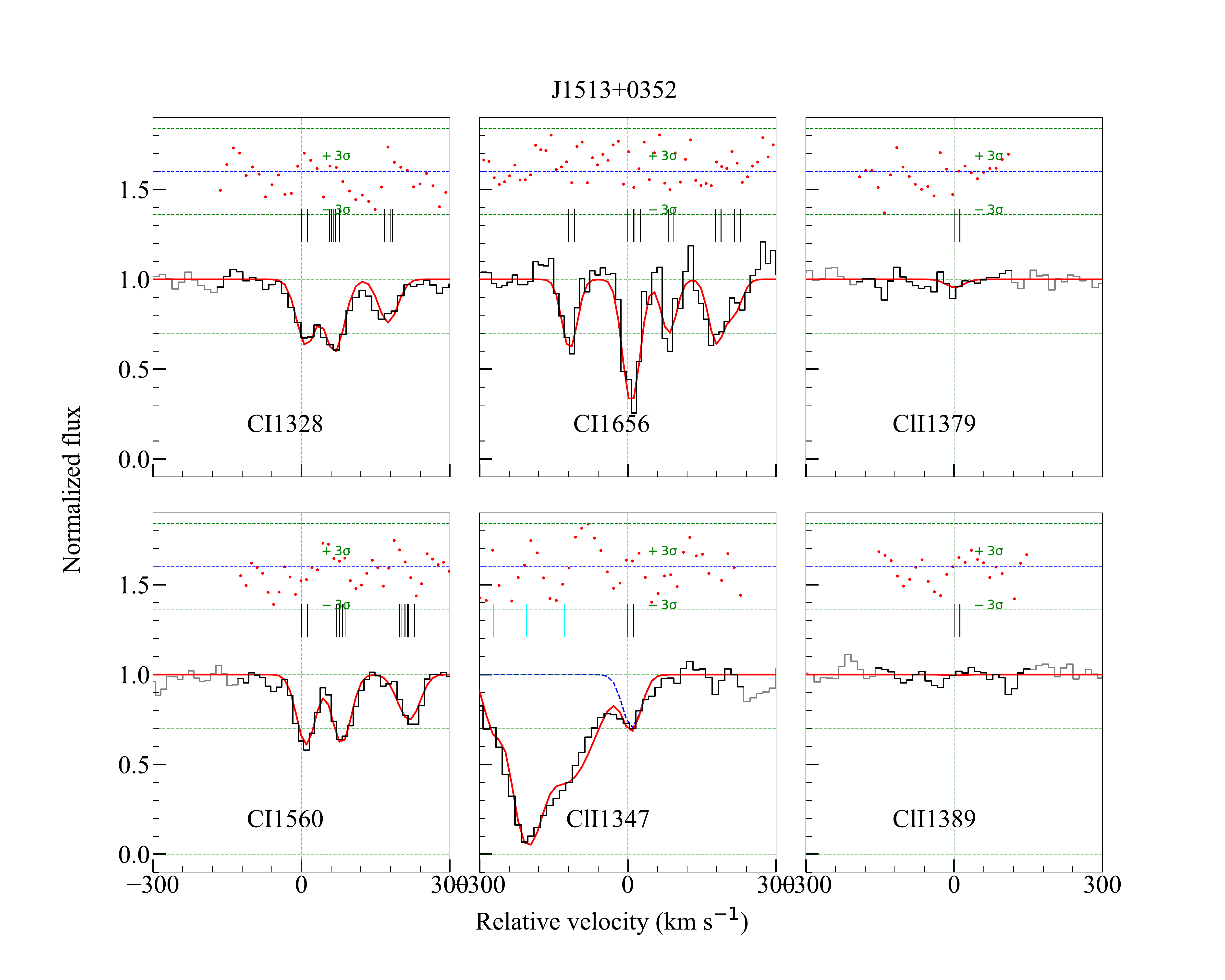}
    \caption{Fit to \CI\, and \ClI\, lines associated with the $\zabs=2.464$ ESDLA system towards QSO SDSS J1513$+$0352. The plot with neutral Chlorine ('\ClI$\lambda$1347' transition) includes velocity components from two other absorption systems,  $\zabs=2.339$ (\SiIV\, absorber) and $\zabs=2.006$ (\CIV\, absorber), fitted together. The total \CI\, column density (logN(\CI) = 15.2$\pm$0.8, logN(\CI$^{*}$) = 15.0$\pm$0.8 and logN(\CI$^{**}$) = 14.6$\pm$0.7) with a multi-component fit is consistent with the single component fit result reported in \citet{Ranjan+2018}.}
    \label{J1513chlorine}
\end{figure*}{}

\begin{figure*}
    \centering
    \includegraphics[trim=0 0 0 0,clip,width=1.1\hsize]{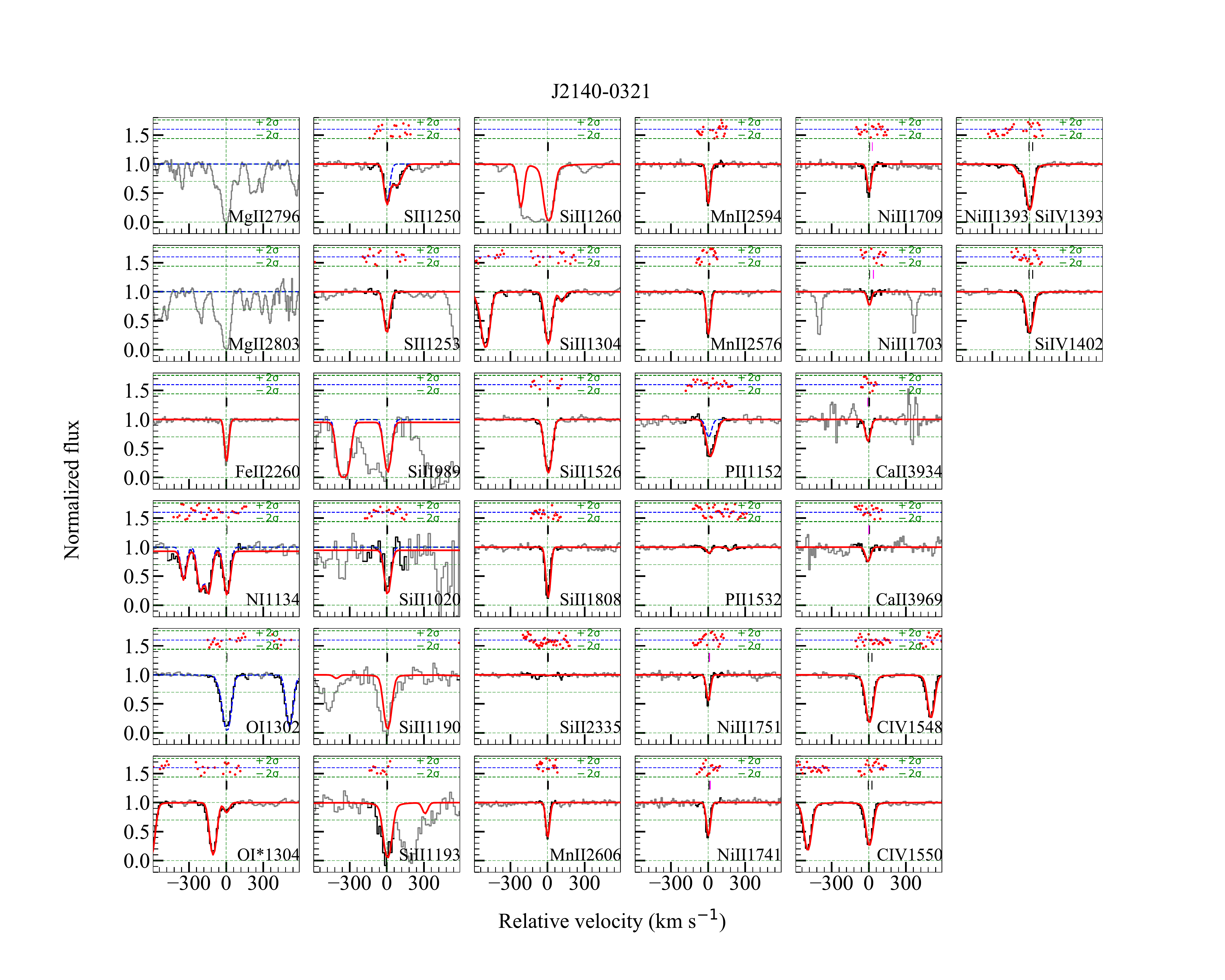}
    \caption{Different ionisation metal lines (\MgII, \FeII, \SiII, \NI, \OI, \OI$^{*}$, \SII, \MnII, \PII, \NiII, \CaII, \CIV,\, and \SiIV) associated with the $\zabs=2.339$ ESDLA system towards QSO SDSS J2140$-$0321. The plots showing '\OI$\lambda$1302' and '\OI$^{*}\,\lambda$1304' include additional components from '\SiII$\lambda$1304' transition. The legends as well as the assumption on sub-component redshifts and $b$-values are the same as in Fig.~\ref{J0017met}.}
    \label{J2140met}
\end{figure*}{}

\begin{figure*}
    \centering
    \includegraphics[trim=0 0 0 0,clip,width=1.1\hsize]{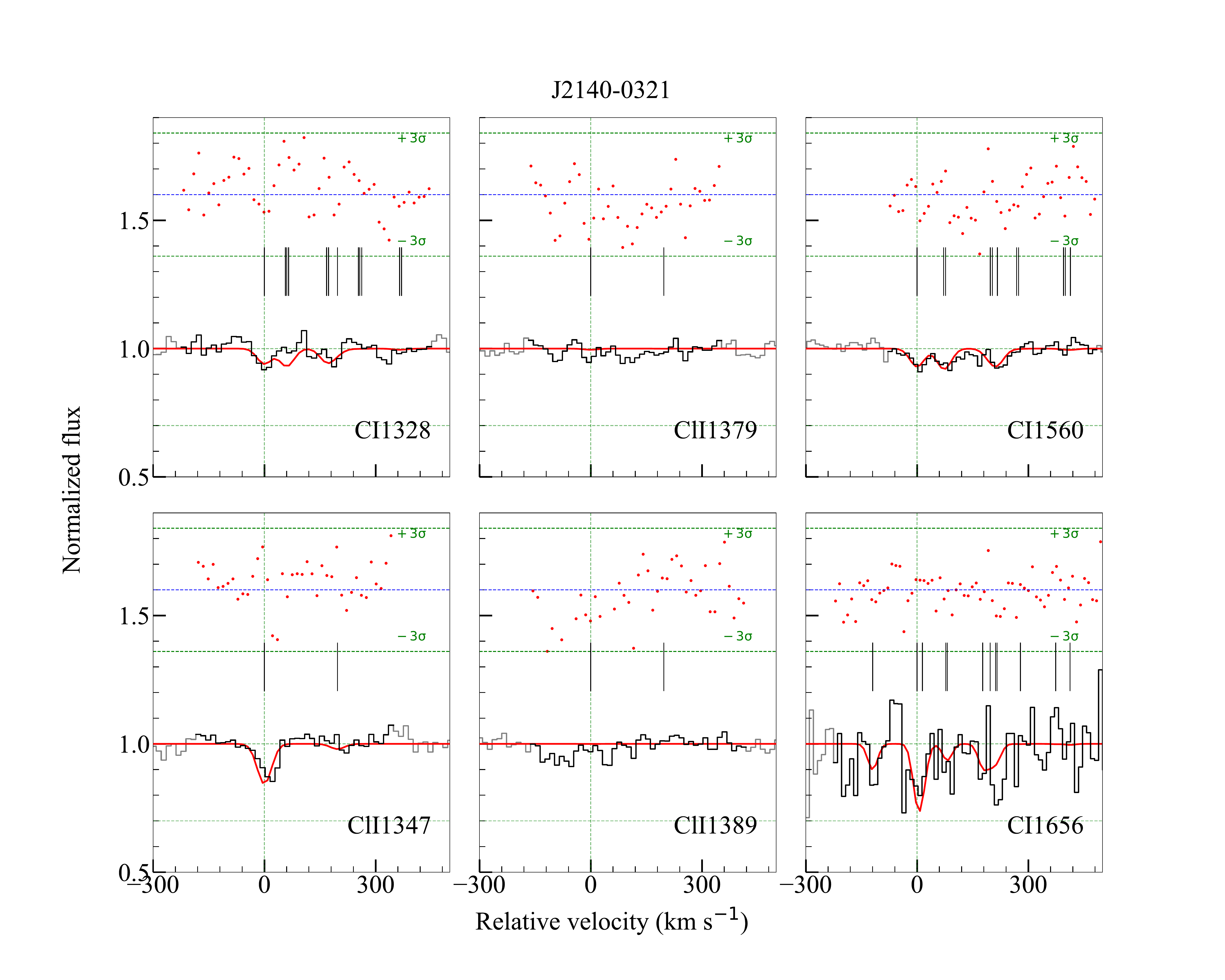}
    \caption{Fit to \CI\, and \ClI\, lines associated with the $\zabs=2.339$ ESDLA system towards QSO SDSS J2140$-$0321. The total \CI\, column density (logN(\CI) = 13.18$\pm$0.04) with a multi-component fit is $\sim$0.3 dex smaller that the single component fit result reported in \citet{Ranjan2020}}
    \label{J2140chlorine}
\end{figure*}{}

\begin{figure*}
    \centering
    \includegraphics[trim=0 0 0 0,clip,width=1.1\hsize]{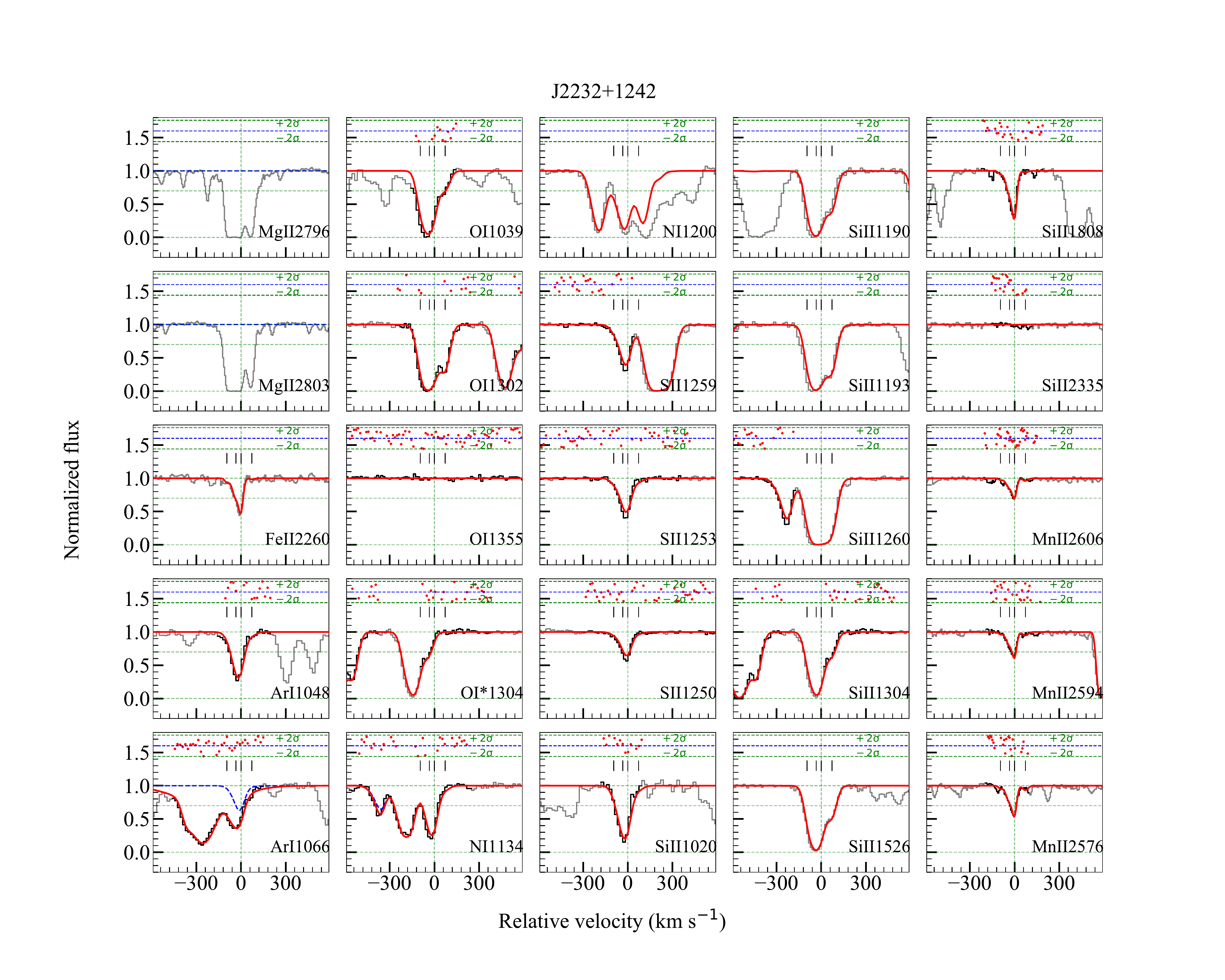}
    \caption{Different ionisation metal lines (\MgII, \FeII, \ArI, \OI\, \OI$^{*}$, \NI, \SII, \SiII, and \MnII) associated with the $\zabs=2.230$ ESDLA system towards QSO SDSS J2232$+$1242. The plots show '\OI$\lambda$1302' and '\OI$^{*}\,\lambda$1304'. The subplots showing '\ArI$\lambda$1066' and '\NI$\lambda$1134' transitions have contamination from the \lya\, forest that were included while fitting. The legends as well as the assumption on sub-component redshifts and $b$-values are the same as in Fig.~\ref{J0017met}.}
    \label{J2232met}
\end{figure*}{}

\begin{figure*}
    \centering
    \includegraphics[trim=0 0 0 0,clip,width=1.1\hsize]{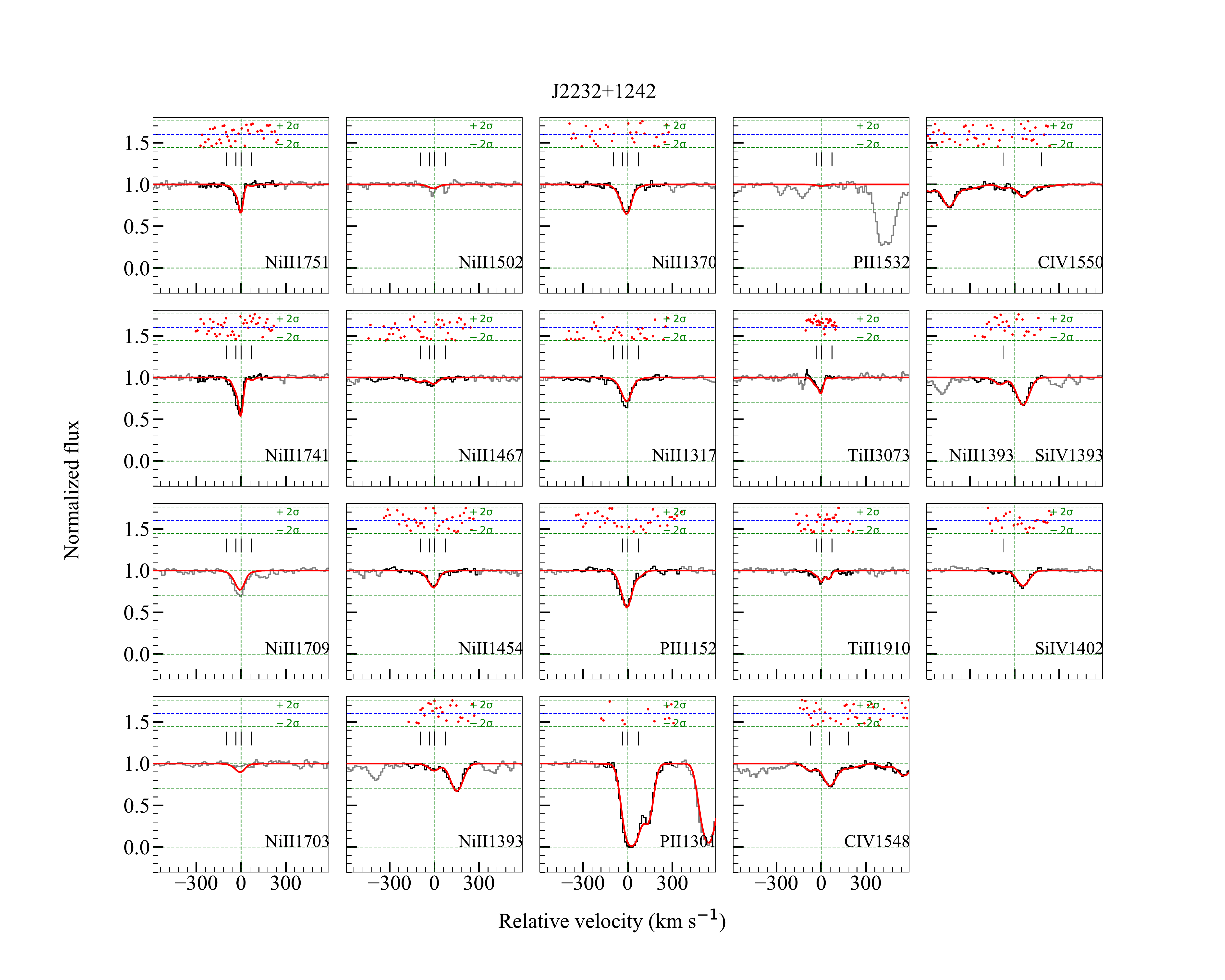}
    \caption{Different ionisation metal lines (\NiII, \PII, \TiII, \CIV,\, and \SiIV) associated with the $\zabs=2.230$ ESDLA system towards QSO SDSS J2232$+$1242. The legends as well as the assumption on sub-component redshifts and $b$-values are the same as in Fig.~\ref{J0017met}.}
    \label{J2232met}
\end{figure*}{}

\begin{figure*}
    \centering
    \includegraphics[trim=0 0 0 0,clip,width=1.1\hsize]{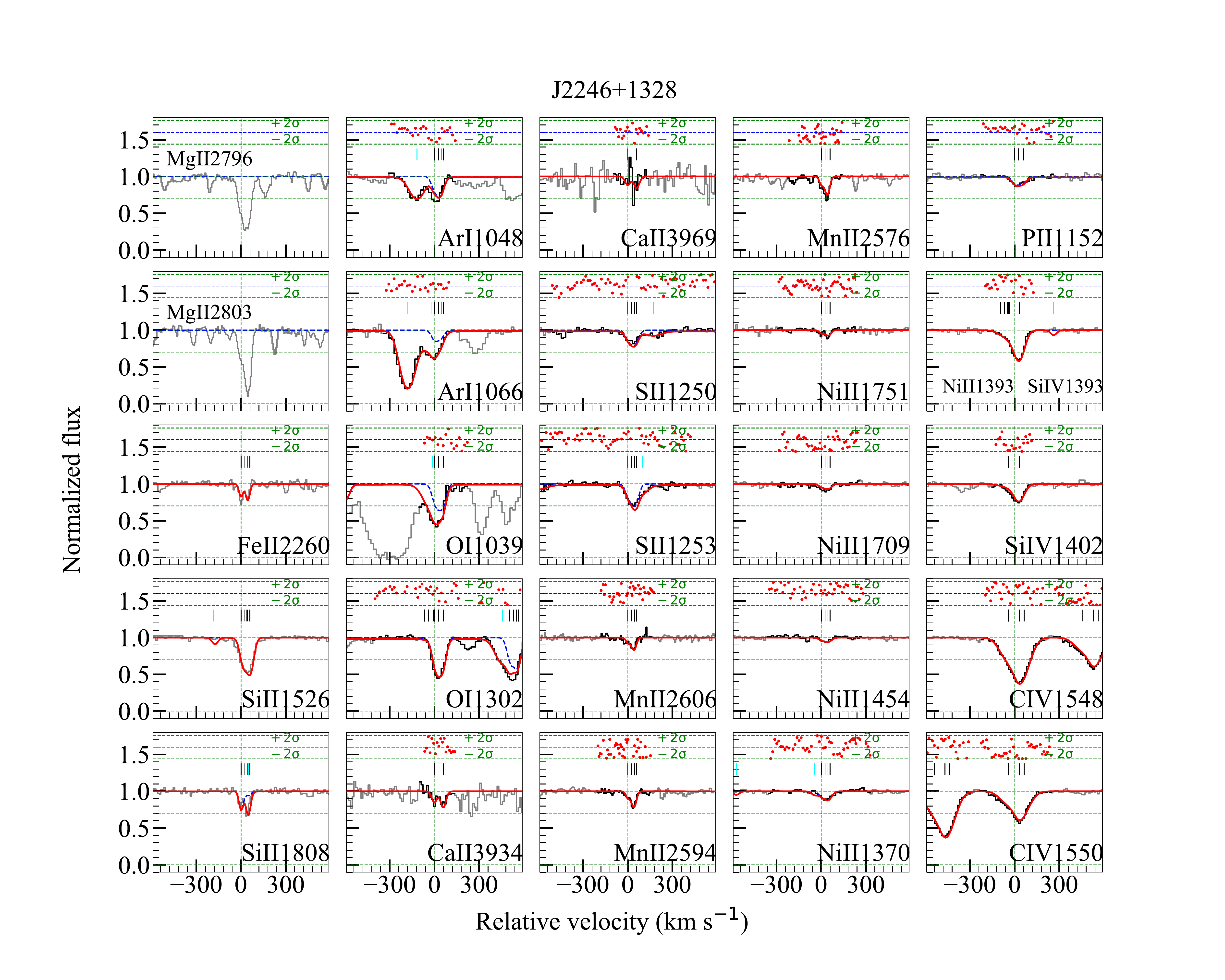}
    \caption{Different ionisation metal lines (\MgII, \FeII, \SiII, \ArI, \OI, \CaII, \SII, \MnII, \NiII, \PII, \SiIV,\, and \CIV) associated with the $\zabs=2.215$ ESDLA system towards QSO SDSS J2246$+$1328. The plots with neutral Argon (\ArI) includes components from the \lya\, forest (\HI) lines. The legends as well as the assumption on sub-component redshifts and $b$-values are the same as in Fig.~\ref{J0017met}.}
    \label{J2246met}
\end{figure*}{}

\begin{figure*}
    \centering
    \includegraphics[trim=0 0 0 0,clip,width=1.1\hsize]{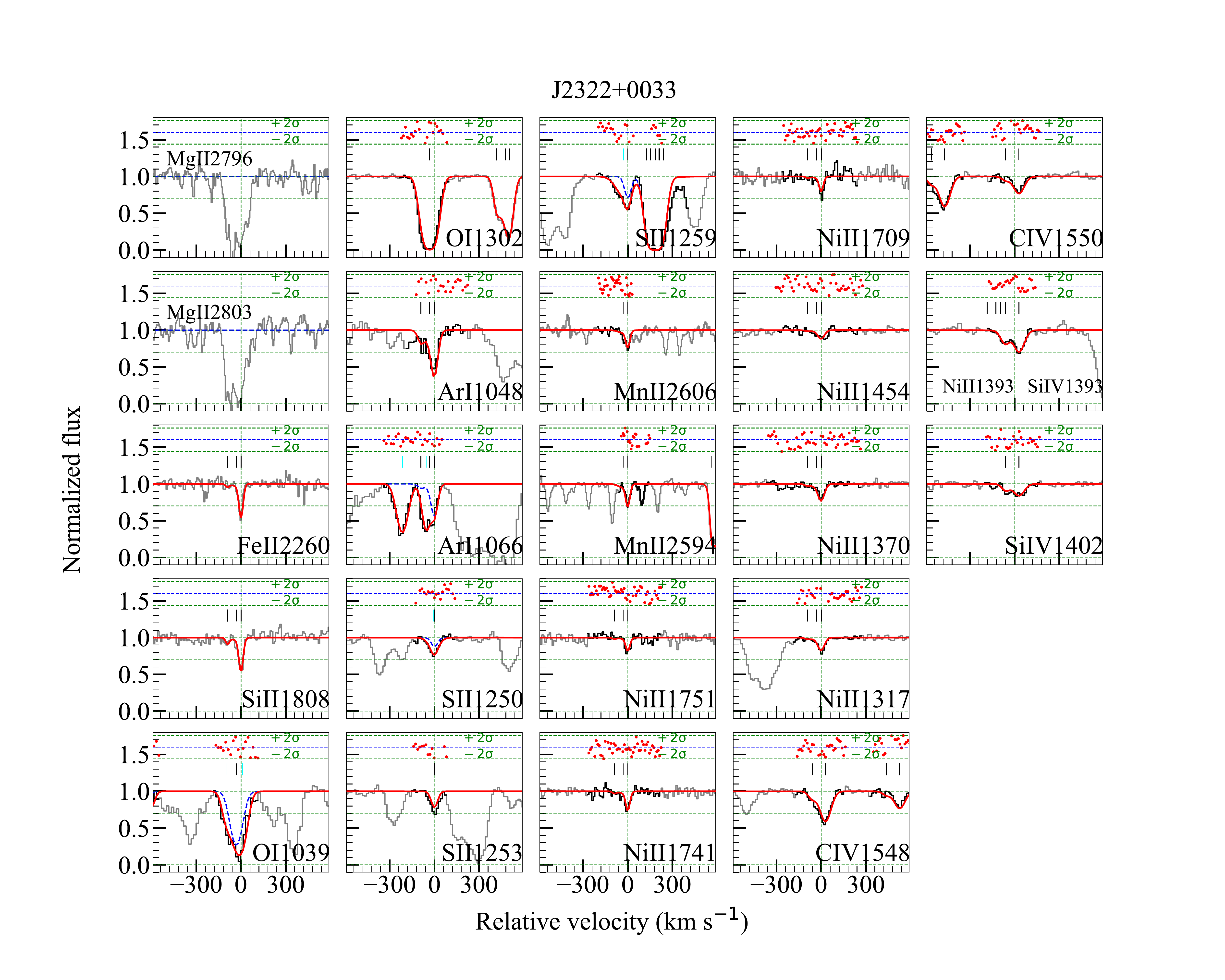}
    \caption{Different ionisation metal lines (\MgII, \FeII, \SiII, \OI, \ArI, \SII, \MnII, \NiII, \CIV,\, and \SiIV) associated with the $\zabs=2.477$ ESDLA system towards QSO SDSS J2322$+$0033. The legends as well as the assumption on sub-component redshifts and $b$-values are the same as in Fig.~\ref{J0017met}.}
    \label{J2322met}
\end{figure*}{}

\end{appendix}

\end{document}